\documentclass[12pt]{article}
\pdfoutput=1
\usepackage{amsmath,amssymb,amsthm,bm,graphicx,float,array,multirow,multicol,rotfloat,caption,subcaption,hyperref,cleveref,enumerate,geometry,mathdots,adjustbox,booktabs,parskip}
\usepackage{pdflscape}
\usepackage[utf8]{inputenc}
\usepackage{csquotes}
\usepackage[para]{threeparttable}
\usepackage[usenames,dvipsnames]{xcolor}
\usepackage[all]{xy}
\usepackage[normalem]{ulem}
\usepackage[numbers,sort&compress]{natbib}
\numberwithin{equation}{section}
\numberwithin{figure}{section}
\makeatletter

\theoremstyle{plain}
\newtheorem*{thm*}{Theorem}
\newtheorem{thm}{Theorem}[section]

\newtheorem{alg}[thm]{Algorithm}

\theoremstyle{definition}

\newtheorem*{defn*}{Definition}

\makeatother

\newcommand{\verteq}{\rotatebox{90}{$\,=$}}
\newcommand{\equalto}[2]{\underset{\scriptstyle\overset{\mkern4mu\verteq}{#2}}{#1}}

\begin{document}

\begin{titlepage}
\vspace*{-3cm} 
\begin{flushright}
{\tt CALT-TH-2021-001}\\
\end{flushright}
\begin{center}
\vspace{2cm}
{\LARGE\bfseries Two 6d origins of 4d SCFTs: \\ class $\mathcal{S}$ and 6d $(1,0)$ on a torus\\}
\vspace{1.2cm}
{\large
Florent Baume$^{1}$, Monica Jinwoo Kang$^{2,3}$, and Craig Lawrie$^1$\\}
\vspace{.7cm}
{ $^1$ Department of Physics and Astronomy, University of Pennsylvania}\par
{Philadelphia, PA 19104, U.S.A.}\par
\vspace{.2cm}
{ $^2$ Walter Burke Institute for Theoretical Physics, California Institute of Technology}\par
{Pasadena, CA 91125, U.S.A.}\par
\vspace{.2cm}
{ $^3$ Department of Physics, Korea Advanced Institute of Science and Technology}\par
{Daejeon 34141, Republic of Korea}\par
\vspace{.2cm}

\vspace{.3cm}

\scalebox{.95}{\tt  fbaume@sas.upenn.edu, monica@caltech.edu, craig.lawrie1729@gmail.com}\par
\vspace{1.2cm}
\end{center}
We consider all 4d $\mathcal{N}=2$ theories of class $\mathcal{S}$ arising from the compactification of exceptional 6d $(2,0)$ SCFTs on a three-punctured sphere with a simple puncture. We find that each of these 4d theories has another origin as a 6d $(1,0)$ SCFT compactified on a torus, which we check by identifying and comparing the central charges and the flavor symmetry. Each 6d theory is identified with a complex structure deformation of $(\mathfrak{e}_n,\mathfrak{e}_n)$ minimal conformal matter, which corresponds to a Higgs branch renormalization group flow. We find that this structure is precisely replicated by the partial closure of the punctures in the class $\mathcal{S}$ construction. We explain how the plurality of origins makes manifest some aspects of 4d SCFTs, including flavor symmetry enhancements and determining if it is a product SCFT. We further highlight the string theoretic basis for this identification of 4d theories from different origins via mirror symmetry.

\vfill 
\end{titlepage}

\tableofcontents
\newpage
\section{Introduction}

Motivated by string theory and its landscape, it has long been a source of deep physical insight to understand how to construct lower-dimensional field theories from higher-dimensional theories via compactification. Utilizing the powerful technique of geometric engineering, the compactifying space provides valuable geometric and topological data that capture some physical aspects of the lower-dimensional theory. 

A quintessential example of such a geometrization is an understanding of the Montonen--Olive $SL(2,\mathbb{Z})$ duality group of 4d $\mathcal{N}=4$ super-Yang--Mills. The duality interchanges strong and weak coupling and this makes any field-theoretic verification difficult due to the lack of computational control in strong-coupling regime. From the 6d perspective, the 4d $\mathcal{N}=4$ theory is obtained by compactifying the 6d $(2,0)$ superconformal field theory (SCFT) on a torus, and the $SL(2,\mathbb{Z})$ is made manifest as the symmetry acting on the complex structure parameter of the torus \cite{Witten:1995zh}. 

A generalization of this example is the class $\mathcal{S}$ construction of 4d $\mathcal{N}=2$ SCFTs. Theories of class $\mathcal{S}$ \cite{Gaiotto:2009hg,Gaiotto:2009we} are those obtained by taking a twisted compactification of the 6d $(2,0)$ SCFT of type $\mathfrak{g}$ on a punctured Riemann surface. Regular punctures are associated to codimension-two defect operators in the 6d theory and can be characterized by nilpotent orbits of $\mathfrak{g}$.\footnote{We focus mainly on regular punctures in this paper.} Let 
\begin{equation}
    \mathcal{S}_\mathfrak{g}\langle C_{g,n}\rangle\{Y_1, \cdots , Y_n\} \,
\end{equation}
denote the 4d $\mathcal{N}=2$ theory obtained from twisted compactifications of the 6d $(2,0)$ SCFT of type $\mathfrak{g}$ on an $n$-punctured Riemann surface of genus $g$, where the punctures are labeled by nilpotent orbits $Y_i$. Physical properties of the 4d theories including the central charges and the dimensions of the moduli space are encoded in the geometry of $C_{g,n}$ and the $Y_i$. When $\mathfrak{g} = \mathfrak{su}_n$, we can think of the class $\mathcal{S}$ construction in terms of the worldvolume theory on a stack of M5-branes wrapping the punctured Riemann surface.

\begin{figure}[H]
    \centering
    \includegraphics[scale=1.8]{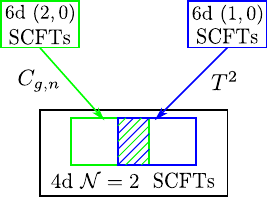}
    \caption{The 6d $(2,0)$ SCFTs compactified on a punctured Riemann surface $C_{g,n}$ give rise to some 4d $\mathcal{N}=2$ SCFTs, as depicted in green. On the other hand, the 6d $(1,0)$ SCFTs compactified on a $T^2$ also give rise to 4d $\mathcal{N}=2$ SCFTs, which is depicted in blue. These two sets of 4d $\mathcal{N}=2$ theories via 6d compactifications may have overlaps, and we shade these 4d theories with two different origins. The shaded area is the core interest of this paper.}\label{fig:introfig}
\end{figure}

Another way of constructing 4d $\mathcal{N}=2$ SCFTs is utilizing the toolkit of F-theory. More specifically, we consider a 6d $(1,0)$ SCFT, via Calabi--Yau compactifications of F-theory with geometric engineering, and further compactify on a torus to get a 4d $\mathcal{N}=2$ theory. Considering string dualities, in this case the duality between M-theory and F-theory, leads to the expectation that some 4d $\mathcal{N}=2$ theories will have realizations as both 6d $(2,0)$ and 6d $(1,0)$ compactifications, as depicted in Figure \ref{fig:introfig}. In this paper, we study theories which have both such origins, and we find that 4d physical properties, such as flavor symmetries, that are manifest in one origin can be obscured in the other. The plurality of 6d origins reveals a richer understanding of the 4d $\mathcal{N}=2$ theories, which we highlight later in the paper.

Let $\mathcal{T}_\mathfrak{g}$ denote the 6d $(1,0)$ SCFT known as the minimal $(\mathfrak{g}, \mathfrak{g})$ conformal matter,\footnote{This theory can also be called $(G, G)$ conformal matter. We use the algebra notation throughout this paper.} which arises as the worldvolume theory of an M5-brane probing an ADE-singularity \cite{DelZotto:2014hpa}. A class of interacting 6d $(1,0)$ fixed points can be obtained by performing Higgs branch renormalization group flows from $\mathcal{T}_\mathfrak{g}$. Certain flows can be specified by a choice of nilpotent orbits $Y_1$ and $Y_2$ inside the $\mathfrak{g} \oplus \mathfrak{g}$ flavor symmetry \cite{Heckman:2016ssk}. Let $\mathcal{T}_\mathfrak{g}\{Y_1, Y_2\}$ be the 6d $(1,0)$ SCFT obtained by Higgsing $\mathcal{T}_\mathfrak{g}$ by the nilpotent orbits $Y_1$ and $Y_2$. The theories $\mathcal{T}_\mathfrak{g}\{Y_1, Y_2\}$ are \emph{very Higgsable}, which means that one can repeatedly Higgs the theory and flow into the infrared until one ends up with a theory of free hypermultiplets.

On the tensor branch, the 6d $(1,0)$ theory $\mathcal{T}_\mathfrak{g}$ consists of a number of tensor multiplets, vector multiplets, and hypermultiplets. The 6d $(1,0)$ tensor multiplet $(B^+, \chi, \sigma)$ contains a self-dual two-form, a Weyl fermion, and a scalar field, whereas the vector multiplet $(A, \lambda)$ contains a gauge potential and a Weyl fermion. The hypermultiplet $(\zeta, \phi)$ contains a Weyl fermion and a scalar. These multiplets are summarized in Table \ref{tb:fieldcontent}. When compactified on a $T^2$, we get a 4d $\mathcal{N}=2$ theory which we denote as $\mathcal{T}_\mathfrak{g}\langle T^2 \rangle$. The complex dimension of the 4d Coulomb branch is given as the number of tensor multiplets plus the ranks of the gauge algebras \cite{Ohmori:2015pua}. 

\begin{figure}[H]
    \centering
    \includegraphics[scale=1.5]{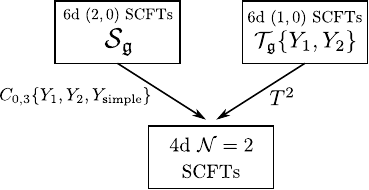}
    \caption{The 4d SCFTs arising from 6d $(2,0)$ SCFTs of type $\mathfrak{g}$ on a Riemann surface $C_{0,3}$, which corresponds to a 3-punctured sphere where one of the punctures is a simple puncture, have another origin from 6d $(1,0)$ SCFTs via compactifying the nilpotent Higgs branch deformations of minimal conformal matter on a $T^2$.}
    \label{fig:introfig2}
\end{figure}

In this paper, we provide an abundance of evidence for the following statements:
\begin{enumerate}
    \item The 4d $\mathcal{N} = 2$ theories obtained from both the 6d $(2,0)$ SCFT and the 6d $(1,0)$ SCFT origins are related via
    \begin{equation}
    \label{eqn:big}
    \mathcal{S}_\mathfrak{g}\langle C_{0,3}\rangle\{Y_1, Y_2, Y_\text{simple}\} = \mathcal{T}_\mathfrak{g}\{Y_1, Y_2\}\langle T^2 \rangle \,
    \end{equation}
    for all possible nilpotent orbits $Y_1$ and $Y_2$ of $\mathfrak{g}$. We take $\mathcal{S}_\mathfrak{g}\langle C_{0,3}\rangle\{Y_1, Y_2, Y_\text{simple}\}$ to be 4d interacting SCFTs arising from the 6d $(2,0)$ construction after subtracting off any free hypermultiplets. The $Y_\text{simple}$ represents the subregular nilpotent orbit of $\mathfrak{g}$, corresponding to the simplest non-trivial puncture.
    \item If $Y_1 > \widetilde{Y}_1$ in the hierarchy of nilpotent orbits, then
    \begin{align}
        \mathcal{T}_\mathfrak{g}\{Y_1, Y_2\} \rightarrow \mathcal{T}_\mathfrak{g}\{\widetilde{Y}_1, Y_2\}
    \end{align}
    is a Higgs branch renormalization group flow between the two 6d SCFTs. A partial closure of punctures between the 4d SCFTs is then given by
    \begin{align}
        \mathcal{S}_\mathfrak{g}\langle C_{0,3}\rangle\{Y_1, Y_2, Y_\text{simple}\} \rightarrow \mathcal{S}_\mathfrak{g}\langle C_{0,3}\rangle\{\widetilde{Y}_1, Y_2, Y_\text{simple}\} .
    \end{align}
\end{enumerate}
This correspondence is depicted in Figure \ref{fig:introfig2}.\\

The two important quantities that we compare between the theories are the central charges $a$ and $c$, defined as the coefficients appearing in the trace of the energy-momentum tensor:
\begin{align}
    \langle T_\mu^\mu \rangle = \frac{c}{16\pi^2} W_{\mu\nu\rho\sigma}W^{\mu\nu\rho\sigma} - \frac{a}{16\pi^2} E_4  + \cdots \,,
\end{align}
where $W_{\mu\nu\rho\sigma}$ is the Weyl tensor, $E_4$ is the Euler density, and $\cdots$ represents terms associated to flavor symmetries. It is often useful to write the linear combinations of these central charges
\begin{equation}
    \begin{cases}
        n_v &= 4(2a - c), \cr
        n_h &= -4(4a - 5c).
    \end{cases}
\end{equation}
If there exists a weakly-coupled description of the theory, then these quantities provide the number of vector multiplets and the number of hypermultiplets in the theory, respectively. A 4d $\mathcal{N}=2$ vector multiplet associated to a simple gauge algebra $\mathfrak{g}$ contains the field content $(A_\mu, \lambda_\alpha, \varphi)$, where $A_\mu$ is the vector potential for the gauge field, $\lambda$ is a spinor transforming in the doublet of the $SU(2)_R$ R-symmetry, and $\varphi$ is a scalar. All three fields transform in the adjoint representation of $\mathfrak{g}$. A hypermultiplet, which is associated to an algebra $\mathfrak{g}$ and a pseudo-real representation $\bm{R}$ of $\mathfrak{g}$, contains the fields $(\psi_\alpha, q)$, where $\psi$ is a spinor and $Q$ a quaternionic scalar field transforming as an $SU(2)_R$ doublet. Each field in the hypermultiplet transforms in the representation $\bm{R}$. If the representation $\bm{R}$ is irreducible, then such a hypermultiplet is called a half-hypermultiplet. If $\bm{R}$ is reducible (for example, when $\bm{R} = \bm{S} \oplus \bm{\overline{S}}$ for a complex representation $\bm{S}$), then the hypermultiplet is called a full hypermultiplet.

\begin{table}[H]
    \centering
    \begin{threeparttable}
    \begin{tabular}{c|ccc}
    \toprule
     & 6d $(2,0)$ & 6d $(1,0)$ & 4d $\mathcal{N}=2$ \\\midrule
     Vector multiplet & --- & $(A, \lambda)$ &  $(A, \lambda, \varphi)$ \\
     Hypermultiplet &  --- & $(\zeta, \phi)$ & $(\psi, q)$ \\
     Tensor multiplet & $(B^+, \psi, \phi)$ & $(B^+, \chi, \sigma)$ & --- \\
     \bottomrule
    \end{tabular}
    \end{threeparttable}
    \caption{The field contents of 6d $(2,0)$, 6d $(1,0)$, and 4d $\mathcal{N}=2$ theories.}
    \label{tb:fieldcontent}
\end{table}

We are specifically focusing on the 4d theories arising as compactifications of the 6d $(2,0)$ theory on $C_{0,3}$, which implies that the space of exactly marginal couplings of such 4d theories is trivial. For each simple factor $\mathfrak{f}_i$ in the flavor symmetry $\mathfrak{f}$, there is a flavor central charge $k_i$ defined as the coefficient appearing in the two-point function of flavor currents $J$ \cite{Osborn:1993cr}. In our normalization, the two point-function of flavor currents is
\begin{equation}
    \left<J_\mu^i(x) J_\nu^j(0)\right> = \frac{3k_i}{4\pi^4} \,\delta^{ij} \frac{x^2\eta_{\mu\nu} - 2x_\mu x_\nu}{x^8}\,.
\end{equation}
We identify the 6d $(1,0)$ origin of each 4d $\mathcal{N}=2$ SCFT from 6d $(2,0)$ theories (from tinkertoy models first described in \cite{Chacaltana:2010ks}) by these quantities: the central charges ($a$, $c$, and $k_i$), and the flavor symmetries.
If the conformal manifold preserving 4d $\mathcal{N}=2$ symmetry is non-trivial, the central charges ($a$, $c$, and $k_i$) remain the same at every point of the conformal manifold and there can be flavor symmetry enhancements only at points where a sector of the theory becomes free \cite{Beem:2013sza, Beem:2014zpa}.

We now turn to the string theoretic motivation for equation \eqref{eqn:big}. The minimal $(\mathfrak{g}, \mathfrak{g})$ conformal matter theory $\mathcal{T}_\mathfrak{g}$ can be obtained as the worldvolume theory of an M5-brane probing an ADE-singularity. Specifically, consider the 11d spacetime
\begin{equation}
    \mathbb{R}^{1,5} \times \mathbb{R}_\perp \times \mathbb{C}^2/\Gamma_\mathfrak{g} \,,
\end{equation}
where we include a single M5-brane, wrapping the $\mathbb{R}^{1,5}$, placed at the origin of the final factor. This gives rise to an effective 7d super-Yang--Mills theory with a gauge algebra $\mathfrak{g}$ that contains a domain wall on which the conformal matter theory lives \cite{DelZotto:2014hpa}. As a domain wall, the 6d theory naively inherits a flavor symmetry of $\mathfrak{g} \oplus \mathfrak{g}$; however, this can be modified by a choice of boundary conditions in the 7d theory. The $SU(2)_R$ triplet of scalars $\Phi^i$ in the vector multiplet of the 7d super-Yang--Mills theory satisfies a Nahm equation
\begin{equation}
    \partial_A \Phi^i = \epsilon^{ijk}[\Phi^j,\Phi^k] \,,
\end{equation}
where $\partial_A$ is the covariant derivative along the semi-infinite interval. The scalar fields can have poles at a boundary $x^6_b$:
\begin{equation}\label{eqn:bcs}
    \Phi^i \sim \pm \frac{t_\rho^i}{x^6 - x^6_b} \,,
\end{equation}
where $t^i_\rho$ are the generators for an $\mathfrak{su}_2$ subalgebra of $\mathfrak{g}$ specified by the embedding
\begin{align}
    \rho: \mathfrak{su}_2 \rightarrow \mathfrak{g}.
\end{align}
The choice of $\rho$ corresponds to the choice of a nilpotent orbit of $\mathfrak{g}$, by the Jacobson--Morozov theorem, and breaks the flavor group on the M5-brane to the commutant of the embedded $\mathfrak{su}_2$.\footnote{See \cite{Gaiotto:2014lca} for a detailed discussion of these boundary conditions in massive type IIA.}$^{,}$\footnote{More precisely, a nilpotent orbit of $\mathfrak{g}_\mathbb{C}$ corresponds to a choice of homomorphism $\rho: \mathfrak{sl}_2 \rightarrow \mathfrak{g}_\mathbb{C}$, and the commutant of $\text{Im}(\rho)$ inside of $\mathfrak{g}_\mathbb{C}$ defines the remnant flavor algebra.}

We can further consider putting the M5-brane theory on a $T^2$, shrink one of the $S^1$s to get type IIA theory, and then T-dualize on the other $S^1$ to obtain type IIB theory on 
\begin{equation}
    \mathbb{R}^{1,3} \times S^1 \times \mathbb{R}_\perp \times \mathbb{C}^2/\Gamma_\mathfrak{g} \,,
    \label{eq:typeIIBsetup}
\end{equation}
with a single D3-brane on $\mathbb{R}^{1,3}$. This is the type IIB realization of the 6d $(2,0)$ SCFT of type $\mathfrak{g}$, together with a codimension-two defect from the D3-brane. This defect corresponds to a simple puncture \cite{Ohmori:2015pua}. Shrinking the $S^1$ from the equation \eqref{eq:typeIIBsetup}, we obtain a 5d gauge theory with a gauge algebra $\mathfrak{g}$, and the Nahm pole specified by the equation \eqref{eqn:bcs} breaks the flavor symmetry seen on the 4d spacetime down to the commutant of the nilpotent orbit. In the perspective of class $\mathcal{S}$, this is the behavior of a puncture associated to that nilpotent orbit. This duality chain between M-theory and type IIB thus leads us to the equation \eqref{eqn:big}, which we explicitly verify in this paper.

We show explicitly that the equation \eqref{eqn:big} holds for the cases of SCFTs of type $\mathfrak{g}=\mathfrak{e}_6,\,\mathfrak{e}_7,\,\mathfrak{e}_8$ holds and maintains through Higgs flow such that
\begin{align}
\label{eqn:HiggsFlown}
    \mathcal{S}_\mathfrak{g}\langle C_{0,3}\rangle\{\widetilde{Y}_1, Y_2, Y_\text{simple}\} = \mathcal{T}_\mathfrak{g}\{\widetilde{Y}_1, Y_2\}\langle T^2 \rangle \,
\end{align}
for all $\widetilde{Y}_1<Y_1$ in the hierarchy of nilpotent orbits.

To determine 4d $\mathcal{N}=2$ SCFTs have both the 6d $(2,0)$ and 6d $(1,0)$ origins, we compare for each of the resulting 4d theories from two 6d SCFT origins the central charges ($a$ and $c$), the flavor algebra, and the flavor central charges ($k_i$). The computation of these quantities from the 6d $(2,0)$ origin is well-known \cite{Chacaltana:2012zy}, though determining the flavor symmetry involves an explicit determination of the Hall--Littlewood index of the 4d theory. From the 6d $(1,0)$ theory perspective, the 4d central charges can be determined from the 6d anomaly polynomial and the flavor symmetries are obtained from studying the tensor branch geometry. More specifically, we study how the choice of nilpotent orbits that Higgs the $\mathfrak{e}_n \oplus \mathfrak{e}_n$ flavor symmetry of $\mathcal{T}_{\mathfrak{e}_n}$ (for $n=6,7,8$) determine the geometric data associated to the resulting 6d $(1,0)$ SCFT. Once we have found the geometric data describing the tensor branch, we can determine the flavor symmetries and the anomaly polynomial, as we describe in detail later in the paper. We find that the 4d $\mathcal{N}=2$ SCFTs data (the central charges, the flavor symmetry, and the flavor central charges) from both the 6d $(2,0)$ SCFTs and the 6d $(1,0)$ SCFTs origins arrange into identical nilpotent hierarchies, as in equation \eqref{eqn:HiggsFlown}, specified by pairs of nilpotent orbits. This provides compelling evidence for identifying the two origins of 4d SCFTs. We further note that the central charges and the flavor symmetries do not necessarily uniquely specify a 4d $\mathcal{N}=2$ SCFT, as it can be seen from some examples studying the global structure of the flavor group in \cite{Distler:2020tub}. However, the fact that the whole hierarchies of nilpotent orbits, or identically the Higgs flow, match affirms that the two 6d origins considered do provide the same set of 4d $\mathcal{N}=2$ SCFTs.

The structure of this paper is as follows. In Section \ref{sec:classsintro}, we review the class $\mathcal{S}$ construction of 4d $\mathcal{N}=2$ SCFTs from the 6d $(2,0)$ theory. In Section \ref{sec:geom-config}, we describe the geometric construction of 6d $(1,0)$ SCFTs from F-theory, and in Section \ref{sec:flavor} we describe how to determine the non-Abelian flavor symmetry of such a 6d $(1,0)$ SCFT. In Section \ref{sec:anom}, we explain how to determine the 6d $(1,0)$ anomaly polynomials from the geometric data. We discuss the compactifications of 6d $(1,0)$ SCFTs on $T^2$, and how to determine the 4d physical data from the 6d anomaly polynomial in Section \ref{sec:4dto6d}. 
In Section \ref{sec:Eesults}, we enumerate and determine the flavor symmetries and anomaly polynomials for every 6d $(1,0)$ SCFT $\mathcal{T}_{\mathfrak{e}_n}\{Y_1, Y_2\}$ and their $T^2$ compactifications, and then we match each of those to the class $\mathcal{S}$ theories $\mathcal{S}_{\mathfrak{e}_n}\langle C_{0,3} \rangle \{Y_1, Y_2, Y_\text{simple}\}$.
Based our results, we explain in Section \ref{sec:fenc} how the flavor symmetry enhancement of the 4d $\mathcal{N}=2$ SCFT is manifest from the 6d $(1,0)$ SCFT origin. We further show that the it can be determined whether the 4d $\mathcal{N}=2$ SCFT is a product SCFT directly from the 6d $(1,0)$ origin in Section \ref{sec:product}. We turn our focus to the implications in Section \ref{sec:agt}; in particular, we discuss how our results can be utilized in the context of the AGT correspondence to understand a potential new relationship between 4d and 2d theories. Finally, we explain in Section \ref{sec:mirror} how the two 6d origins for a given 4d $\mathcal{N}=2$ SCFT are related via mirror symmetry.

\section{4d \texorpdfstring{\boldmath{$\mathcal{N}=2$}}{N=2} SCFTs from 6d \texorpdfstring{\boldmath{$(2,0)$}}{(2,0)}: class \texorpdfstring{\boldmath{$\mathcal{S}$}}{S}}\label{sec:classsintro}

A large class of 4d $\mathcal{N}=2$ superconformal field theories, known as class $\mathcal{S}$ \cite{Gaiotto:2009we,Gaiotto:2009hg}, can be constructed as a partially twisted compactification of the 6d $(2,0)$ SCFT of type $\mathfrak{g}$ on a Riemann surface of genus $g$ with punctures, which are codimension-two defects. Each type of regular puncture is associated to a nilpotent orbit of $\mathfrak{g}$.\footnote{We shall consider only untwisted punctures in this paper.} An $n$-punctured Riemann surface of genus $g$ has a (non-unique) pair-of-pants decomposition as the gluing along punctures of three-punctured spheres, as long as 
\begin{align}
    3g - 3 + n \geq 0\, ,
\end{align}
and these three-punctured spheres are known as \emph{fixtures}. Similarly, we slightly abuse notation and refer to the 4d $\mathcal{N}=2$ SCFTs that arise from compactifications of the $(2,0)$ theory of type $\mathfrak{g}$ on a fixture also as fixtures.

The class $\mathcal{S}$ construction gives rise to 4d $\mathcal{N}=2$ theories, where much of the physical data of the 4d theory depends on the geometric properties of $C_{g,n}$. To determine the central charges and the flavor algebras of the relevant fixtures we require the following facts. For a compactification of the 6d $(2,0)$ theory of type $\mathfrak{g}$ on a genus $g$ Riemann surface with untwisted regular punctures $\mathcal{O}_1, \cdots, \mathcal{O}_p$ the relevant 4d field theoretic data can be split up into a global contribution and local contributions from each of the individual punctures \cite{Chacaltana:2012zy}. The dimension of the Coulomb branch of the resulting 4d SCFT is 
\begin{equation}
    \text{dim}(\mathcal{C}) = (g-1)\, \text{dim}(\mathfrak{g}) + \sum_i \delta d(\mathcal{O}_i) \,,
\end{equation}
where $\delta d(\mathcal{O}_i)$ is the contribution from each individual puncture \cite{Chacaltana:2012zy}. To denote a puncture we use $\mathcal{O}_i$ and the associated nilpotent orbit $Y_i$ interchangeably. Of course, $\text{dim}(\mathcal{C})$ is required to be positive\footnote{In fact, the Coulomb branch is graded and we require the dimensions of each graded piece to be non-negative.} for the compactification to give rise to a consistent theory. Similarly, the central charges $(a, c)$ or, equivalently, the effective numbers of vector and hypermultiplets can be determined additively from the individual puncture data:
\begin{equation}\label{eqn:nvnhS}
    \begin{aligned}
        n_v &= (g - 1)\left( \frac{4}{3}h_{\mathfrak{g}}^\vee \text{dim}(\mathfrak{g}) + \text{rank}(\mathfrak{g})\right) + \sum_i \delta n_v(\mathcal{O}_i) \,, \\
        n_h &= (g - 1)\left( \frac{4}{3}h_{\mathfrak{g}}^\vee \text{dim}(\mathfrak{g})\right) + \sum_i \delta n_h(\mathcal{O}_i) \,.
    \end{aligned}
\end{equation}
Furthermore, the manifest flavor symmetry $\mathfrak{f}^{(\text{manifest})}$ can be determined additively from the punctures $\mathcal{O}_1, \cdots, \mathcal{O}_p$ such that
\begin{equation}
    \mathfrak{f} \supseteq \mathfrak{f}^{(\text{manifest})} = \mathfrak{f}(\mathcal{O}_1) \oplus \cdots \oplus \mathfrak{f}(\mathcal{O}_p) \,,
    \label{eqn:manifest-flavor}
\end{equation}
where $\mathfrak{f}(\mathcal{O}_i)$ is the flavor symmetry associated to the $i$th puncture. The actual flavor symmetry of the 4d theory $\mathfrak{f}$ may be larger than the manifest flavor symmetry (i.e. $\mathfrak{f} \supseteq \mathfrak{f}^{(\text{manifest})}$) and this can be determined by the computation of the first terms of the Hall--Littlewood index \cite{Gaiotto:2012uq}. 

\begin{table}[H]
    \centering
    \renewcommand{\arraystretch}{1.1}
    \begin{threeparttable}
        $\begin{array}{cccc}
            \toprule
            \text{Bala--Carter Label} & (\delta n_v, \delta n_h) & \delta d & \text{Flavor Symmetry} \\\midrule
            0 & (588, 624) & 36 & (\mathfrak{e}_6)_{24} \\
            A_1 & (565, 590) & 35 & (\mathfrak{su}_6)_{18} \\
            2A_1 & (548, 568) & 34 & (\mathfrak{so}_7)_{16} \oplus \mathfrak{u}_1 \\
            3A_1 & (533, 549) & 33 & (\mathfrak{su}_3)_{24} \oplus (\mathfrak{su}_2)_{13} \\ 
            A_2 & (521, 536) & 33 & (\mathfrak{su}_3)_{12} \oplus (\mathfrak{su}_3)_{12} \\
            A_2 + A_1 & (510, 523) & 32 & (\mathfrak{su}_3)_{12} \oplus \mathfrak{u}_1 \\
            2A_2 & (484, 496) & 30 & (\mathfrak{g}_2)_{12} \\
             (0, (F_4)_{12}) & (614, 624) & 13 & (\mathfrak{f}_4)_{12} \\\midrule
            E_6(a_1) & (167, 168) & 11 & \varnothing \\
            \bottomrule
        \end{array}$
    \end{threeparttable}
    \caption{The untwisted punctures for the 6d $(2,0)$ theory of type $\mathfrak{e}_6$ that can appear in a fixture together with a simple puncture. The last row is the simple puncture.}
    \label{tbl:e6punct}
\end{table}

The untwisted punctures for the 6d $(2,0)$ theory of type $\mathfrak{e}_6$ were first enumerated in \cite{Chacaltana:2014jba}, for type $\mathfrak{e}_7$ was done in \cite{Chacaltana:2017boe}, and for type $\mathfrak{e}_8$ has been studied in \cite{Chcaltana:2018zag}.\footnote{We only consider the untwisted punctures. If we were to include the twisted punctures, they only exist for $\mathfrak{e}_6$, not $\mathfrak{e}_7$ or $\mathfrak{e}_8$.} 
The punctures were studied in the context of the ``tinkertoy'' program \cite{Chacaltana:2010ks}; all 4d $\mathcal{N}=2$ SCFTs obtained from compactifying the 6d $(2,0)$ SCFT of type $\mathfrak{g}$ on a three-punctured sphere are enumerated in the tinkertoy models for all three types $\mathfrak{e}_6$, $\mathfrak{e}_7$, and $\mathfrak{e}_8$.
We extract the punctures that can appear in a fixture together with a simple puncture such that the 4d $\mathcal{N}=2$ SCFT obtained after compactification of the 6d $(2,0)$ SCFT on this fixture is consistent. These punctures are labeled with the Bala--Carter notation for nilpotent orbits \cite{MR417306,MR417307} and we list them together with their relevant physical data in Tables \ref{tbl:e6punct}, \ref{tbl:e7punct}, and \ref{tbl:e8punct} for the different types $\mathfrak{e}_6$, $\mathfrak{e}_7$, and $\mathfrak{e}_8$, respectively.\\

\begin{table}[H]
    \centering
    \renewcommand{\arraystretch}{1.1}
    \begin{threeparttable}
        $\begin{array}{cccc}
            \toprule
            \text{Bala--Carter Label} & (\delta n_v, \delta n_h) & \delta d & \text{Flavor Symmetry} \\\midrule
            0 & (1533, 1596) & 63 & (\mathfrak{e}_7)_{36} \\
            A_1 & (1498, 1544) & 62 & (\mathfrak{so}_{12})_{28} \\
            2A_1 & (1471, 1508) & 61 & (\mathfrak{so}_9)_{24} \oplus (\mathfrak{su}_2)_{20} \\
            3A_1^{\prime\prime} & (1452, 1488) & 60 & (\mathfrak{f}_4)_{24} \\
            3A_1^\prime & (1448, 1479) & 60 & (\mathfrak{usp}_6)_{20} \oplus (\mathfrak{su}_2)_{19} \\
            A_2 & (1430, 1460) & 60 & (\mathfrak{su}_6)_{20} \\ 
            4A_1 & (1429, 1457) & 59 & (\mathfrak{usp}_6)_{19} \\
            A_2 + A_1 & (1411, 1436) & 59 & (\mathfrak{su}_4)_{18} \oplus \mathfrak{u}_1 \\
            A_2 + 2A_1 & (1394, 1416) & 58 & (\mathfrak{su}_2)_{16} \oplus (\mathfrak{su}_2)_{28} \oplus (\mathfrak{su}_2)_{84} \\
            A_3 & (1343, 1364) & 57 & (\mathfrak{so}_7)_{16} \oplus (\mathfrak{su}_2)_{12} \\
            2A_2 & (1367, 1388) & 57 & (\mathfrak{g}_2)_{16} \oplus (\mathfrak{su}_2)_{36} \\
            A_2 + 3A_1 & (1379, 1400) & 57 & (\mathfrak{g}_2)_{28} \\
            (A_3 + A_1)^{\prime\prime} & (1332, 1352) & 56 & (\mathfrak{so}_7)_{16} \\
            2A_2 + A_1 & (1352, 1370) & 56 & (\mathfrak{su}_2)_{36} \oplus (\mathfrak{su}_2)_{38} \\
            (A_3 + A_1)^{\prime} & (1328, 1345) & 56 & (\mathfrak{su}_2)_{13} \oplus (\mathfrak{su}_2)_{24} \oplus (\mathfrak{su}_2)_{12} \\
            D_4(a_1) & (1316, 1332) & 56 & (\mathfrak{su}_2)_{12} \oplus (\mathfrak{su}_2)_{12} \oplus (\mathfrak{su}_2)_{12} \\
            A_3 + 2A_1 & (1317, 1333) & 55 & (\mathfrak{su}_2)_{13} \oplus (\mathfrak{su}_2)_{24} \\
            D_4(a_1) + A_1 & (1305, 1320) & 55 & (\mathfrak{su}_2)_{12} \oplus (\mathfrak{su}_2)_{12} \\
            A_3 + A_2 & (1294, 1308) & 54 & (\mathfrak{su}_2)_{12} \oplus \mathfrak{u}_1 \\
            A_4 & (1239, 1252) & 53 & (\mathfrak{su}_3)_{12}\oplus \mathfrak{u}_1 \\
            A_5^{\prime\prime} & (1132, 1144) & 48 & (\mathfrak{g}_2)_{12} \\
            (0, E_6) & (1588, 1596) & 64 & (\mathfrak{e}_6)_{12} \\
            (3A_1^{\prime\prime}, F_4) & (1695, 1704) & 69 & (\mathfrak{f}_4)_{12} \\\midrule
            E_7(a_1) & (383, 384) & 17 & \varnothing \\
            \bottomrule
        \end{array}$
    \end{threeparttable}
    \caption{The punctures for the 6d $(2,0)$ theory of type $\mathfrak{e}_7$ that can appear in a fixture together with a simple puncture. The last row is the simple puncture itself.}
    \label{tbl:e7punct}
\end{table}

\begin{table}[H]
    \centering
    \footnotesize
    \begin{threeparttable}
        \resizebox{!}{0.47\textheight}{
        $\begin{array}{cccc}
            \toprule
            \text{Bala--Carter Label} & (\delta n_v, \delta n_h) & \delta d & \text{Flavor Symmetry} \\\midrule
            0 & (4840, 4960) & 120 & (\mathfrak{e}_8)_{60} \\
            A_1 & (4781, 4872) & 119 & (\mathfrak{e}_7)_{48} \\
            2A_1 & (4734, 4808) & 118 & (\mathfrak{so}_{13})_{40} \\
            3A_1 & (4695, 4759) & 117 & (\mathfrak{f}_4)_{36} \oplus (\mathfrak{su}_2)_{31} \\
            A_2 & (4665, 4728) & 117 & (\mathfrak{e}_6)_{36} \\
            4A_1 & (4660, 4716) & 116 & (\mathfrak{usp}_{8})_{31} \\
            A_2 + A_1 & (4630, 4682) & 116 & (\mathfrak{su}_6)_{30} \\
            A_2 + 2A_1 & (4601, 4648) & 115 & (\mathfrak{so}_{7})_{28} \oplus (\mathfrak{su}_2)_{144} \\
            A_2 + 3A_1 & (4574, 4617) & 114 & (\mathfrak{g}_2)_{48} \oplus (\mathfrak{su}_2)_{25} \\
            2A_2 & (4550, 4592) & 114 & (\mathfrak{g}_2)_{24} \oplus (\mathfrak{g}_2)_{24} \\
            A_3 & (4514, 4560) & 114 & (\mathfrak{so}_{11})_{28} \\
            2A_2 + A_1 & (4527, 4566) & 113 & (\mathfrak{g}_2)_{24} \oplus (\mathfrak{su}_2)_{62} \\
            A_3 + A_1 & (4487, 4525) & 113 & (\mathfrak{so}_{7})_{24} \oplus (\mathfrak{su}_2)_{21} \\
            D_4(a_1) & (4467, 4504) & 113 & (\mathfrak{so}_{8})_{24} \\
            2A_2 + 2A_1 & (4504, 4540) & 112 & (\mathfrak{usp}_{4})_{62} \\
            A_3 + 2A_1 & (4464, 4498) & 112 & (\mathfrak{usp}_{4})_{21} \oplus (\mathfrak{su}_2)_{40} \\
            D_4(a_1) + A_1 & (4444, 4476) & 112 & (\mathfrak{su}_2)_{20} \oplus  (\mathfrak{su}_2)_{20} \oplus  (\mathfrak{su}_2)_{20} \\
            D_4 & (4236, 4272) & 108 & (\mathfrak{f}_4)_{24} \\
            A_3 + A_2 & (4425, 4456) & 111 & (\mathfrak{usp}_{4})_{20} \oplus \mathfrak{u}_1 \\
            A_4 & (4330, 4360) & 110 & (\mathfrak{su}_{5})_{20} \\
            A_3 + A_2 + A_1 & (4406, 4435) & 110 & (\mathfrak{su}_2)_{384} \oplus (\mathfrak{su}_2)_{19} \\
            D_4(a_1) + A_2 & (4388, 4416) & 110 & (\mathfrak{su}_3)_{96} \\
            A_4 + A_1 & (4311, 4337) & 109 & (\mathfrak{su}_3)_{18} \oplus \mathfrak{u}_1 \\
            D_4 + A_1 & (4213, 4241) & 107 & (\mathfrak{usp}_6)_{19} \\
            D_5(a_1) & (4195, 4220) & 107 & (\mathfrak{su}_4)_{18} \\
            2A_3 & (4324, 4350) & 108 & (\mathfrak{usp}_{4})_{31} \\
            A_4 + 2A_1 & (4294, 4318) & 108 & (\mathfrak{su}_2)_{30} \oplus \mathfrak{u}_1 \\
            A_4 + A_2 & (4265, 4288) & 107 & (\mathfrak{su}_2)_{16} \oplus (\mathfrak{su}_2)_{200} \\
            D_5(a_1) + A_1 & (4178, 4200) & 106 & (\mathfrak{su}_2)_{16} \oplus (\mathfrak{su}_2)_{112} \\
            A_4 + A_2 + A_1 & (4250, 4272) & 106 & (\mathfrak{su}_2)_{200} \\
            D_4 + A_2 & (4163, 4184) & 105 & (\mathfrak{su}_3)_{28} \\
            A_5 & (4127, 4149) & 105 & (\mathfrak{g}_2)_{16} \oplus (\mathfrak{su}_2)_{13} \\
            E_6(a_3) & (4115, 4136) & 105 & (\mathfrak{g}_2)_{16} \\
            D_5 & (3884, 3904) & 100 & (\mathfrak{so}_{7})_{16} \\
            A_4 + A_3 & (4184, 4204) & 104 & (\mathfrak{su}_2)_{124} \\
            A_5 + A_1 & (4112, 4131) & 104 & (\mathfrak{su}_2)_{38} \oplus (\mathfrak{su}_2)_{13} \\
            D_5(a_1) + A_2 & (4136, 4155) & 104 & (\mathfrak{su}_2)_{75} \\
            D_6(a_2) & (4088, 4106) & 104 & (\mathfrak{su}_2)_{13} \oplus (\mathfrak{su}_2)_{13} \\
            E_6(a_3) + A_1 & (4100, 4118) & 104 & (\mathfrak{su}_2)_{38} \\
            E_7(a_5) & (4076, 4093) & 104 & (\mathfrak{su}_2)_{13} \\
            E_8(a_7) & (4064, 4080) & 104 & \varnothing \\
            A_6 & (3905, 3920) & 99 & (\mathfrak{su}_2)_{12} \oplus (\mathfrak{su}_2)_{60} \\
            D_5 + A_1 & (3869, 3885) & 99 & (\mathfrak{su}_2)_{13} \oplus (\mathfrak{su}_2)_{24} \\
            D_6(a_1) & (3857, 3872) & 99 & (\mathfrak{su}_2)_{12} \oplus (\mathfrak{su}_2)_{12} \\
            E_7(a_4) & (3846, 3860) & 98 & (\mathfrak{su}_2)_{12} \\
            E_6(a_1) & (3675, 3688) & 95 & (\mathfrak{su}_3)_{12} \\
            E_6 & (3220, 3232) & 84 & (\mathfrak{g}_2)_{12} \\
            (A_1, E_7) & (5014, 5048) & 122 & (\mathfrak{e}_7)_{24} \\
            (D_4, F_4) & (5640, 5648) & 136 & (\mathfrak{f}_4)_{12} \\
            (A_2, E_6) & (5185, 5192) & 125 & (\mathfrak{e}_6)_{12} \\
            (0, E_7) & (4955, 4960) & 121 & (\mathfrak{e}_7)_{12} \\\midrule
            E_8(a_1) & (1079, 1080) & 29 & \varnothing \\
            \bottomrule
        \end{array}$
        }
    \end{threeparttable}
    \caption{The punctures for the 6d $(2,0)$ theory of type $\mathfrak{e}_8$ that can appear in a fixture together with a simple puncture. The simple puncture is written in the final row.}
    \label{tbl:e8punct}
\end{table}

There exists a partial ordering on the nilpotent orbits for a given semi-simple Lie algebra, which translates into a partial ordering on the possible punctures.\footnote{See \cite{MR1251060} for more details on nilpotent orbits of Lie algebras.} If $Y_1$ and $\widetilde{Y}_1$ are related as $Y_1 > \widetilde{Y}_1$ under this partial ordering, then one can consider the partial closure of the puncture $Y_1$ to $\widetilde{Y}_1$. That is, we can begin with a class $\mathcal{S}$ theory
\begin{equation}
    \mathcal{S}_\mathfrak{g}\langle  C_{g,n}\rangle \{Y_1, \cdots\} \,,
\end{equation}
and give a nilpotent vacuum expectation value to the moment-map operator, the scalar, highest-weight component of the flavor supermultiplet, and then flow into the infrared to obtain the class $\mathcal{S}$ theory
\begin{equation}
    \mathcal{S}_\mathfrak{g}\langle  C_{g,n}\rangle \{\widetilde{Y}_1, \cdots\} \,.
\end{equation}
This procedure is described in \cite{Tachikawa:2015bga} and we review it briefly here.

Consider a 4d $\mathcal{N}=2$ theory of class $\mathcal{S}$ where one of the punctures has an associated flavor symmetry $\mathfrak{f}$. There exists a scalar field $\phi$, transforming in the triplet representation of the $SU(2)_R$, which is the superconformal primary of the multiplet in which the flavor current of $\mathfrak{f}$ lives. Let $\phi^+$ be the chiral operator corresponding to the highest weight of the $\bm{3}$. We can give a nilpotent vacuum expectation value to $\phi^+$,
\begin{equation}\label{eqn:vev}
    \langle \phi^+ \rangle = J_Y \,,
\end{equation}
where $J_Y$ is the Jordan normal form associated to nilpotent element $Y$. Each nilpotent element of a simple Lie algebra $\mathfrak{f}$ is associated to a homomorphism $\rho_Y: \mathfrak{su}_2 \rightarrow \mathfrak{f}$. Giving $\phi^+$ the vev as in the equation \eqref{eqn:vev} breaks the flavor symmetry 
\begin{equation}
    \mathfrak{f} \rightarrow \mathfrak{f}^\prime \,,
\end{equation}
where $\mathfrak{f}^\prime$ is the centralizer of $\rho_Y$. The breaking of the flavor symmetry leaves behind Goldstone bosons, and their superpartners, however we can then flow into the infrared and end up at a new interacting fixed point with flavor symmetry $\mathfrak{f}^\prime$.\footnote{There is no guarantee that one ends up with an interacting SCFT after this process, however the cases that we consider do. See \cite{Tachikawa:2015bga} for more details.}

\section{Geometric construction of 6d \texorpdfstring{\boldmath{$(1,0)$}}{(1,0)} SCFTs}\label{sec:geom-config}

We discussed the construction of 4d $\mathcal{N} = 2$ theories arising as compactifications of the 6d $(2,0)$ SCFTs of type $\mathfrak{g}$ in Section \ref{sec:classsintro}. There are two infinite series of such 6d $(2,0)$ SCFTs and they are respectively associated to $\mathfrak{su}_n$ and $\mathfrak{so}_{2n}$ Lie algebras; similarly, there are three sporadic 6d $(2,0)$ SCFTs that are associated to $\mathfrak{e}_6$, $\mathfrak{e}_7$, and $\mathfrak{e}_8$, respectively. The 4d $\mathcal{N} = 2$ theories obtained from these 6d $(2,0)$ SCFTs depend on this Lie algebra $\mathfrak{g}$ together with the genus $g$ of the Riemann surface and the data describing the punctures $\mathcal{O}_i$\,:
\begin{equation}
    (\mathfrak{g}, g, \{\mathcal{O}_i\}) \,.
\end{equation}

Generally speaking, 4d $\mathcal{N}=2$ SCFTs may have an alternate construction via a $T^2$ compactification from 6d $(1,0)$ SCFTs, which carry half the amount of supersymmetry than the 6d $(2,0)$ SCFTs. In this case, the properties of the 4d SCFTs obtained from the compactification on $T^2$ depends principally on the data of their 6d $(1,0)$ SCFT origins.\footnote{One may turn on certain kinds of 4d $\mathcal{N}=2$ supersymmetry-preserving Wilson lines along the $T^2$ directions; however, these are not considered in this paper.} The world of 6d $(1,0)$ theories is far more expansive than that of 6d $(2,0)$ theories, and it has a great number of infinite series. To understand which 4d $\mathcal{N}=2$ SCFTs can be obtained from 6d $(1,0)$ SCFTs, we first explain the geometric construction of the 6d $(1,0)$ theories.

More precisely, we take the F-theory \cite{Vafa:1996xn,Morrison:1996na,Morrison:1996pp} perspective and build a geometric construction for these 6d $(1,0)$ theories. By utilizing a geometric-engineering point of view, F-theory compactified on a non-compact elliptically-fibered Calabi--Yau threefold gives rise to a tensor branch description of a six-dimensional field theory with eight supercharges. In particular, when the base of the Calabi--Yau threefold is non-compact (and thus gravity is decoupled), then the resulting theory can be a 6d $(1,0)$ superconformal field theory. 

The 6d $(1,0)$ SCFTs constructed from F-theory compactifications are related via renormalization group flows corresponding to complex structure deformations of the non-compact Calabi--Yau threefolds.
In this paper, we show that these complex structure deformations are related to similar deformations between the 4d $\mathcal{N}=2$ SCFTs obtained by a $T^2$ compactification. These deformations reproduce those associated to partial closure of the punctures, which is visible from the class $\mathcal{S}$ construction of the same 4d SCFTs.

\subsection{Geometric configurations and 6d SCFTs}

Each 6d $(1,0)$ SCFT is associated to a configuration of compact rational curves $C_i$ together with a choice of a (possibly trivial) gauge algebra $\mathfrak{g}_i$ attached to each curve \cite{Heckman:2013pva,Heckman:2015bfa}. The intersection matrix of the curves must be negative-definite
\begin{equation}
    A^{ij} = C^{i} \cdot C^j \prec 0 \,.
\end{equation}
This follows from the requirement that all curves can simultaneously be shrunk to zero-volume, which is a necessity for the existence of an interacting SCFT. Each curve has a self-intersection number 
\begin{equation}
    C^i \cdot C^i = -n_i \,,
\end{equation}
where $n_i$ is strictly positive, and each curve $C_i$ can only intersect a different curve $C_{j \neq i}$  in at most one point. The data of a 6d SCFT can then be represented by writing the $n_i$ of each $C_i$ such that they are adjacent if the two curves intersect, and writing each associated gauge algebra above each $n_i$. As an example, we can consider three curves $C_1$, $C_2$, and $C_3$ with the intersection matrix
\begin{equation}
A=
    \begin{pmatrix}
      -1 & 1 & 0 \\
      1 & -3 & 1 \\
      0 & 1 & -1 
    \end{pmatrix} \,,
\end{equation}
and take the gauge algebras over each curve to be 
\begin{align}
    \mathfrak{g}_1 = \mathfrak{g}_3 = \varnothing\,,\quad \mathfrak{g}_2 = \mathfrak{su}_3. 
\end{align}
This SCFT can then be compactly written via the notation
\begin{equation}
    1\,\overset{\mathfrak{su}_3}{3}\,1 \,,
\end{equation}
where we have dropped trivial gauge algebras from the notation. In addition to the negative-definiteness requirement, it is necessary that the singular fiber above the points where any two distinct $C_i$ intersect is minimal. The latter is required for one to be at a generic point on the tensor branch. We explain how to construct these curve configurations which satisfy the negative-definiteness and minimality conditions in Section 
\ref{sec:buildbase}.

One way to construct an F-theory realization of these six-dimensional SCFTs is using configurations with intersecting curves whose adjacency matrix is negative-definite so that such configurations can be used to represent the base of a minimal elliptic fibration. The beauty of this geometric presentation of a SCFT is that many of the interesting properties of a SCFT, such as the central charges and anomaly polynomials, can be calculated directly in terms of the geometric data.

\subsection{How to build and decorate a base}\label{sec:buildbase}

Configurations of $\mathbb{P}^1$s, together with the singular fibers above them, that give rise to a 6d SCFT in the limit where 
\begin{equation}
    \text{vol}(\mathbb{P}^1) \rightarrow 0 \,
\end{equation}
for all $\mathbb{P}^1$s are highly constrained. All such configurations can be constructed out of ``building blocks'' that are subjected to the necessary conditions for the resulting theory to be superconformal.

We first form a putative curve configuration by gluing non-Higgsable clusters to E-strings. The non-Higgsable clusters are given as the following configurations of curves \cite{Morrison:2012np}:
\begin{equation}
    \begin{gathered}
      \overset{\mathfrak{su}_3}{3} \,, \quad 
      \overset{\mathfrak{so}_8}{4} \,, \quad 
      \overset{\mathfrak{f}_4}{5} \,, \quad 
      \overset{\mathfrak{e}_6}{6} \,, \quad 
      \overset{\mathfrak{e}_7}{7} \,, \quad 
      \overset{\mathfrak{e}_7}{8} \,, \quad
      \overset{\mathfrak{e}_8}{(12)} \,, \quad
      \overset{\mathfrak{su}_2}{2}\overset{\mathfrak{g}_2}{3} \,, \quad 
      2\overset{\mathfrak{su}_2}{2}\overset{\mathfrak{g}_2}{3} \,, \quad 
      \overset{\mathfrak{su}_2}{2}\overset{\mathfrak{so}_7}{3}\overset{\mathfrak{su}_2}{2} \,,
      \cr 
     \begin{aligned}
         &\phantom{2} \\[-10pt]
         22\cdots 2&2 \,,
    \end{aligned} \quad 
       \begin{aligned}
         &2 \\[-10pt]
         22\cdots 22&22 \,,
    \end{aligned} \quad
      \begin{aligned}
         &2 \\[-10pt]
         22&222\,,
    \end{aligned} \quad \begin{aligned}
         &2 \\[-10pt]
         22&2222\,,
    \end{aligned} \quad \begin{aligned}
         &2 \\[-10pt]
         22&22222\,.
    \end{aligned}
    \end{gathered}
\end{equation}
These non-Higgsable clusters can be glued together by E-strings in the following ways. Let $I_Q$ denote the rank $Q$ E-string, which is associated to the following curve configuration:
\begin{equation}\label{eqn:rankqestring}
    1 \overbrace{2\,2\cdots\,2\,2}^{Q-1}  \,.
\end{equation} 
Gluing together non-Higgsable clusters with an E-string leads to curve configurations of the form
\begin{equation}\label{eqn:glueone}
    \cdots \overset{\mathfrak{g}}{n}I_Q \overset{\mathfrak{h}}{m}\cdots \,.
\end{equation}
If non-Higgsable clusters are glued to both the left and the right of an E-string, then it is a requirement that $Q \leq 2$. There are two choices for $Q$ satisfying this requirement: $Q=1$ or $Q=2$. If $Q = 1$, then we must have
\begin{equation}
    \mathfrak{g} \oplus \mathfrak{h} \subseteq \mathfrak{e}_8 \,,
\end{equation}
whereas if $Q = 2$, then
\begin{equation}
    \mathfrak{g} \subseteq \mathfrak{e}_8 \,, \quad m = 2 \,, \quad \mathfrak{h} = \mathfrak{su}_2 \,.
\end{equation}
We note that these two conditions also must be satisfied (with $\mathfrak{g} = \varnothing$) when the curves to the left of the E-string in the equation \eqref{eqn:glueone} are absent. When the curves to the right of the E-string in the equation \eqref{eqn:glueone} are absent, then gluings are of the form
\begin{equation}\label{eqn:gluetwo}
    \cdots \overset{\mathfrak{g}}{n}I_Q
\end{equation}
and we are now allowed to (a priori) consider arbitrary values of $Q$, as long as $\mathfrak{g} \subseteq \mathfrak{e}_8$ is satisfied. These gluings can be repeated ad nauseam subject to the constraint that the intersection matrix, $A^{ij}$, between the compact curves be negative-definite:
\begin{equation}
    A^{ij} \prec 0 \,.
\end{equation}

In addition to specifying the curves (i.e.~the tensor multiplets) and the algebras (i.e.~the vector multiplets), one must also specify the number of hypermultiplets transforming in each representation $\bm{R}$ of each gauge algebra. Interestingly, it turns out that these are almost uniquely fixed by the choice of $n$ and $\mathfrak{g}$ with rare exceptions.\footnote{There is an example of $\mathfrak{su}_6$ on a $(-1)$-curve where it is not unique.} The prescribed types of hypermultiplets appearing on each curve are summarized in Table 3 of \cite{Heckman:2018jxk}.

Once a curve configuration has been constructed by gluing together the above building blocks, it may admit further ``tunings'' that also lead to an interacting SCFT. A tuning involves a procedure such as
\begin{equation}
    \cdots \overset{\mathfrak{h}}{n}\cdots \quad\rightarrow\quad \cdots \overset{\mathfrak{g}}{n}\cdots
\end{equation}
for $\mathfrak{h} \subset \mathfrak{g}$. The value of $n$ constrains what $\mathfrak{g}$ are permitted, and these tunings are subject to the $\mathfrak{e}_8$ gluing conditions described above. Furthermore, the constraints on the numbers of hypermultiplets at each curve restricts which pairs of gauge algebras can be on adjacent curves; two adjacent curves with gauge algebras $\mathfrak{g}$ and $\mathfrak{h}$ on them must have a single bifundamental (half-)hypermultiplet between them.\footnote{In fact, the hypermultiplet charged under both gauge algebras does not need to be in the fundamental representation of each algebra, but can be in other representations prescribed by the algebras and the self-intersections of the curves supporting them.} 
When tuning a configuration involving a $(-2)$-curve which has no gauge algebra over and is not part of an E-string, another condition must be satisfied: the $(-2)$ curve can only be connected to another $(-2)$-curve with an $\mathfrak{su}_2$ gauge algebra over it.\footnote{These conditions, and the overall construction of such geometric data associated to 6d SCFTs, are reviewed in \cite{Heckman:2018jxk}.}

\subsection{Conformal matter and Higgs branch deformations}\label{sec:def}

When constructing 6d SCFTs following the method described in Section \ref{sec:buildbase}, it becomes apparent that the same patterns of curves and algebras appear abundantly in the geometric data. These are the configurations
\begin{equation}\label{eqn:cms}
    1\overset{\mathfrak{su}_{3}}{3}1 \,,\qquad 1\overset{\mathfrak{su}_{2}}{2}\overset{\mathfrak{so}_{7}}{3}\overset{\mathfrak{su}_{2}}{2}1 \,, \qquad 1\ 2\overset{\mathfrak{su}_{2}}{2}\overset{\mathfrak{g}_{2}}{3}1\overset{\mathfrak{f}_{4}}{5}1\overset{\mathfrak{g}_{2}}{3}\overset{\mathfrak{su}_{2}}{2}2\ 1 \,.
\end{equation}
In their own right, these configurations correspond to nothing other than the minimal $(\mathfrak{e}_n, \mathfrak{e}_n)$ conformal matter \cite{DelZotto:2014hpa}, which we denote as $\mathcal{T}_{\mathfrak{e}_n}$. 
 These 6d $(1,0)$ SCFTs have an $\mathfrak{e}_n \oplus \mathfrak{e}_n$ flavor symmetry which can be determined from the configurations given in equation \eqref{eqn:cms} following the logic described in Section \ref{sec:flavor}.

These theories can also be obtained from M-theory perspective by considering the worldvolume theory on an M5-brane probing $\mathbb{C}^2/\Gamma_{ADE}$. These are the theories of interest in this paper, as it is known that $\mathcal{T}_\mathfrak{g}$ compactified on a $T^2$ gives rise to the same 4d $\mathcal{N}=2$ theory as the 6d $(2,0)$ theory of type $\mathfrak{g}$ compactified on a sphere with two full punctures and a simple puncture \cite{Ohmori:2015pua}. In this paper, we compare Higgs branch deformations of $\mathcal{T}_\mathfrak{g}$ to class $\mathcal{S}$ theories.

It was observed in \cite{Heckman:2016ssk} that a particular class of Higgs branch deformations of 6d conformal matter are specified by a pair of nilpotent orbits $Y_L$ and $Y_R$ of $\mathfrak{g}$. Then, part of the flavor symmetry group of the theory is given by the commutant of the image of the nilpotent orbits in $\mathfrak{g}$. Furthermore, two theories $\mathcal{T}_\mathfrak{g}\{Y_L,Y_R\}$ and $\mathcal{T}_\mathfrak{g}\{\widetilde{Y}_L,Y_R\}$ are related by a Higgs branch deformation if they satisfy $Y_L > \widetilde{Y}_L$ under the standard partial ordering of nilpotent orbits. For long-enough quivers, the flows are in one-to-one correspondence with the nilpotent hierarchy. However, in the case of short quivers such as minimal conformal matter, not all nilpotent orbits can be used, as the breaking patterns on both sides of the quiver become correlated.

\begin{figure}[H]
  \centering
  \begin{subfigure}[b]{\textwidth}
    \centering
    \includegraphics[scale=2.2]{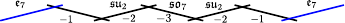}
    \vspace{2mm}
    \caption{The geometric configuration at a generic point of the tensor branch for $(\mathfrak{e}_7, \mathfrak{e}_7)$ conformal matter.}
  \end{subfigure}
  \newline\vspace{0mm}
  \begin{subfigure}[b]{\textwidth}
    \centering
    \includegraphics[scale=2.2]{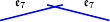}
    \vspace{3mm}
    \caption{The geometric configuration at the origin of the tensor branch for $(\mathfrak{e}_7, \mathfrak{e}_7)$ conformal matter. At the intersection of the two non-compact curves there is a singularity.}
  \end{subfigure} 
  \newline\vspace{2mm}
  \begin{subfigure}[b]{\textwidth}
    \centering
    \includegraphics[scale=2.2]{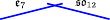}
    \vspace{3mm}
    \caption{A complex structure deformation of the configuration in (b), where the III$^*$ fiber over the right non-compact curve is deformed into an I$_2^*$ fiber. This configuration corresponds to a new SCFT obtained by a Higgs branch RG flow from $(\mathfrak{e}_7, \mathfrak{e}_7)$ conformal matter. This is the origin of the tensor branch, and again,  there is a singularity at the intersection of the two non-compact curves.}
  \end{subfigure}
  \newline\vspace{2mm}
  \begin{subfigure}[b]{\textwidth}
    \centering
    \includegraphics[scale=2.2]{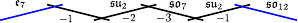}
    \vspace{3mm}
    \caption{The geometric configuration at a generic point of the tensor branch, obtained by blowing up at the singular point where the two non-compact curves intersected, for the singular configuration in (c).}
  \end{subfigure}
  \caption{A depiction of the four steps describing the Higgs branch flow that moves between the two 6d SCFTs,
$
      1\overset{\mathfrak{su}_2}{2}\overset{\mathfrak{so}_7}{3}\overset{\mathfrak{su}_2}{2}1 \rightarrow 1\overset{\mathfrak{su}_2}{2}\overset{\mathfrak{so}_7}{3}\overset{\mathfrak{su}_2}{1} \,,
$
  where we have written the generic tensor branch configurations in (a) and (d) and the singular configuration when all the compact curves are shrunk in (b) and (c). The black curves are compact and the blue curves are non-compact; the Lie algebra denotes the type of singular fiber tuned over that curve, and the number beneath the compact curves is the self-intersection.}\label{fig:FlavHiggsDef}
\end{figure}

The Higgs branch renormalization group flows are triggered by complex structure deformations of the singularities of the associated non-compact elliptically-fibered Calabi--Yau threefold. These deformations are carried out at the origin of the tensor branch, where all of the compact curves have shrunk to zero-volume. An example of such a complex structure deformation and its effect on the geometric data at a generic point of the tensor branch is represented in Figure \ref{fig:FlavHiggsDef}.

The complex structure deformations that we consider have the following effects on the tensor branch. We can think of a class of deformations as Higgsing a gauge algebra over one of the compact curves, which changes the tensor branch as 
\begin{equation}\label{eqn:gt}
    \cdots \overset{\mathfrak{g}}{n}\cdots \quad\rightarrow\quad \cdots\overset{\mathfrak{g}^\prime}{n}\cdots \,,
\end{equation}
where $\mathfrak{g}^\prime \subset \mathfrak{g}$. Another common way in which the tensor branch can be modified by a complex structure deformation is
\begin{equation}\label{eqn:edis1}
    \cdots\overset{\mathfrak{g}}{n}1\overset{\mathfrak{h}}{m}\cdots \quad\rightarrow\quad \cdots\overset{\mathfrak{g}^\prime}{(n-1)}\overset{\mathfrak{h}^\prime}{(m-1)}\cdots\,,
\end{equation}
where $\mathfrak{g}^\prime \subseteq \mathfrak{g}$ and $\mathfrak{h}^\prime \subseteq \mathfrak{h}$. We can consider a similar case where the curve supporting $\mathfrak{h}$ is non-compact. For such a case, we simply observe the tensor branch modification to be the following:
\begin{equation}\label{eqn:edis2}
    \cdots\overset{\mathfrak{g}}{n}1 \quad\rightarrow\quad \cdots\overset{\mathfrak{g}^\prime}{(n-1)}\,.
\end{equation}

The final class of complex structure deformations that we consider involves nucleating E-strings. In this case, we have a rank $Q$ E-string attached to a $(-n)$-curve, and there exists a complex structure deformation partitioning it into lower-rank E-strings. In the heterotic description, this type of deformation corresponds to moving in different branches of the instanton moduli space \cite{Heckman:2016ssk}. For example, we can transform the rank $Q$ E-string into two E-strings of ranks $k$ and $Q-k$:
  \begin{equation}\label{eqn:enuc}
          \overset{\mathfrak{g}}{n}\overbrace{12\cdots 2}^{Q} \quad\rightarrow\quad \overbrace{2\cdots 21}^{k}\overset{\mathfrak{g}}{n}\overbrace{12\cdots 2}^{Q-k}\,.
\end{equation}
The number of possibilities is equal to the number of partitions of $Q$.

The Higgs branch renormalization group flows of 6d $(1,0)$ SCFTs and their connection to the complex structure deformations of the associated Calabi--Yau geometry have been studied in \cite{Heckman:2016ssk,Mekareeya:2016yal,Heckman:2018pqx,Hassler:2019eso}. For the nilpotent Higgsings of minimal conformal matter $\mathcal{T}_\mathfrak{g}\{Y_L, Y_R\}$, the geometric data of the tensor branches all have the form of the modifications shown in the equations \eqref{eqn:gt}, \eqref{eqn:edis1}, \eqref{eqn:edis2}, and \eqref{eqn:enuc} applied repeatedly to the configurations for $\mathcal{T}_\mathfrak{g}$ given in the equation \eqref{eqn:cms}.

\section{Non-Abelian flavor symmetries for 6d \texorpdfstring{\boldmath{$(1,0)$}}{(1,0)} SCFTs}\label{sec:flavor}

In this section, we focus on understanding when non-Abelian flavor symmetries arise for 6d SCFTs and how to determine them. There are two different origins of non-Abelian flavor algebras:\footnote{Here, and throughout this section, we assume that we are considering \emph{strict} $(1,0)$ SCFTs. This removes the need to discuss a third origin of flavor symmetries that only appears for the geometric configurations associated to the 6d $(2,0)$ SCFTs.} either from any E-string building blocks of the 6d SCFT or from the classical flavor symmetry that rotates $k$ $(1,0)$ hypermultiplets transforming in a representation $\bm{R}$ of the gauge algebra.

While in this paper we shall only concern ourselves with non-Abelian flavor symmetries, we note that continuous Abelian flavor symmetries of 6d SCFTs have been explored in \cite{Lee:2018ihr, Apruzzi:2020eqi}. Naively, there appears to be many $U(1)$ factors in the flavor symmetry group. If we have $k$ hypermultiplets transforming in a complex representation of the symmetry group, then the classical flavor symmetry rotating those hypermultiplets is $U(k)$. However, as discussed in the above references, these $U(1)$s are often broken, either entirely or to a particular diagonal subgroup, by the presence of ABJ anomalies. Due to the more subtle nature of such flavor symmetries, we shall leave their discussion for future work.

\subsection{The E-string origin of flavor symmetries}\label{sec:estrf}

One of the geometric building blocks of 6d SCFTs is the E-string, originally studied in \cite{Ganor:1996mu,Seiberg:1996vs,Witten:1996qb,Morrison:1996pp,Witten:1996qz,Cheung:1997id}. The rank $Q$ E-string is associated to the compact curve configuration given in equation \eqref{eqn:rankqestring}:
\begin{equation}
    1 \overbrace{2\,2\cdots\,2\,2}^{Q-1} \,.
\end{equation}
For all ranks, the E-string has an $\mathfrak{e}_8$ flavor symmetry. This arises from a non-compact curve which intersects the $(-1)$-curve and has a II$^*$ singular fiber over it. When the $(-1)$-curve is glued together with another compact curve, this involves gauging a part of the $\mathfrak{e}_8$ flavor symmetry.\footnote{Note that a $(-1)$-curve can intersect at most two other compact curves \cite{Heckman:2015bfa}.} Let us consider the following compact curve fragment:
\begin{equation}
    \cdots \overset{\mathfrak{g}}{n}1\overset{\mathfrak{h}}{m}\cdots \,,
\end{equation}
where we allow the possibility for $\mathfrak{h}$ to be trivial (in which case $m=2$ is required), or indeed for the curves to the right of the $(-1)$-curve to be absent entirely. One flavor node is attached to the $(-1)$-curve for each simple non-Abelian factor appearing in\footnote{In fact, we must also specify how the flavor node is attached. In Section \ref{sec:GS}, we see that anomaly cancellation requires that the number of intersections be equal to the Dynkin embedding index of that subalgebra.}
\begin{equation}
    \text{Commutant}(\rho, \mathfrak{e}_8, \mathfrak{g} \oplus \mathfrak{h}) \,,
\end{equation}
where $\rho$ is an embedding given by:
\begin{align}
    \rho :\ \mathfrak{g} \oplus \mathfrak{h} \rightarrow \mathfrak{e}_8 \,.
\end{align}
The relevant subalgebras of $\mathfrak{e}_8$, and the Dynkin indices of the embeddings for each irreducible component of the subalgebras, required for the determination of the commutants are given in Table \ref{tb:e8breakdowns}.\footnote{The Dynkin indices have recently been explored and found to be relevant in the context of F-theory in \cite{Esole:2020tby} where embeddings are considered for determining matter representations with a perspective of Dynkin. We refer the reader there for the relevant definitions.}

\begin{table}[H]
\renewcommand{\arraystretch}{1.1}
    \centering
    \footnotesize
    \begin{threeparttable}
    $\begin{array}{c}
    \toprule
    \text{Subalgebras of $\mathfrak{e}_8$}\\\midrule
    \mathfrak{su}_2^1\oplus\mathfrak{e}_7^1\\
    \mathfrak{su}_3^1\oplus\mathfrak{e}_6^1\\
    \mathfrak{su}_4^1\oplus\mathfrak{so}_{10}^1\\
    \mathfrak{su}_5^1\oplus\mathfrak{su}_5^1\\
    \mathfrak{su}_9^1\\
    \mathfrak{so}_7^1\oplus\mathfrak{so}_9^1\\
    \mathfrak{so}_8^1\oplus\mathfrak{so}_8^1\\
    \mathfrak{so}_{16}^1\\
    \mathfrak{g}_2^1\oplus\mathfrak{f}_4^1\\
    \mathfrak{g}_2^1\oplus\mathfrak{g}_2^1\oplus\mathfrak{su}_2^8\\
    \mathfrak{g}_2^1\oplus\mathfrak{su}_3^1\oplus\mathfrak{su}_3^2\\
    \mathfrak{so}_7^1\oplus\mathfrak{su}_3^1\oplus\mathfrak{su}_2^2\oplus\mathfrak{u}_1\\
    \mathfrak{so}_8^1\oplus\mathfrak{su}_2^1\oplus\mathfrak{su}_2^1\oplus\mathfrak{su}_2^1\oplus\mathfrak{su}_2^1\\
    \mathfrak{su}_3^1\oplus\mathfrak{su}_3^1\oplus\mathfrak{su}_3^1\oplus\mathfrak{su}_3^1\\
    \mathfrak{so}_7^1\oplus\mathfrak{su}_2^1\oplus\mathfrak{su}_2^1\oplus\mathfrak{usp}_4^1\\
    \mathfrak{su}_3^1\oplus\mathfrak{su}_2^1\oplus\mathfrak{su}_6^1\\
    \mathfrak{g}_2^1\oplus\mathfrak{su}_2^1\oplus\mathfrak{usp}_6^1\\
    \mathfrak{g}_2^1\oplus\mathfrak{so}_7^1\oplus\mathfrak{u}_1\\
    \mathfrak{so}_9^1\oplus\mathfrak{su}_2^1\oplus\mathfrak{su}_2^1\oplus\mathfrak{su}_2^2\\
    \mathfrak{su}_3^1\oplus\mathfrak{so}_9^1\oplus\mathfrak{u}_1\\
    \mathfrak{su}_3^1\oplus\mathfrak{so}_8^1\oplus\mathfrak{u}_1\\
    \mathfrak{su}_3^1\oplus\mathfrak{f}_4^1\\
    \mathfrak{su}_2^1\oplus\mathfrak{f}_4^1\oplus\mathfrak{su}_2^3\\
    \mathfrak{g}_2^1\oplus\mathfrak{so}_9^1\\
    \mathfrak{g}_2^1\oplus\mathfrak{so}_8^1\\
    \bottomrule
    \end{array}$
    \end{threeparttable}
    \caption{The subalgebras of $\mathfrak{e}_8$ relevant for determining the non-Abelian global symmetries arising from E-string components in the 6d SCFTs considered in this paper. The superscript denotes the Dynkin index of the embedding.}
    \label{tb:e8breakdowns}
\end{table}

For $Q > 1$, the E-string theory also has an $\mathfrak{su}_2$ global symmetry. This $\mathfrak{su}_2$ can be thought of as a flavor node which is attached to \emph{each} of the $(-2)$-curves of the E-string, with intersection number one. In addition to the option of gluing compact curves to the $(-1)$-curve of the E-string, one can glue onto one of the $(-2)$-curves; however, this is highly constrained. Such a gluing requires that the E-string is of rank two, i.e. $Q = 2$, and the curve to which it is glued must support an $\mathfrak{su}_2$ gauge algebra; this procedure functions as a gauging of the $\mathfrak{su}_2$ flavor. As such, this flavor symmetry only persists to a symmetry of the full SCFT when the $(-2)$-curves of the E-string are not connected to any other compact curves.

\subsection{The classical origin of flavor symmetries}
\label{sec:classicalFlavor}

For each curve decorated by a non-minimal gauge algebra, an inspection of the geometry reveals the presence of hypermultiplets transforming in various representations $\bm{R}$. To satisfy the gauge anomaly cancellation condition, there must be a particular number, $k$, of full hypermultiplets \cite{Heckman:2015bfa}. Then, there is a rotation of these $k$ hypermultiplets which gives rise to a classical flavor symmetry, whose specific action depends on the nature of the representation:
\begin{equation}
    \mathfrak{f} = \begin{cases}
      \mathfrak{su}_k\, &\text{if } \bm{R} \text{ is complex}\, ,\cr
      \mathfrak{so}_{2k}\, &\text{if } \bm{R} \text{ is quaternionic}\, ,\cr
      \mathfrak{usp}_{2k}\, &\text{if } \bm{R} \text{ is real} \,.
    \end{cases}
\end{equation}

We expect these classical flavor symmetries to be flavor symmetries of the SCFT that exists at the origin of the tensor branch, except in a few special cases where we have an $\mathfrak{su}_2$ gauge algebra with fewer than eight fundamental hypermultiplets. For instance, in the case of an $\mathfrak{su}_2$ gauge group on top of a $(-2)$-curve, the cancellation of gauge anomalies requires the presence of eight half-hypermultiplets. This indicates that the hypermultiplets fall in the vector representation of $\mathfrak{so}_8$. However, in the geometric realization, the fiber on top of the $(-2)$-curve has a singularity of type IV$^{\text{ns}}$ and therefore it is associated to a gauge algebra $\mathfrak{usp}_2 \cong\mathfrak{su}_2$. The type IV$^{\text{ns}}$ fiber  on the $(-2)$-curve meets an I$_0^{*\text{ss}}$ fiber which is associated to the flavor algebra $\mathfrak{so}_7$. The eight half-hypermultiplets in the fundamental of $\mathfrak{su}_2$ transform in the spinor representation of the $\mathfrak{so}_7$. This can also be understood from a purely field theoretic perspective: there is an inconsistency with theories involving an $\mathfrak{su}_2$ gauge algebra with fundamental hypermultiplets transforming in the vector of an $\mathfrak{so}_8$ flavor algebra, which makes the flavor symmetry algebra to be $\mathfrak{so}_7$ instead \cite{Ohmori:2015pia}. A similar reasoning applied to different numbers of half-hypermultiplets charged under an $\mathfrak{su}_2$ algebra leads to \cite{Heckman:2015bfa,Ohmori:2015pia,Morrison:2016djb}:
\begin{center}
\begin{tabular}{c|ccccc}
     $\frac{1}{2}$-hypermultiplets & 3 & 4 & 6 & 7 & 8\\\hline
     non-Abelian flavor $\mathfrak{f}$ & $\varnothing$ & $\mathfrak{su}_{2}$ & $\mathfrak{su}_{3}$ & $\mathfrak{g}_2$ & $\mathfrak{so}_7$
\end{tabular} \,.
\end{center}
In F-theory, these are modeled by non-compact elliptic fibrations endowed with Lie algebras $\mathfrak{su}_2 \oplus \mathfrak{f}$. Some of these models have been studied explicitly from geometric perspective: $\mathfrak{su}_2 \oplus \mathfrak{su}_2$ in \cite{Esole:2018csl}, $\mathfrak{su}_2 \oplus \mathfrak{su}_3$ in \cite{Esole:2019asj}, $\mathfrak{su}_2 \oplus \mathfrak{g}_2$ in \cite{Esole:2018mqb}, and $\mathfrak{so}_7$ in \cite{Esole:2017qeh}. 

 \begin{figure}[H]
  \centering
  \begin{subfigure}[b]{0.9\textwidth}
    \centering
    \includegraphics[scale=1.8]{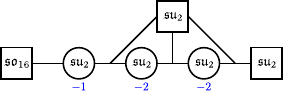}
    \caption{The quiver associated to the configuration $\overset{\mathfrak{su}_2}{1}\overset{\mathfrak{su}_2}{2}\overset{\mathfrak{su}_2}{2}$.}
  \end{subfigure} \newline
  \begin{subfigure}[b]{0.9\textwidth}
    \centering
    \includegraphics[scale=1.8]{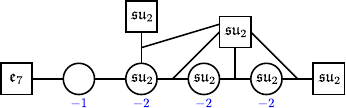}
    \caption{The quiver associated to the configuration $1\overset{\mathfrak{su}_2}{2}\overset{\mathfrak{su}_2}{2}\overset{\mathfrak{su}_2}{2}$.}
  \end{subfigure} \newline
  \begin{subfigure}[b]{0.9\textwidth}
    \centering
    \includegraphics[scale=1.8]{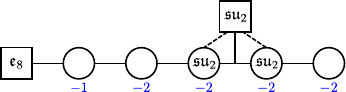}
    \caption{The quiver associated to the configuration $1\,2\overset{\mathfrak{su}_2}{2}\overset{\mathfrak{su}_2}{2}2$.}
  \end{subfigure}
  \caption{Three examples of quivers involving baryonic $\mathfrak{su}_2$ global symmetries. The circular nodes are the gauge nodes associated to the compact curves, and the square nodes are the flavor symmetries added following Section \ref{sec:flavor}. The numbers in blue below each gauge node are the self-intersection numbers of that compact curve. The trivalent vertices indicate the presence of matter in the trifundamental representation and the dashed lines indicate matter transforming in the fundamental representation of the gauge node and the adjoint representation of the flavor node.}
  \label{fig:baryonic}
\end{figure}

Finally, there is the special case of a single hypermultiplet charged under the $(\bm{2}, \bm{2})$ bifundamental representation of an $\mathfrak{su}_2 \oplus \mathfrak{su}_2$ algebra. As this representation is real, there is an additional $\mathfrak{usp}_2 \cong \mathfrak{su}_2$ flavor symmetry rotating the hypermultiplet. If there are more than two $\mathfrak{su}_2$ algebras linked together, only the diagonal subgroup rotating the baryonic operators made out of the bifundamental hypermultiplets survives \cite{Mekareeya:2017jgc, Apruzzi:2020eqi}. Examples of quivers exhibiting this special baryonic symmetry and the associated trifundamental representations are pictured in Figure \ref{fig:baryonic}.

Before closing this section, we note that in \cite{Bertolini:2015bwa,Merkx:2017jey} it was argued that a few flavor algebras appear inconsistent with a geometric construction via F-theory. An example  is $\mathfrak{usp}_k$ rotating the hypermultiplets in the fundamental representation of $\mathfrak{f}_4$. It may therefore be expected that the apparent flavor symmetry does not persist at the SCFT point. However, as we see in the remainder of this work, some of these apparently-inconsistent realizations do appear in 4d SCFTs obtained by compactifying a 6d $(1,0)$ theory on the torus, and their central charges match that of their class $\mathcal{S}$ avatars. To emphasize this point, we shall now consider an explicit example. The logic of this section dictates that the 6d geometric configuration 
\begin{equation}\label{eqn:f4eg}
    \overset{\mathfrak{f}_4}{1} \,,
\end{equation}
should have a $\mathfrak{usp}_8$ flavor symmetry rotating the four fundamental hypermultiplets of the $\mathfrak{f}_4$ gauge algebra. From the perspective of the elliptic fibration in \cite{Bertolini:2015bwa} the singular fiber associated to that flavor algebra is not apparent. This is not surprising and is expected as $\mathfrak{f}_4 \oplus \mathfrak{usp}_8$ is not a subalgebra of $\mathfrak{e}_8$ and hence the Katz--Vafa method of matter representations from F-theory model does not apply \cite{Katz:1996xe,Esole:2020tby}. However, the 4d $\mathcal{N}=2$ SCFT obtained from the compactification of the 6d $(2,0)$ theory, which is also obtained as the $T^2$ compactification of the configuration in equation \eqref{eqn:f4eg} as can be seen in Table \ref{tbl:E8E8}, does have a $\mathfrak{usp}_8$ flavor symmetry. It therefore seems strained to argue that, only in these few cases, the symmetry does not exist for the 6d SCFT, but that the exact same symmetry algebra re-emerges in the infrared of the flow to the 4d SCFT. We are led to the conclusion that these flavor symmetry algebras may be hidden in apparent F-theory constructions and is yet to be resolved.

\section{Anomaly polynomials for 6d \texorpdfstring{\boldmath{$(1,0)$}}{(1,0)} SCFTs}\label{sec:anom}

Theories of class $\mathcal{S}$ are obtained from a twisted compactification of the 6d $(2,0)$ theories on a punctured Riemann surface $C_{g,n}$. We wish to compare these 4d $\mathcal{N}=2$ theories with those obtained by compactifying 6d $(1,0)$ SCFTs on a torus. Our main tools for determining whether two theories are the same are their 4d central charges, namely $a$ and $c$, and the central charges of the non-Abelian flavor symmetries $k_i$. From the 6d $(1,0)$ SCFT origin, such central charges are determined from the 't Hooft coefficients of the 6d anomaly polynomial $I_8$. 

In this section, we explain the procedure required to determine the anomaly polynomial for a 6d $(1,0)$ SCFT with the given geometric data. The anomaly polynomial $I_8$ is an eight-form built out of the characteristic classes of the algebras associated to the symmetries of the theory, such as the Lorentz, R-, gauge, and flavor symmetries. We first take the perspective of \cite{Ohmori:2014kda, Intriligator:2014eaa} and extend further to compute the anomaly polynomial on the tensor branch. This anomaly polynomial is unaltered in the SCFT limit where all curves are shrunk to zero-volume. The anomaly polynomial can be expressed as the following general form
\begin{equation}\label{eqn:I8general}
  \begin{aligned}
    I_8 &=  \frac{\alpha}{24} c_2(R)^2+ \frac{\beta}{24}  c_2(R) p_1(T) + \frac{\gamma}{24}  p_1(T)^2 + \frac{\delta}{24} p_2(T) \cr &\quad + \sum_a \text{Tr}F_a^2 \left(\kappa_a p_1(T) + \nu_a c_2(R) + \sum_b \rho_{ab} \text{Tr}  F_b^2\right) + \sum_a \mu_a \text{Tr}F_a^4  \,,
  \end{aligned}
\end{equation}
where each summation over $a$ or $b$ runs over the simple non-Abelian flavor symmetries of the theory. The anomaly polynomial in equation \eqref{eqn:I8general} has been worked out from the geometric perspective of F-theory for a variety of supergravity  \cite{Grassi:2011hq,Park:2011wv,Park:2011ji,Grimm:2012yq,DelZotto:2014fia,Esole:2015xfa,Arras:2016evy,Esole:2017rgz,Esole:2017qeh,Monnier:2017oqd,Esole:2017hlw,Esole:2018csl,Esole:2018mqb,Monnier:2018nfs,Esole:2019asj} and superconformal \cite{Bershadsky:1997sb,Cvetic:2012xn,Ohmori:2014pca,Ohmori:2014kda,Intriligator:2014eaa,DelZotto:2014hpa,Cordova:2015fha,Cremonesi:2015bld,Beccaria:2015ypa,Bobev:2016phc,Apruzzi:2016nfr,Yankielowicz:2017xkf,Mekareeya:2017jgc, Mekareeya:2017sqh,Shimizu:2017kzs,Bobev:2017uzs} theories; we give a sample of the literature here. 

The anomaly polynomial has two contributions: the one-loop contribution and the Green--Schwarz--West--Sagnotti \cite{Green:1984sg, Green:1984bx,Sagnotti:1992qw} contribution.\footnote{The latter is often referred to simply as the Green--Schwarz or ``GS'' contribution.} The one-loop contribution is, roughly speaking, the contribution from the free multiplets that live at a generic point of the tensor branch. The Green--Schwarz contribution, which was understood in the context of F-theory in \cite{Sadov:1996zm}, arises from the tensionful string-like degrees of freedom also living at a generic point of the tensor branch. The strings, which arise as D3-branes wrapping the curves of the geometric configuration, become tensionless at the origin of the tensor branch, reflecting the presence of an interacting SCFT \cite{Witten:1995zh}. We write the anomaly polynomial as
\begin{equation}
    I_8 = I_8^\text{1-loop} + I_8^\text{GS} \,,
\end{equation}
and compute each term in turn.

\subsection{One-loop contribution to the anomaly polynomial}\label{sec:oneloop}

The one-loop contribution is determined directly from the geometric configuration at a generic point of the tensor branch. To compute the one-loop term, we first need to know the contributions to the anomaly polynomial of single tensor multiplets, hypermultiplets, and vector multiplets, which were first computed in \cite{AlvarezGaume:1983ig}. They are given by:
\begin{align}
    I_8^\text{tensor} &= \frac{1}{24}c_2(R)^2 + \frac{1}{48}c_2(R)p_1(T) + \frac{1}{5760}\left( 23p_1(T)^2 - 116p_2(T) \right)\,,\\
        I_8^\text{vector}(\mathfrak{g}, F) &= -\frac{1}{24}\left( \text{Tr}_{\bm{adj}}F^4 + 6 c_2(R)\text{Tr}_{\bm{adj}}F^2 + \text{dim}(\mathfrak{g})c_2(R)^2 \right) \cr
        &\quad -\frac{1}{48}p_1(T)\left(\text{Tr}_{\bm{adj}}F^2 + \text{dim}(\mathfrak{g})c_2(R) \right) - \frac{\text{dim}(\mathfrak{g})}{5760}\left(7p_1(T)^2 - 4p_2(T)\right)\,,\\
	I_8^\text{hyper}(\mathfrak{g}, \bm{R}, F) &= \frac{1}{24} \text{Tr}_{\bm{R}} F^4 + \frac{1}{48}p_1(T)  \text{Tr}_{\bm{R}} F^2 + \frac{\text{dim}(\bm{R})}{5760}\left( 7p_1(T)^2 - 4p_2(T) \right) \,,
\end{align}
where $\mathfrak{g}$ is a semi-simple algebra, either gauge or flavor, $\bm{R}$ is a representation of $\mathfrak{g}$, and $F$ is the curvature associated to $\mathfrak{g}$. Furthermore, the various traces are taken over the representation written as a subscript. We convert all of the traces into the one-instanton-normalized traces, which we denote as $\text{Tr}F^2$ and $\text{Tr}F^4$. This conversion is given in terms of Casimir invariants of the Lie algebra:
\begin{equation}
    \begin{aligned}
        \text{Tr}_{\bm{R}}F^2 &= A_{\bm{R}}\, \text{Tr}F^2 \,,\\
        \text{Tr}_{\bm{R}}F^4 &= B_{\bm{R}}\, \text{Tr}F^4 + C_{\bm{R}}\, (\text{Tr}F^2)^2 \,.
    \end{aligned}
\end{equation}
The coefficients $A_{\bm{R}}\,,B_{\bm{R}}\,,C_{\bm{R}}$ for the  representations relevant to F-theory constructions can be found in appendix F of \cite{Heckman:2018jxk}.\footnote{There are various conventions for the coefficients $A_{\bm{R}}$, $B_{\bm{R}}$, and $C_{\bm{R}}$ in the physics literature. In \cite{Heckman:2018jxk}, they are denoted by $h_{\bm{R}}$,  $x_{\bm{R}}$, and $y_{\bm{R}}$ respectively.} We commonly have hypermultiplets charged under bifundamental representations, or more exotic bi-representations, hence we note that if a hypermultiplet is charged under the representation $(\bm{R}_1,\bm{R}_2)$ of two algebras $\mathfrak{g}_1\oplus\mathfrak{g}_2$, the traces can be written in terms of the data of each factor:                      
\begin{equation}
    \begin{aligned}
	    \text{Tr}_{(\bm{R}_1,\bm{R}_2)}F^2 &= \text{dim}(\bm{R}_1)\,\text{Tr}_{\bm{R}_2}F^2 + \text{dim}(\bm{R}_2)\,\text{Tr}_{\bm{R}_1}F^2\,,\\
	    \text{Tr}_{(\bm{R}_1,\bm{R}_2)}F^4 &= \text{dim}(\bm{R}_1)\,\text{Tr}_{\bm{R}_2}F^4 +6\,\text{Tr}_{\bm{R}_1}F^2\,\text{Tr}_{\bm{R}_2}F^2 + \text{dim}(\bm{R}_2)\,\text{Tr}_{\bm{R}_1}F^4 \,.
    \end{aligned}
\end{equation}

Beyond the standard multiplets, we also need to know the contribution of rank $Q$ E-string, as well as $\mathcal{N}=(2,0)$ tensor multiplets. For the E-string, the anomaly polynomial is given by \cite{Ohmori:2014pca}:
\begin{align}\label{eqn:estrinc}
    I^{\text{E-string}}_8(Q, \text{Tr}F^2_{\mathfrak{e}_8}, c_2(L)) =&\,\frac{Q^3}{6}(c_2(L)-c_2(R))^2 + \frac{Q^2}{8}(c_2(L)-c_2(R))I_{(4)} \cr &\quad+ Q(\frac{1}{2}I_{(4)}^2-I_{(8)})-I_\text{free}\,,
\end{align}
where
\begin{align}
    I_{(4)} =&\, \frac{1}{2}(\text{Tr}F^2_{\mathfrak{e}_8}+p_1(T)-2c_2(L)-2c_2(R))\,,\\
    I_{(8)} =&\, \frac{1}{192}\left(4p_2(T)-\left(4c_2(L)+p_1(T)\right)\left(4c_2(R)+p_1(T)\right)\right)\,,\\
    I_\text{free} =&\,\frac{1}{24}c_2(L)^2 +\frac{1}{48}c_2(L)p_1(T)+\frac{7}{5760}p_1(T)^2 - \frac{1}{1440}p_2(T)\,.
\end{align}
The trace $\text{Tr}F^2_{\mathfrak{e}_8}$ is associated to the $\mathfrak{e}_8$ flavor symmetry of the E-string. For the rank one E-string the $c_2(L)$ terms are absent. For $Q>1$ we have  $c_2(L)=\frac{1}{4}\text{Tr}F^2_{\mathfrak{su}_2}$, which corresponds to the curvature of the $\mathfrak{su}_2$ flavor. Finally, the anomaly polynomial of an $\mathcal{N}=(2,0)$ tensor multiplet is \cite{AlvarezGaume:1983ig}:
\begin{align}
    I_8^{(2,0)} = \frac{1}{24}\left(c_2(L)^2 + c_2(R)^2\right) + \frac{1}{48}\left(c_2(L) + c_2(R)\right)p_1(T)+\frac{1}{192}p_1(T)^2-\frac{1}{48}p_2(T) \,.
\end{align}

Now that we have the contributions $I_8^\text{tensor}$, $I_8^\text{vector}$, $I_8^\text{hyper}$, $I_8^\text{E-string}$, and $I_8^{(2,0)}$ to hand we can describe the algorithm to construct the one-loop anomaly contribution. We state the algorithm as the following.

\begin{alg}\label{alg:oneloop}
The one-loop anomaly polynomial is determined from a geometric configuration of curves and algebras via the following five steps.
\begin{enumerate}
    \item For each curve with an associated gauge algebra include an $I_8^\text{tensor}$.
    \item For each simple gauge algebra $\mathfrak{g}$ include an $I_8^\text{vector}(\mathfrak{g}, F_{\mathfrak{g}})$ where $F_{\mathfrak{g}}$ is the curvature of $\mathfrak{g}$.
    \item For each hypermultiplet in a representation $\bm{R}$ of a (possibly semi-simple) Lie algebra $\mathfrak{h}$, where $\mathfrak{h}$ may contain both gauge and flavor factors, include an $I_8^\text{hyper}(\mathfrak{h}, \bm{R}, F_\mathfrak{h})$. The curvature $F_\mathfrak{h}$ of $\mathfrak{h}$ is reducible if $\mathfrak{h}$ is not simple.
    \item For each rank $Q$ E-string in the curve configuration include an $I_8^\text{E-string}(Q)$. Write $\mathfrak{h}=\oplus_i \mathfrak{h}_i$, where $\mathfrak{h}_i$ denotes the algebras on the compact and non-compact curves intersecting the $(-1)$-curve of the E-string. Then the embedding $\rho : \mathfrak{h} \rightarrow \mathfrak{e}_8$ describes how the $\mathfrak{e}_8$ flavor symmetry of the E-string decomposes. For each irreducible component $\mathfrak{h}_i$, we denote the corresponding embedding part as $\rho_i$ and the Dynkin index of each factor as $\ell_{\rho_i}$. The curvature of the $\mathfrak{e}_8$ flavor symmetry decomposes via
    \begin{equation}
        \text{Tr}F_{\mathfrak{e}_8}^2 \rightarrow \sum_i \ell_{\rho_i} \text{Tr}F_{\mathfrak{h}_i}^2 
    \end{equation}
    in $I_8^\text{E-string}(Q)$. If a $(-2)$-curve of the E-string intersects another curve, then we must have $Q = 2$ and the algebra on the neighboring curve must be $\mathfrak{su}_2$; in this case we must also replace $c_2(L)$ in $I_8^\text{E-string}(Q)$ with $\frac{1}{4}\text{Tr}F_{\mathfrak{su}_2}^2$, where $F_{\mathfrak{su}_2}$ is the curvature of the neighboring $\mathfrak{su}_2$.
    \item For each $(-2)$-curve with no associated algebra, that is not part of an E-string, include an $I_8^{(2,0)}$. Replace $c_2(L)$ in $I_8^{(2,0)}$ with $\frac{1}{4}\text{Tr}F_{\mathfrak{su}_2}^2$, where $F_{\mathfrak{su}_2}$ is the curvature of the $\mathfrak{su}_2$ algebra on the adjacent curve.\footnote{As we are considering theories which have strict $(1,0)$ supersymmetry, each such $(-2)$-curve has exactly one neighbor, which has an $\mathfrak{su}_2$ algebra over it.}
\end{enumerate}
Summing each of these contributions across the entire geometric configuration, we obtain the one-loop contribution to the anomaly polynomial $I_8^\text{one-loop}$. 
\end{alg}

To highlight Algorithm \ref{alg:oneloop}, we give a simple but illustrative example with the following geometric configuration:
\begin{equation}
    1\,2\,\overset{\mathfrak{su}_2}{2}\overset{\mathfrak{g}_2}{3}1\overset{\mathfrak{g}_2}{3} \,.
\end{equation}
We label the curvatures of the three gauge algebras, left-to-right, as $F_1$, $F_2$, and $F_3$, respectively. First, we find the flavor symmetries associated to this curve configuration as
\begin{itemize}
    \item an $\mathfrak{e}_8$ attached to the leftmost $(-1)$-curve. 
    \item an $\mathfrak{su}_2$ attached to the rightmost $(-1)$-curve.
    \item a $\mathfrak{usp}_2$ attached to the rightmost $(-3)$-curve.
\end{itemize}
We denote the curvatures of these flavor algebras $F_4$, $F_5$, and $F_6$, respectively.
Then the one-loop anomaly is composed of their contributions as
\begin{equation}\label{eqn:oneloopex}
  \begin{aligned}
    I_8^\text{one-loop} &= 3I_8^\text{tensor} + I_8^\text{vector}(\mathfrak{su}_2, F_1) + I_8^\text{vector}(\mathfrak{g}_2, F_2) + I_8^\text{vector}(\mathfrak{g}_2, F_3) \cr
    &\quad + \frac{1}{2}I_8^\text{hyper}(\mathfrak{su}_2 \oplus \mathfrak{g}_2, (\bm{2,7}), F_1 \oplus F_2) + \frac{1}{2}I_8^\text{hyper}(\mathfrak{g}_2 \oplus \mathfrak{usp}_2, (\bm{7,2}), F_3 \oplus F_6) \cr 
    &\quad + I_8^\text{E-string}(2, \text{Tr}F_{\mathfrak{e}_8}^2 \rightarrow  \text{Tr}F_{4}^2, c_2(L) \rightarrow \frac{1}{4}\text{Tr}F_{1}^2) \cr 
    &\quad + I_8^\text{E-string}(1, \text{Tr}F_{\mathfrak{e}_8}^2 \rightarrow  \text{Tr}F_{2}^2 + \text{Tr}F_{3}^2 + 8\text{Tr}F_{5}^2) \,.
  \end{aligned}
\end{equation}
The factor $8$ in the final line of the equation \eqref{eqn:oneloopex} is given by the Dynkin embedding index of the $\mathfrak{su}_2$ flavor symmetry factor inside $\mathfrak{e}_8$, which is $8$; this is listed in Table \ref{tb:e8breakdowns} where we summarized all the subalgebras of $\mathfrak{e}_8$ relevant for our purposes with their Dynkin indices for each irreducible component of the subalgebras. For the E-string contributions, we use the notation $I_8^\text{E-string}(Q, \cdots)$ as in equation \eqref{eqn:estrinc}, where we have written explicitly the replacements for the $\mathfrak{e}_8$ and $\mathfrak{su}_2$ curvatures as described in step 4 of Algorithm \ref{alg:oneloop}.

The one-loop anomaly generally contains pure gauge anomalies, as well as mixed-gauge-flavor, mixed-gauge-gravity, and mixed-gauge-R-symmetry anomalies. These are cancelled by the Green--Schwarz contribution to the anomaly polynomial.

\subsection{Green--Schwarz anomaly contribution}\label{sec:GS}

A prescription to find the Green--Schwarz contribution to the anomaly polynomial $I_8^\text{GS}$ directly from the generalized quiver and its intersection data was given in \cite{Ohmori:2014kda}. The GS contribution is determined at a non-generic point of the tensor branch where all of the curves in the curve configuration associated to E-strings are shrunk. 

Shrinking one of the E-strings is equivalent to recursively shrinking the corresponding $(-1)$-curve that does not have an associated gauge algebra, which requires a blow-down operation modifying the intersection matrix. Let $\widetilde{A}^{ij}$ be the intersection matrix obtained at the non-generic point of the tensor branch after shrinking all E-strings and $A^{ij}$ be the intersection matrix at a generic point.

At this non-generic point of the tensor branch, the Green--Schwarz anomaly contribution is given by \cite{Ohmori:2014kda,Intriligator:2014eaa}
\begin{equation}\label{eqn:GSterm}
    I_8^\text{GS} = -\frac{1}{2}\widetilde{A}_{ij}I^i I^j \,,
\end{equation}
where the four-form $I^i$ takes the following form
\begin{equation}
        I^i = \frac{1}{4}\left(-\widetilde{A}^{ij} \text{Tr}F_j^2 - B^{ia} \text{Tr} F_a^2 - (2 + \widetilde{A}^{ii}) p_1(T)\right) + y^i c_2(R) \,,
\end{equation}
with $i,j$ running over compact curves, $\widetilde{A}_{ij}= (\widetilde{A}^{-1})_{ij}$ is the inverse of the associated adjacency matrix, and the index $a$ runs over the simple non-Abelian flavor algebras of the theory. The coefficient $y^i$ is $h_{\mathfrak{g}_i}^\vee$  at a generic point of the tensor branch and its modification must be tracked when shrinking the E-strings. The coefficients of all but the $\text{Tr} F_a^2$ term are determined from the tensor branch configuration in \cite{Ohmori:2014kda}, where they found $B^{ia}$ for the considered examples to be either zero or one. 

For more generic cases, we find that anomaly cancellation for the mixed-gauge-flavor anomaly, which is required for a consistent 6d SCFT \cite{Cordova:2020tij}, enforces that we take
\begin{equation}\label{eqn:BiaPre}
    B^{ia} = \begin{cases}
      4\, \eta\,  A_{\bm{R}_i}  A_{\bm{R}_a} \quad &\text{ if } \mathfrak{f}_a \text{ is classical flavor,} \cr
      \ell_{\rho_a} \quad &\text{ if } \mathfrak{f}_a \text{ is E-string flavor,}
    \end{cases}
\end{equation}
when we have a compact curve labeled by $i$ that intersects a flavor curve labeled by $a$, where $\eta$ is given by
\begin{align}
    \eta=\begin{cases}
        \frac{1}{2}\quad &\text{for a half-hypermultiplet,}\\
        1\quad &\text{else.} 
    \end{cases}
\end{align}
If the flavor algebra $\mathfrak{f}_a$ arises from an E-string (see Section \ref{sec:estrf} for its detail), then it is associated to an embedding $\rho_a : \mathfrak{f}_a \rightarrow \mathfrak{e}_8$ and with the Dynkin index of this embedding $\ell_{\rho_a}$.
When the flavor symmetry originates from a non-compact curve $\Sigma^a$, we recover the expected result that $B^{ia}$ is the intersection number between the compact and non-compact curves. 

\subsection{A single tensor theory with flavor}

To illustrate the cancellation of mixed-gauge-flavor anomalies using the prescription given in the equation \eqref{eqn:BiaPre}, we determine the anomaly polynomials for all 6d SCFTs with a single tensor associated to a $(-n)$-curve and arbitrary gauge group:
\begin{equation}
    \overset{\mathfrak{g}}{n} \,.
\end{equation}
For simplicity, we consider cases where there is only one type of hypermultiplet representation, and thus only a single flavor algebra $\mathfrak{f}$.\footnote{It is a straightforward generalization to include other matter representations and/or other flavor symmetry algebras.} In view of the matter content, this adds only a single (possibly half-)hypermultiplet in the $(\bm{R_g}, \bm{R_f})$ representation to the rest of the spectrum, which are the tensor multiplet and the vector multiplet transforming in the adjoint representation of $\mathfrak{g}$. The gauge anomaly cancellation condition fixes the number of hypermultiplets, and the flavor symmetry is the classical symmetry rotating those hypermultiplets, as described in Section \ref{sec:classicalFlavor}. Thus the flavor symmetry is fixed by the values of $n$ and $\mathfrak{g}$. The multiplicity $\eta$ of the hypermultiplet depends on whether the representation $(\bm{R_g}, \bm{R_f})$ is quaternionic:
\begin{equation}
    \eta = \begin{cases}
    \frac{1}{2} \quad \text{ if } \, (\bm{R_g}, \bm{R_f}) \, \text{ is quaternionic, } \cr
    1 \quad \text{ otherwise. }
    \end{cases}
\end{equation}

The one-loop contribution to the anomaly polynomial can be straightforwardly worked out from the algorithm given in Section \ref{sec:oneloop}. It is simply
\begin{equation}
    I_8^\text{one-loop} = I_8^\text{tensor}
    + I_8^\text{vector}(\mathfrak{g}, F_\mathfrak{g})
    + \eta\, I_8^\text{hyper}(\mathfrak{g}\oplus\mathfrak{f},(\bm{R_g},\bm{R_f}) ,F_\mathfrak{g}\oplus F_\mathfrak{f})\,.
\end{equation}

The Green--Schwarz term is defined by equation \eqref{eqn:GSterm} and in this case it is
\begin{equation}
    I_8^\text{GS} = \frac{1}{2n}(I^1)^2\,,\qquad
    I^1 = \frac{1}{4}\left(b_1 \text{Tr}F_\mathfrak{g}^2 + b_2 \text{Tr} F_\mathfrak{f}^2 + b_3 p_1(T)\right) + b_4\, c_2(R) \,.
\end{equation}
In this expression we have left the coefficients appearing in $I^1$ arbitrary. We show that  cancellation of all gauge and mixed-gauge anomalies requires that the coefficients must be as described in Section \ref{sec:GS}.

Combining the one-loop and GS contributions one finds that the resulting anomaly polynomial contains the terms 
\begin{equation}
  \begin{aligned}
    I_8 &\supset \frac{1}{48}\left(\eta A_{\bm{R_g}}d_{\bm{R}_f} - h_G^\vee + 3b_1b_3{16n} \right)p_1(T) \text{Tr}F_\mathfrak{g}^2 
    + \frac{1}{4}\left(\frac{b_1b_4}{n} - h_G^\vee \right)c_2(R) \text{Tr}F_\mathfrak{g}^2 \cr &\qquad 
    + \frac{1}{24}\left(\frac{3b_1^2}{4n} + \eta d_{\bm{R_g}}C_{\bm{R_g}} - C_{\bm{adj_g}} \right) \left(\text{Tr}F_\mathfrak{g}^2\right)^2
    + \frac{1}{24}\left( \eta d_{\bm{R_f}}B_{\bm{R_g}} - B_{\bm{adj_g}} \right) \text{Tr}F_\mathfrak{g}^4 \cr &\qquad 
    + \frac{1}{4}\left( \frac{b_1b_2}{4n}  + \eta A_{\bm{R_g}}A_{\bm{R_f}} \right) \text{Tr}F_\mathfrak{g}^2 \text{Tr}F_\mathfrak{f}^2 \,.
  \end{aligned}
\end{equation}
To cancel all of the (mixed-)gauge anomalies we must have
\begin{equation}\label{eq:GSoneTensorResult}
    b_1 = n \,, \quad b_2 = -4 \eta A_{\bm{R_g}}A_{\bm{R_f}} \,, \quad b_3 = -\frac{1}{3}\left(\eta A_{\bm{R_g}}d_{\bm{R}_f} - h_G^\vee \right) \,, \quad b_4 = h_G^\vee \,.
\end{equation}
As we can see the coefficient of the $\text{Tr}F_\mathfrak{f}^2$ term, $b_2$, is as stated in equation \eqref{eqn:BiaPre}. The types of representations and the number of hypermultiplets transforming therein are highly prescribed by the choice of $\mathfrak{g}$ and $n$. In Table \ref{tbl:bireps}, we document the coefficients appearing in the Green--Schwarz term for 6d SCFTs of this form (i.e.~a single tensor with an associated gauge algebra and matter all transforming under one flavor algebra), and we can see that in each case the equation \eqref{eqn:BiaPre} is verified.

\begin{table}[H]
    \centering
    \begin{threeparttable}
        $\begin{array}{ccccc}
            \toprule
            n & \mathfrak{g} & \mathfrak{f} & \text{Rep} & (b_1, b_2, b_3, b_4) \cr\midrule
            1 & \mathfrak{su}_2 & \mathfrak{so}_{20} &  \frac{1}{2}(\bm{2}, \bm{20}) & (1, -1, -1, 2) \cr
            2 & \mathfrak{su}_2 & \mathfrak{so}_7 & \frac{1}{2}(\bm{2}, \bm{8}) & (2, -1, 0, 2) \cr
            1,2 & \mathfrak{su}_3 & \mathfrak{su}_{18 - 6n} &  (\bm{3}, \bm{18-6n}) & (n, -1, n - 2, 3) \cr
            2 & \mathfrak{su}_{k \geq 4} & \mathfrak{su}_{2k} & (\bm{k}, \bm{2k}) & (2, -1, 0, k) \cr
            1 & \mathfrak{usp}_{2k} & \mathfrak{so}_{4k+16} & \frac{1}{2}(\bm{2k}, \bm{4k+16}) & (1, -1, -1, k + 1) \cr
            3 & \mathfrak{so}_7 & \mathfrak{usp}_{4} & \frac{1}{2}(\bm{8}, \bm{4}) & (3, -1, 1, 5) \cr
            4 & \mathfrak{so}_{k \geq 9} & \mathfrak{usp}_{2k - 16} & \frac{1}{2}(\bm{k}, \bm{2k - 8}) & (4, -1, -2, k - 2) \cr
            1, 2, 3 & \mathfrak{g}_2 & \mathfrak{usp}_{20-6n} & \frac{1}{2}(\bm{7}, \bm{20-6n}) & (n, -1, n - 2, 4) \cr
            1, 2, 3, 4 & \mathfrak{f}_4 & \mathfrak{usp}_{10-2n} & \frac{1}{2}(\bm{26}, \bm{10-2n}) & (n, -3, n - 2, 9) \cr
            1, 2, 3, 4 & \mathfrak{e}_6 & \mathfrak{su}_{6-n} & (\bm{27}, \bm{6-n}) & (n, -6, n - 2, 12) \cr
            1, 2, 3, 4, 5 & \mathfrak{e}_7 & \mathfrak{so}_{8 - n} & \frac{1}{2}(\bm{56}, \bm{8-n}) & (n, -12, n-2, 18) \cr\bottomrule
        \end{array}$
    \end{threeparttable}
    \caption{Green--Schwarz coefficients for SCFTs with one tensor multiplet and only one flavor algebra. The flavor algebra is determined by the values of $n$ and $\mathfrak{g}$. We have not written the configurations where the flavor algebra rotating the hypermultiplets is Abelian.}
    \label{tbl:bireps}
\end{table}

\section{4d \texorpdfstring{\boldmath{$\mathcal{N}=2$}}{N=2} central charges from 6d \texorpdfstring{\boldmath{$(1,0)$}}{(1,0)} anomalies}\label{sec:4dto6d}

Compactifications of 6d $(1,0)$ SCFTs on $T^2$ in such a way that one obtains a 4d $\mathcal{N}=2$ SCFT have been much studied in recent years, in particular in \cite{Ohmori:2015pua,DelZotto:2015rca,Ohmori:2015pia,Ohmori:2015tka,Mekareeya:2017jgc,Mekareeya:2017sqh,Kim:2018bpg,Apruzzi:2018oge,Kim:2018lfo,Zafrir:2018hkr,Ohmori:2018ona,Pasquetti:2019hxf,Giacomelli:2020jel,Baume:2020ure, Heckman:2020otd}. To understand the nature of the resulting 4d $\mathcal{N}=2$ SCFTs it is necessary to utilize different techniques depending on the particular compactification and on the particular class of $(1,0)$ SCFTs considered. In this paper, we study examples of a particularly amenable class known as the very Higgsable 6d SCFTs.

A 6d SCFT is called very Higgsable if there exist Higgs branch RG flows that break the theory down to a collection of free hypermultiplets \cite{Ohmori:2015pua}. All very Higgsable theories have a trivial endpoint configuration, meaning that if one takes the geometric data at a generic point of the tensor branch and repeatedly blow-down the $(-1)$-curves then one is left with no compact curves. The 6d $(1,0)$ theories considered in this paper, $\mathcal{T}_\mathfrak{g}\{Y_1, Y_2\}$, are obtained from Higgs branch flows of minimal conformal matter. 
It is easily checked from the tensor branch configurations in equation \eqref{eqn:cms} that the theories $\mathcal{T}_\mathfrak{g}$ are very Higgsable, and thus the theories $\mathcal{T}_\mathfrak{g}\{Y_1, Y_2\}$ are also very Higgsable.

If a 6d SCFT is very Higgsable, then one can use similar arguments to those that appear in \cite{Shapere:2008zf} to determine the central charges and flavor central charges of the 4d SCFT that arises from the $T^2$ compactification. We write the relevant terms of the anomaly polynomial of the 6d SCFT as
\begin{equation}
    I_8 =\frac{\beta}{24}  c_2(R) p_1(T) + \frac{\gamma}{24}  p_1(T)^2 + \frac{\delta}{24} p_2(T) + \sum_i \kappa_i p_1(T) TrF_i^2 + \cdots \,, 
\end{equation}
where the sum runs over the simple non-Abelian flavor algebras of the SCFT, and the ellipses contain terms unrelated to (mixed-)gravitational anomalies. The central charges of the 4d SCFT are then given by \cite{Ohmori:2015pua}:
\begin{equation}
    (a, c) = \left(\gamma - \frac{1}{2}\beta - \frac{3}{4}\delta\,, \frac{8}{3}\gamma - \frac{1}{2}\beta - \frac{1}{3}\delta\right) \,.
\end{equation}
These can be converted to the effective numbers of vector multiplets and hypermultiplets via
\begin{equation}\label{eqn:nvnh}
  \begin{aligned}
    (n_v, n_h) &= (4(2a - c), -4(4a - 5c)) \cr
    &= \left(-2\beta - \frac{2}{3}(4\gamma + 7\delta), -2\beta + \frac{16}{3}(4\gamma + 16\delta)\right) \,.
  \end{aligned}
\end{equation}
Similarly, the central charge of the flavor symmetry algebra is
\begin{equation}\label{eqn:ki}
    k_i = 192 \kappa_i \,.
\end{equation}

In this way, the central charges of the 4d $\mathcal{N}=2$ theories $\mathcal{T}_\mathfrak{g}\{Y_1, Y_2\}\langle T^2 \rangle$ can be determined directly from the anomaly polynomials of the 6d $(1,0)$ theories $\mathcal{T}_\mathfrak{g}\{Y_1, Y_2\}$.

\section{Matching 4d SCFTs from different 6d origins}\label{sec:Eesults}

We have introduced all the necessary tools to verify the following relationship for the 4d $\mathcal{N}=2$ theories with two different 6d origins, where the top line is from 6d $(2,0)$ SCFTs and the bottom line is from 6d $(1,0)$ SCFTs:
\begin{align}
\begin{aligned}
\equalto{\mathcal{S}_\mathfrak{g}\langle C_{0,3}\rangle\{Y_1, Y_2, Y_\text{simple}\}}{} 
&\xrightarrow{\ \text{Partial closure of punctures}\ }
\equalto{\mathcal{S}_\mathfrak{g}\langle C_{0,3}\rangle\{\widetilde{Y}_1, Y_2, Y_\text{simple}\}}{}\\[-9pt]
\mathcal{T}_\mathfrak{g}\{Y_1, Y_2\}\qquad\ &\ \xrightarrow{\quad \text{Higgs flow / RG flow}\quad }\qquad\ \mathcal{T}_\mathfrak{g}\{\widetilde{Y}_1, Y_2\}\qquad\qquad .
\end{aligned}
\label{eq:2origins6D}
\end{align}
Here, the equals signs denote the equality of 4d theories after compactifying the 6d $(1,0)$ SCFTs on the bottom row on a torus. We reviewed in Section \ref{sec:classsintro} how to obtain the central charges, flavor symmetries, and other physical data of the 4d $\mathcal{N}=2$ SCFTs obtained by compactifying the 6d $(2,0)$ theory of type $\mathfrak{g}$ on a punctured Riemann surface. These provide the data necessary for the top line in equation \eqref{eq:2origins6D}. In contrast, we get the constructions of 6d $(1,0)$ SCFTs from F-theory and further compactify on a torus to yield the 4d $\mathcal{N}=2$ theories, corresponding to the bottom line of equation \eqref{eq:2origins6D}. We summarized such construction of 6d $(1,0)$ SCFTs from F-theory and how to determine their flavor symmetries and anomaly polynomials in Sections \ref{sec:geom-config}, \ref{sec:flavor}, and \ref{sec:anom}. We explained how to determine the central charges, flavor symmetries, and other physical data of the 4d $\mathcal{N}=2$ SCFTs obtained by compactifying the 6d $(1,0)$ theories on a torus in Section \ref{sec:4dto6d}. 

Utilizing all of these data and methodology, in this section, we compare the 4d $\mathcal{N}=2$ theories obtained from the 6d $(2,0)$ constructions and those from the 6d $(1,0)$ constructions when $\mathfrak{g}$ is an exceptional Lie algebra. It was proposed that the minimal $(\mathfrak{g}, \mathfrak{g})$ conformal matter compactified on a $T^2$ gives rise to the same 4d $\mathcal{N}=2$ SCFT as the 6d $(2,0)$ SCFT of type $\mathfrak{g}$ compactified on a sphere with two full punctures and one simple puncture \cite{Ohmori:2015pua}. This was extended to rank $N$ $(\mathfrak{g}, \mathfrak{g})$ conformal matter compactified on $T^2$ being equivalent to the 6d $(2,0)$ SCFT of type $\mathfrak{g}$ compactified on a sphere with two full and $N$ simple punctures \cite{DelZotto:2015rca}. In this section, we show that this identification extends to all nilpotent Higgs branch deformations of minimal $(\mathfrak{g}, \mathfrak{g})$ conformal matter and all partial closures of the full punctures of the fixture, as presented in equation \eqref{eq:2origins6D}. 

We first discuss the verification of equation \eqref{eq:2origins6D} for the case of $\mathfrak{g}=\mathfrak{e}_6$. Let us start from the 6d $(1,0)$ origin with $(\mathfrak{e}_6, \mathfrak{e}_6)$ conformal matter. As explained in Section \ref{sec:geom-config}, the associated non-compact elliptically fibered Calabi--Yau geometry can be written as
\begin{equation}
    1\overset{\mathfrak{su}_3}{3}1 \,,
\end{equation}
and this theory has a flavor symmetry $\mathfrak{e}_6\oplus\mathfrak{e}_6$. We can give a nilpotent vacuum expectation value to break one of the $\mathfrak{e}_6$ flavor symmetries:
\begin{equation}
    \mathfrak{e}_6 \rightarrow \mathfrak{su}_{6} \,.
\end{equation}
This particular nilpotent orbit is associated to the Bala--Carter label $A_1$. After doing such a complex structure deformation the geometric data at a generic point of the tensor branch is
\begin{equation}
    1\overset{\mathfrak{su}_3}{2} \,.
\end{equation}
This configuration has two further possible deformations. One of them corresponds to giving a nilpotent vacuum expectation value in the same orbit $A_1$ to the other $\mathfrak{e}_6$ factor. This would naively lead to an $\mathfrak{su}_6 \oplus \mathfrak{su}_6$ flavor symmetry, but this is not the case. The configuration at a generic point of the tensor branch after this deformation is
\begin{equation}
    \overset{\mathfrak{su}_3}{1} \,,
\end{equation}
and this shows apparently that the flavor symmetry enhances to an $\mathfrak{su}_{12}$ instead. It is important to note that, when the flavor symmetry is enhanced beyond the naive flavor symmetry associated to each nilpotent orbit, one can read the enhanced flavor symmetry directly off from the geometric configuration, as demonstrated in this simple case. The cancellation of anomalies requires than an $\mathfrak{su}_3$ algebra over a $(-1)$-curve has twelve fundamental hypermultiplets, and thus there is an $\mathfrak{su}_{12}$ flavor symmetry rotating those hypermultiplets, as described in Section \ref{sec:flavor}.

\begin{figure}[H]
    \centering
    \vspace{5mm}
    \includegraphics[scale=1.3]{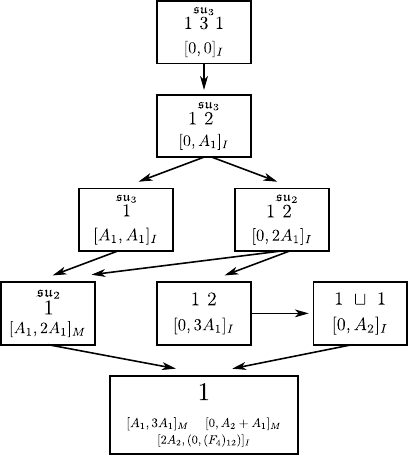}
    \vspace{3mm}
    \caption{Here we depict the Hasse diagram for both the 6d $(1,0)$ theories $\mathcal{T}_{\mathfrak{e}_6}\{Y_1, Y_2\}$ and the 4d $\mathcal{N}=2$ theories $\mathcal{S}_{\mathfrak{e}_6}\langle C_{0,3}\rangle\{Y_1, Y_2, Y_\text{simple}\}$. In each box we write the geometric configuration for $\mathcal{T}_{\mathfrak{e}_6}\{Y_1, Y_2\}$ and $[Y_1, Y_2]_A$, where $Y_1$, $Y_2$ are the Bala--Carter labels for the nilpotent orbits and $A = I\text{(nteracting)}, M\text{(ixed)}$ indicates whether the class $\mathcal{S}$ fixture $\mathcal{S}_{\mathfrak{e}_6}\langle C_{0,3}\rangle\{Y_1, Y_2, Y_\text{simple}\}$ contains a free sector. A sequence of arrows relates boxes associated to $[Y_1, Y_2]$ and $[\widetilde{Y}_1, Y_2]$ if $Y_1 > \widetilde{Y}_1$ in the partial ordering of nilpotent orbits. The central charges and flavor symmetries of the 4d $\mathcal{N}=2$ theories shown here appear in Table \ref{tbl:E6E6}.}\label{fig:E66ddef}
\end{figure}

The other deformation leads to the tensor branch configuration
\begin{equation}
    1\overset{\mathfrak{su}_2}{2} \,,
\end{equation}
where the $\mathfrak{su}_6$ broke further down. The flavor symmetry for this theory is then $\mathfrak{e}_7 \oplus \mathfrak{so}_7$. Again, we see that the enhancement and recombination of the flavor symmetry is manifest in the configuration.
Continuing this process, it is easy to find all possible deformations of the $(\mathfrak{e}_6, \mathfrak{e}_6)$ conformal matter, and they are summarized in the Hasse diagram depicted in Figure \ref{fig:E66ddef}.

We can obtain 4d $\mathcal{N}=2$ SCFTs by compactifying these 6d $(1,0)$ theories on a torus. It is straightforward to compute their central charges using the equation \eqref{eqn:nvnh}, as well as their flavor symmetries which can be readily obtained from the geometric data. This information for the case of the $(\mathfrak{e}_6,\mathfrak{e}_6)$ conformal matter is gathered in Table \ref{tbl:E6E6}.

\begin{table}[H]
\centering
\renewcommand{\arraystretch}{1.4}
\footnotesize
\begin{threeparttable}
\resizebox{\textwidth}{!}{
$\begin{array}{c c c c l}
\toprule
\phantom{dd}\text{dim}(\mathcal{C})\phantom{dd} & \phantom{Fix}\text{Fixture}\phantom{Fix} & \phantom{CFT}\text{6d SCFT}\phantom{CFT} & \phantom{n_vn_h}(n_v, n_h)\phantom{n_vn_h} & \text{Flavor Symmetry}\phantom{6d} \\
\midrule
\multirow{1}{*}{5} & [0, 0]_I & 1\overset{\mathfrak{su}_3}{3}1 & (89, 168) & (\mathfrak{e}_6)_{24}\oplus(\mathfrak{e}_6)_{24}  \\\midrule
\multirow{1}{*}{4} & [0, A_1]_I & 1\overset{\mathfrak{su}_3}{2} & (66, 134) & (\mathfrak{e}_6)_{24}\oplus(\mathfrak{su}_6)_{18}  \\\midrule
\multirow{2}{*}{3} & [0, 2A_1]_I & 1\overset{\mathfrak{su}_2}{2} & (49, 112) & (\mathfrak{e}_7)_{24}\oplus(\mathfrak{so}_7)_{16} \\\cline{2-5}
 & [A_1, A_1]_I & \overset{\mathfrak{su}_3}{1} & (43, 100) & (\mathfrak{su}_{12})_{18} \\\midrule
\multirow{4}{*}{2} & [0, 3A_1]_I & 1\overset{\phantom{\mathfrak{su}_2}}{2} & (34, 93) & (\mathfrak{e}_8)_{24}\oplus(\mathfrak{su}_2)_{13} \\\cline{2-5}
 & [A_1, 2A_1]_M & \overset{\mathfrak{su}_2}{1} & (26, 72) & (\mathfrak{so}_{20})_{16} \\\cline{2-5}
 & [0, A_2]_I & \overset{\phantom{\mathfrak{su}_2}}{1}\sqcup\overset{\phantom{\mathfrak{su}_2}}{1} & (22, 80) & (\mathfrak{e}_8)_{12}\oplus(\mathfrak{e}_8)_{12} \\\midrule
\multirow{1}{*}{1} & \begin{gathered}[] [0, A_2 + A_1]_M \cr [A_1, 3A_1]_M \cr [2A_2, (0, (F_4)_{12})]_I \end{gathered} & 1 & (11,40) & (\mathfrak{e}_8)_{12} \\\bottomrule
\end{array}$}
\end{threeparttable}
\vspace{3mm}
\caption{This table demonstrates the relationship between the 4d $\mathcal{N}=2$ SCFTs obtained as $T^2$ compactifications of nilpotent Higgsings of minimal $(\mathfrak{e}_6, \mathfrak{e}_6)$ conformal matter $\mathcal{T}_{\mathfrak{e}_6}\{Y_1, Y_2\}\langle T^2\rangle$ and the class $\mathcal{S}$ theories $\mathcal{S}_{\mathfrak{e}_6}\langle C_{0,3}\rangle\{Y_1, Y_2, Y_\text{simple}\}$. In the fixture column, we write the Bala--Carter labels for the nilpotent orbits $Y_1$ and $Y_2$. The subscript is $M$ if the 4d $\mathcal{N} = 2$ theory $\mathcal{S}_{\mathfrak{e}_6}\langle C_{0,3}\rangle\{Y_1, Y_2, Y_\text{simple}\}$ has both an interacting sector and additional free hypermultiplets; the subscript $I$ means that there are no free hypermultiplets. In the central charge column, we list the central charges as obtained from the $T^2$ compactification of the 6d $(1,0)$ SCFT via the equation \eqref{eqn:nvnh}, and equally those obtained from the class $\mathcal{S}$ construction using the equation \eqref{eqn:nvnhS}, after subtracting the contributions from any free hypermultiplets. The flavor symmetry and flavor central charge, denoted as the subscript, are determined from the 6d $(1,0)$ SCFT perspective following Section \ref{sec:flavor} and the equation \eqref{eqn:ki}. They are equally determined from the class $\mathcal{S}$ perspective via the computation of the $\tau^2$ term of the Hall--Littlewood index, which was performed in \cite{Chacaltana:2014jba}.}
\label{tbl:E6E6}
\end{table}

We now turn to the 6d $(2,0)$ origin when $\mathfrak{g}=\mathfrak{e}_6$. The 4d $\mathcal{N}=2$ theories
\begin{equation}
    \mathcal{S}_{\mathfrak{e}_6}\langle C_{0,3} \rangle \{Y_1, Y_2, Y_\text{simple}\} \,,
\end{equation}
are specified by a pair of $\mathfrak{e}_6$ nilpotent orbits, $[Y_1, Y_2]$. These pairs can be arranged into a Hasse diagram under the partial ordering of nilpotent orbits, where we identify pairs that lead to the same interacting 4d SCFT, and this is also displayed in Figure \ref{fig:E66ddef}.

The 6d $(2,0)$ SCFT when compactified on a three-punctured sphere may give rise to a 4d SCFT that is a product of an interacting SCFT and a collection of free hypermultiplets. The number of free hypermultiplets can be determined by computing the $\mathcal{O}(\tau)$ term of the Hall--Littlewood index, and thus it is straightforward to subtract those and obtain the central charges for the interacting sectors utilizing equation \eqref{eqn:nvnhS}. These central charges, which are the same as the central charges of the theories $\mathcal{T}_{\mathfrak{e}_6}\{Y_1, Y_2\}\langle T^2 \rangle$, are collated in Table \ref{tbl:E6E6}.

The determination of the flavor symmetries and the flavor central charges requires additional care, as the $C_{0,3}$ in the 6d $(2,0)$ origin only makes manifest the flavor symmetries associated to the individual punctures. To determine the full flavor symmetry one must compute the coefficient of the $\tau^2$ term in the Hall--Littlewood index, which is given in terms of characters of the representations of the manifest flavor algebras, and then recombine those into the summands of a branching rule for the adjoint representation of some larger algebra $\mathfrak{f}$. This is then the superconformal flavor symmetry. The flavor symmetry enhancements were worked out for all $\mathfrak{e}_6$ fixtures with untwisted punctures in \cite{Chacaltana:2014jba}. These non-Abelian flavor symmetries, which are identical to the flavor symmetries manifest in the theories $\mathcal{T}_{\mathfrak{e}_6}\{Y_1, Y_2\}\langle T^2 \rangle$ are listed in the final column of Table \ref{tbl:E6E6}.

\begin{sidewaysfigure}[H]
    \centering
    \includegraphics[width=\textheight]{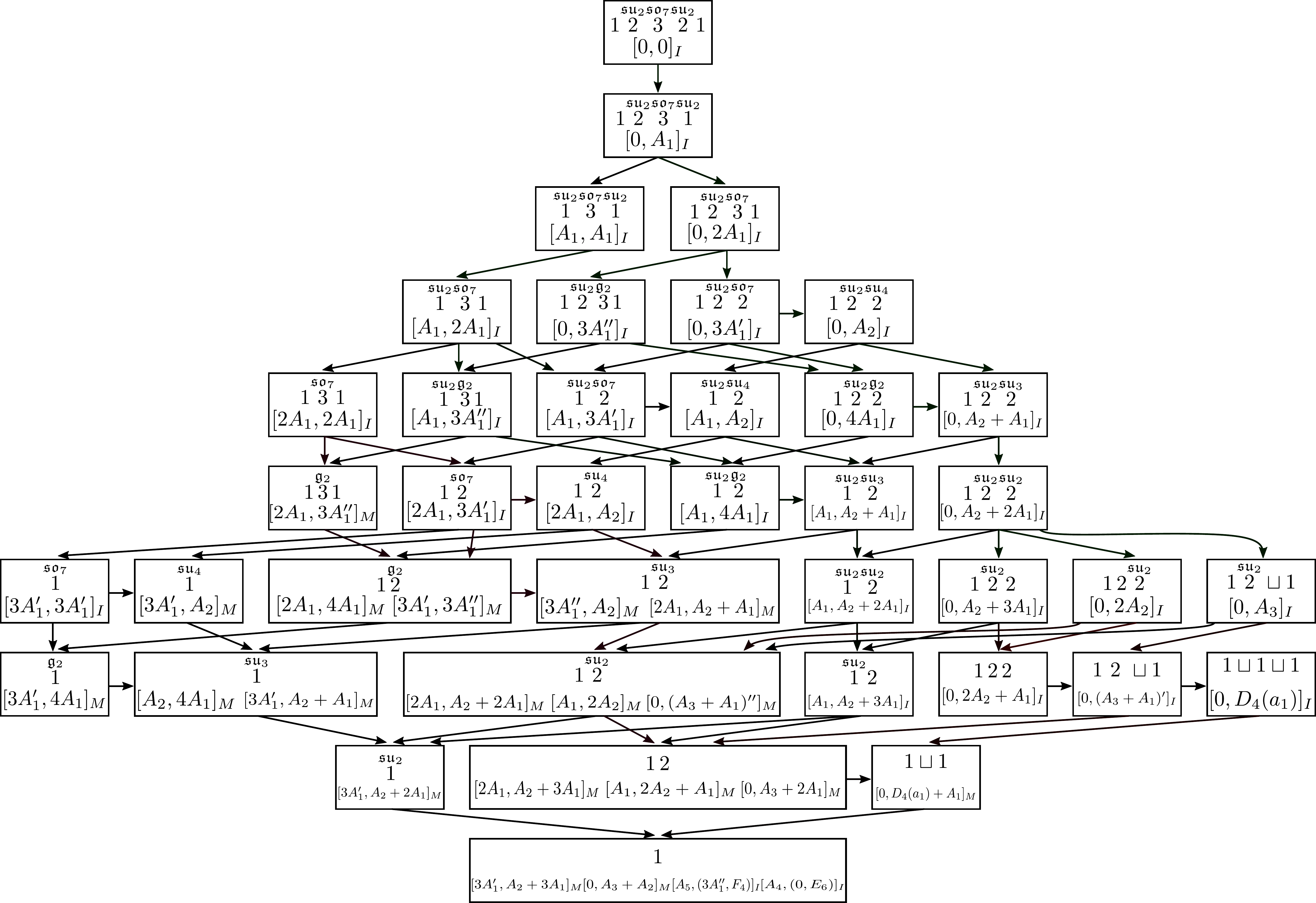}
    \caption{The Hasse diagram for both the 6d $(1,0)$ theories $\mathcal{T}_{\mathfrak{e}_7}\{Y_1, Y_2\}$ and the 4d $\mathcal{N}=2$ theories $\mathcal{S}_{\mathfrak{e}_7}\langle C_{0,3}\rangle\{Y_1, Y_2, Y_\text{simple}\}$. The notation is the same as described in Figure \ref{fig:E66ddef}. The physical properties of the 4d $\mathcal{N}=2$ SCFTs are collected in Table \ref{tbl:E7E7}.} 
    \label{fig:E7Tree}
\end{sidewaysfigure}

Similar correspondences can be established mutatis mutandis when 
$\mathfrak{g}=\mathfrak{e}_7$ and $\mathfrak{g}=\mathfrak{e}_8$. In those cases,
we again observe that the constructions $\mathcal{T}_{\mathfrak{g}}\{Y_1, Y_2\}\langle T^2 \rangle$ and $\mathcal{S}_{\mathfrak{g}}\langle C_{0,3} \rangle \{ Y_1, Y_2, Y_\text{simple}\}$ yield the same set of 4d $\mathcal{N}=2$ SCFTs. The number of consistent fixtures is much larger in these cases and the Hasse diagrams representing the flows are consequently more involved. However, the procedures to find the central charges and flavor symmetries from both the 6d $(1,0)$ and 6d $(2,0)$ points of view are algorithmic. Verification of the equation \eqref{eqn:big} for $\mathfrak{g}=\mathfrak{e}_7$ is given in Table \ref{tbl:E7E7} and the Hasse diagram in Figure \ref{fig:E7Tree}; for $\mathfrak{g}=\mathfrak{e}_8$ the results appear in Table \ref{tbl:E8E8} and Figures \ref{fig:E8Tree1} and \ref{fig:E8Tree2}.

\newpage
\begin{table}[H]
\centering
\renewcommand{\arraystretch}{1.3}
\footnotesize   
\begin{threeparttable}
$\begin{array}{ccccl}
\toprule
\phantom{dd}\text{dim}(\mathcal{C})\phantom{dd} & \phantom{Fix}\text{Fixture}\phantom{Fix} & \phantom{FT}\text{6d SCFT}\phantom{FT} & \phantom{nv}(n_v, n_h)\phantom{nv} & \text{Flavor Symmetry}\phantom{6d} 
\\
\midrule
\multirow{1}{*}{10} & [0, 0]_I & 1\overset{\mathfrak{su}_{2}}{2}\overset{\mathfrak{so}_{7}}{3}\overset{\mathfrak{su}_{2}}{2}1 & (250, 384) & (\mathfrak{e}_7)_{36} \oplus (\mathfrak{e}_7)_{36}   
\\\midrule
\multirow{1}{*}{9} & [0, A_{1}]_I & 1\overset{\mathfrak{su}_{2}}{2}\overset{\mathfrak{so}_{7}}{3}\overset{\mathfrak{su}_{2}}{1} & (215, 332) & (\mathfrak{e}_7)_{36} \oplus (\mathfrak{so}_{12})_{28}   
\\\midrule
\multirow{2}{*}{8} & [0, 2A_{1}]_I & 1\overset{\mathfrak{su}_{2}}{2}\overset{\mathfrak{so}_{7}}{3}1 & (188, 296) & (\mathfrak{e}_7)_{36} \oplus (\mathfrak{so}_{9})_{24} \oplus (\mathfrak{su}_{2})_{20}   
\\\cline{2-5}
 & [A_{1}, A_{1}]_I & \overset{\mathfrak{su}_{2}}{1}\overset{\mathfrak{so}_{7}}{3}\overset{\mathfrak{su}_{2}}{1} &  (180, 280) & (\mathfrak{so}_{12})_{28} \oplus (\mathfrak{so}_{12})_{28}    
 \\\midrule
\multirow{4}{*}{7} & [0, 3A_{1}'']_I & 1\overset{\mathfrak{su}_{2}}{2}\overset{\mathfrak{g}_{2}}{3}1 &  (169, 276) &  (\mathfrak{e}_7)_{36} \oplus (\mathfrak{f}_4)_{24}   
\\\cline{2-5}
 & [A_{1}, 2A_{1}]_I & \overset{\mathfrak{su}_{2}}{1}\overset{\mathfrak{so}_{7}}{3}1 &  (153, 244) & (\mathfrak{so}_{12})_{28} \oplus (\mathfrak{so}_{9})_{24} \oplus (\mathfrak{su}_{2})_{20}   
 \\\cline{2-5}
 & [0, 3A_1']_I &  1\overset{\mathfrak{su}_{2}}{2}\overset{\mathfrak{so}_{7}}{2} & (165, 267) & (\mathfrak{e}_7)_{36} \oplus (\mathfrak{usp}_{6})_{20} \oplus (\mathfrak{su}_{2})_{19}   
 \\\cline{2-5}
 & [0, A_{2}]_I & 1\overset{\mathfrak{su}_{2}}{2}\overset{\mathfrak{su}_{4}}{2} &  (147, 248) & (\mathfrak{e}_7)_{36} \oplus (\mathfrak{su}_{6})_{20}   
 \\\midrule
\multirow{6}{*}{6} & [0, 4A_{1}]_I & 1\overset{\mathfrak{su}_{2}}{2}\overset{\mathfrak{g}_{2}}{2} &  (146, 245) & (\mathfrak{e}_7)_{36} \oplus (\mathfrak{usp}_{6})_{19}   
\\\cline{2-5}
 & [0, A_{2} + A_{1}]_I & 1\overset{\mathfrak{su}_{2}}{2}\overset{\mathfrak{su}_{3}}{2} &  (128, 224) &  (\mathfrak{e}_7)_{36} \oplus (\mathfrak{su}_{4})_{18}  
 \\\cline{2-5}
 & [A_{1}, 3A_{1}'']_I & \overset{\mathfrak{su}_{2}}{1}\overset{\mathfrak{g}_{2}}{3}1  & (134, 224) &  (\mathfrak{f}_4)_{24} \oplus (\mathfrak{so}_{13})_{28}   
 \\\cline{2-5}
 & [2A_{1}, 2A_{1}]_I & 1\overset{\mathfrak{so}_{7}}{3}1 &  (126, 208) & (\mathfrak{so}_{9})_{24} \oplus (\mathfrak{so}_{9})_{24} \oplus (\mathfrak{usp}_{2})_{20}   
 \\\cline{2-5}
 & [A_{1}, 3A_1']_I & \overset{\mathfrak{su}_{2}}{1}\overset{\mathfrak{so}_{7}}{2} &  (130, 215) &  (\mathfrak{so}_{12})_{28} \oplus (\mathfrak{usp}_{6})_{20} \oplus (\mathfrak{su}_{2})_{19}   
 \\\cline{2-5}
 & [A_{1}, A_{2}]_I & \overset{\mathfrak{su}_{2}}{1}\overset{\mathfrak{su}_{4}}{2} &  (112, 196) &  (\mathfrak{so}_{12})_{28} \oplus (\mathfrak{su}_{6})_{20} 
 \\\midrule
\multirow{6}{*}{5} & [0, A_{2} + 2A_{1}]_I & 1\overset{\mathfrak{su}_{2}}{2}\overset{\mathfrak{su}_{2}}{2} &  (111, 204) & (\mathfrak{e}_7)_{36} \oplus (\mathfrak{su}_{2})_{28} \oplus (\mathfrak{su}_{2})_{16} \oplus (\mathfrak{su}_{2})_{84}  
\\\cline{2-5}
 & [2A_1, 3A_1'']_M & 1\overset{\mathfrak{g}_{2}}{3}1 &  (107, 187) & (\mathfrak{f}_4)_{24} \oplus (\mathfrak{f}_4)_{24} \oplus (\mathfrak{su}_{2})_{19}   
 \\\cline{2-5}
 & [A_{1}, 4A_{1}]_I & \overset{\mathfrak{su}_{2}}{1}\overset{\mathfrak{g}_{2}}{2} &  (111, 193) &  (\mathfrak{so}_{13})_{28} \oplus (\mathfrak{usp}_{6})_{19}   
 \\\cline{2-5}
 & [A_{1}, A_{2} + A_{1}]_I & \overset{\mathfrak{su}_{2}}{1}\overset{\mathfrak{su}_{3}}{2} &  (93, 172) &  (\mathfrak{so}_{14})_{28} \oplus (\mathfrak{su}_{4})_{18}  
 \\\cline{2-5}
 & [2A_{1}, 3A_1']_I & 1\overset{\mathfrak{so}_{7}}{2} &  (103, 179) & (\mathfrak{so}_{9})_{24} \oplus (\mathfrak{usp}_{8})_{20} \oplus (\mathfrak{su}_{2})_{19}   
 \\\cline{2-5}
 & [2A_{1}, A_{2}]_I & 1\overset{\mathfrak{su}_{4}}{2} & (85, 160) & (\mathfrak{so}_{10})_{24} \oplus (\mathfrak{su}_{8})_{20}   
 \\\midrule
\multirow{10}{*}{4} & [0, A_{2} + 3A_{1}]_I & 1\overset{\mathfrak{su}_{2}}{2}2 &  (96, 188) & (\mathfrak{e}_7)_{36} \oplus (\mathfrak{g}_2)_{28}   
\\\cline{2-5}
 & [0, 2A_{2}]_I & 12\overset{\mathfrak{su}_{2}}{2} &  (84, 176) & (\mathfrak{e}_8)_{36} \oplus (\mathfrak{g}_2)_{16}   
 \\\cline{2-5}
 & \begin{gathered}[] [2A_1, 4A_1]_M \cr [3A_1', 3A_1'']_M \end{gathered} & 1\overset{\mathfrak{g}_{2}}{2} &  (84, 156) & (\mathfrak{f}_4)_{24} \oplus (\mathfrak{usp}_{8})_{19}    
 \\\cline{2-5}
 & \begin{gathered}[] [2A_1, A_2+A_1]_M \cr [3A_1'', A_2]_M \end{gathered} & 1\overset{\mathfrak{su}_{3}}{2} &  (66, 134) & (\mathfrak{e}_6)_{24} \oplus (\mathfrak{su}_{6})_{18}    
 \\\cline{2-5}
 & [A_{1}, A_{2} + 2A_{1}]_I & \overset{\mathfrak{su}_{2}}{1}\overset{\mathfrak{su}_{2}}{2} &  (76, 152) & (\mathfrak{so}_{16})_{28} \oplus (\mathfrak{su}_{2})_{16} \oplus (\mathfrak{su}_{2})_{56}  
 \\\cline{2-5}
 & [3A_1', 3A_1']_I & \overset{\mathfrak{so}_{7}}{1} &  (80, 150) & (\mathfrak{usp}_{12})_{20} \oplus (\mathfrak{usp}_{2})_{19}   
 \\\cline{2-5}
 &  [3A_1', A_2]_M &  \overset{\mathfrak{su}_{4}}{1} & (62, 130) & (\mathfrak{su}_{12})_{20} \oplus (\mathfrak{su}_{2})_{18}   
 \\\cline{2-5}
  & [0, A_3]_I & 1\overset{\mathfrak{su}_{2}}{2} \sqcup 1 &  (60, 152) & (\mathfrak{e}_8)_{12} \oplus (\mathfrak{e}_7)_{24} \oplus (\mathfrak{so}_{7})_{16}   
  \\\bottomrule
\end{array}$
\end{threeparttable}
\end{table}
\newpage
\begin{table}[H]
\centering
\renewcommand{\arraystretch}{1.3}
\footnotesize
\begin{threeparttable}
$\begin{array}{ccccl}
\toprule
\phantom{dd}\text{dim}(\mathcal{C})\phantom{dd} & \phantom{Fixt}\text{Fixture}\phantom{Fixt} & \phantom{CFT}\text{6d SCFT}\phantom{CFT} & \phantom{nv}(n_v, n_h)\phantom{nv} & \text{Flavor Symmetry}\phantom{6dflavor} 
\\
\midrule
\multirow{12}{*}{3} & [0, 2A_{2}+ A_{1}]_I & 1\overset{\phantom{\mathfrak{su}_2}}{2}2 &  (69, 158) & (\mathfrak{e}_8)_{36} \oplus (\mathfrak{su}_{2})_{38}   
\\\cline{2-5}
& [0,(A_3+A_1)']_I  & \overset{\phantom{\mathfrak{su}_2}}{1} \overset{\phantom{\mathfrak{su}_2}}{2}\sqcup\overset{\phantom{\mathfrak{su}_2}}{1} & (45, 133) & (\mathfrak{e}_8)_{12} \oplus (\mathfrak{e}_8)_{24} \oplus (\mathfrak{su}_{2})_{13} 
\\\cline{2-5}
 & [A_{1}, A_{2} + 3A_{1}]_I & \overset{\mathfrak{su}_{2}}{1}2 &  (61, 136) & (\mathfrak{so}_{19})_{28} 
 \\\cline{2-5}
 & \begin{gathered}[] [0, {(A_3+A_1)''}]_M \cr [A_1, 2A_2]_M \cr [2A_1, A_2+2A_1]_M \end{gathered} & 1\overset{\mathfrak{su}_{2}}{2} &  (49, 112) & (\mathfrak{e}_7)_{24} \oplus (\mathfrak{so}_{7})_{16} 
 \\\cline{2-5}
 & [3A_1', 4A_1]_M & \overset{\mathfrak{g}_{2}}{1} &  (61, 125) & (\mathfrak{usp}_{14})_{19} 
 \\\cline{2-5}
 & \begin{gathered}[] [3A_1', A_2+A_1]_M \cr [A_2, 4A_1]_M \end{gathered} & \overset{\mathfrak{su}_{3}}{1} &  (43, 100) & (\mathfrak{su}_{12})_{18} 
 \\\cline{2-5}
 & [0,D_4(a_1)]_I & \overset{\phantom{\mathfrak{su}_2}}{1}\sqcup \overset{\phantom{\mathfrak{su}_2}}{1}\sqcup \overset{\phantom{\mathfrak{su}_2}}{1} &  (33, 120) &  (\mathfrak{e}_8)_{12} \oplus (\mathfrak{e}_8)_{12} \oplus (\mathfrak{e}_8)_{12} 
 \\\midrule
\multirow{4}{*}{2}  & \begin{gathered}[]
[0, A_3+2A_1]_M \cr [A_1, 2A_2+A_1]_M \cr [2A_1, A_2+3A_1]_M
\end{gathered} & 1\overset{\phantom{\mathfrak{su}_2}}{2} &  (34, 93) & (\mathfrak{e}_8)_{24} \oplus (\mathfrak{su}_{2})_{13} 
\\\cline{2-5}
 & [3A_1', A_2+2A_1]_M & \overset{\mathfrak{su}_{2}}{1} & (26, 72) & (\mathfrak{so}_{20})_{16} 
 \\\cline{2-5}
 & [0,D_4(a_1)+A_1]_M & \overset{\phantom{\mathfrak{su}_2}}{1}\sqcup \overset{\phantom{\mathfrak{su}_2}}{1} &  (22, 80) & (\mathfrak{e}_8)_{12} \oplus (\mathfrak{e}_8)_{12} 
 \\\midrule
1 & \begin{gathered}[] [0, A_3+A_2]_M \cr [3A_1', A_2+3A_1]_M \cr [A_4, (0, E_6)]_I \cr [A_5^{\prime\prime}, (3A_1^{\prime\prime}, F_4)]_I \end{gathered} & \overset{\phantom{\mathfrak{su}_2}}{1} &  (11, 40) & (\mathfrak{e}_8)_{12} 
\\\bottomrule
\end{array}$
\end{threeparttable}
\vspace{2mm}
\caption{In this table, we show the relationship between the 4d $\mathcal{N}=2$ SCFTs obtained as $T^2$ compactifications of nilpotent Higgsing of minimal $(\mathfrak{e}_7, \mathfrak{e}_7)$ conformal matter, $\mathcal{T}_{\mathfrak{e}_7}\{Y_1, Y_2\}\langle T^2\rangle$, and the class $\mathcal{S}$ theories $\mathcal{S}_{\mathfrak{e}_7}\langle C_{0,3}\rangle\{Y_1, Y_2, Y_\text{simple}\}$. The information contained in each column is as described in the caption of Table \ref{tbl:E6E6}. From the class $\mathcal{S}$ perspective the determination of the flavor symmetry requires knowledge of the Hall--Littlewood index, which was determined for $\mathfrak{e}_7$ fixtures in \cite{Chacaltana:2017boe}.}
\label{tbl:E7E7}
\end{table}

\newpage
\newgeometry{bottom=1.1cm}
\begin{sidewaysfigure}[H]
    \centering
    \includegraphics[width=1.1\textheight]{E8tree_part1.pdf}
        \caption{The Hasse diagram for both the 6d $(1,0)$ theories $\mathcal{T}_{\mathfrak{e}_8}\{Y_1, Y_2\}$ and the 4d $\mathcal{N}=2$ theories $\mathcal{S}_{\mathfrak{e}_8}\langle C_{0,3}\rangle\{Y_1, Y_2, Y_\text{simple}\}$. The notation is the same as described in Figure \ref{fig:E66ddef}. The physical properties of the 4d $\mathcal{N}=2$ SCFTs are collected in Table \ref{tbl:E8E8}. This figure is continued in Figure \ref{fig:E8Tree2}.} 
    \label{fig:E8Tree1}
\end{sidewaysfigure}
\restoregeometry

\newpage
\newgeometry{bottom=1.1cm}
\begin{sidewaysfigure}[H]
    \centering
    \includegraphics[width=1.1\textheight]{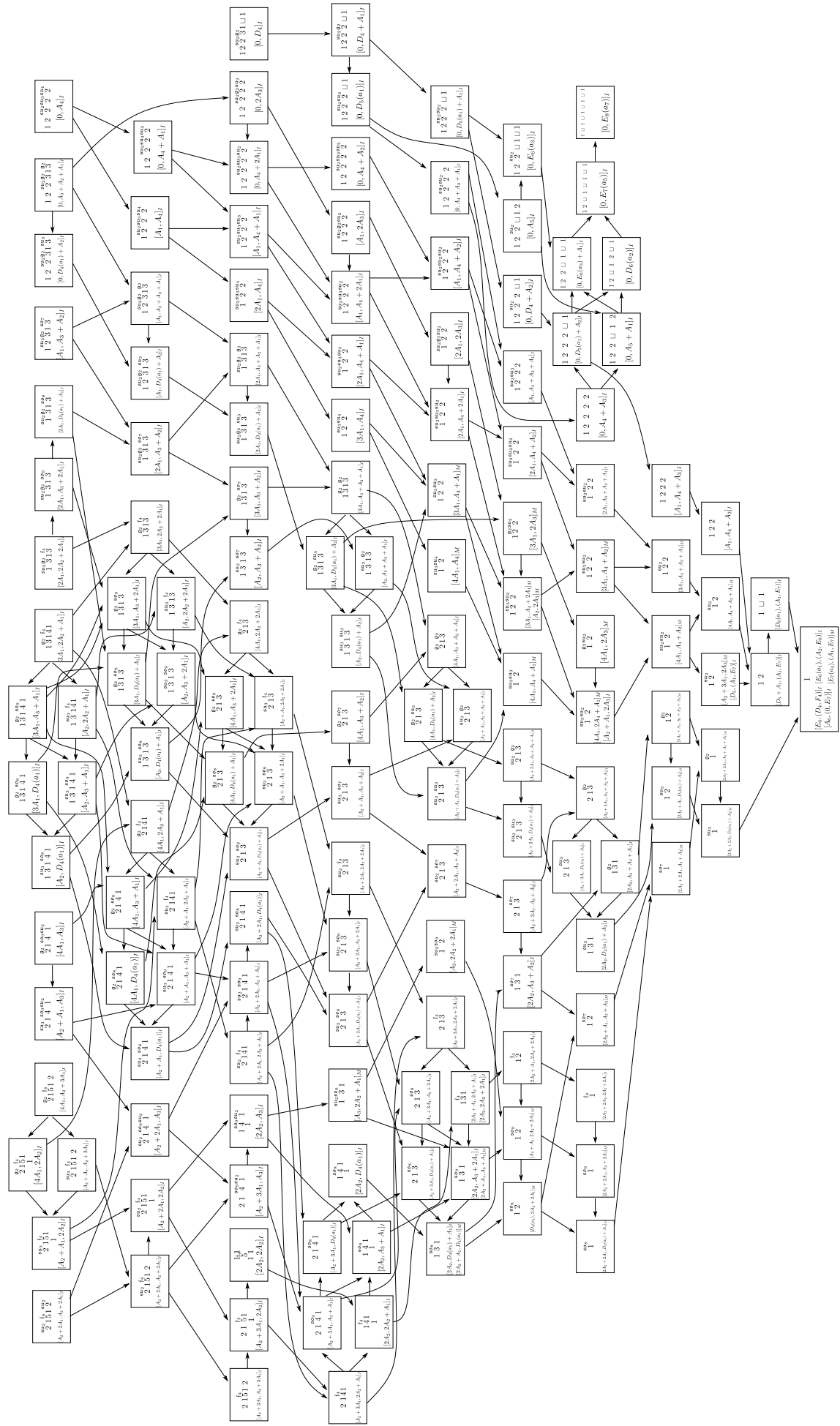}
    \caption{The second half of Figure \ref{fig:E8Tree1}. The final row of Figure \ref{fig:E8Tree1} is repeated here for clarity.}
    \label{fig:E8Tree2}
\end{sidewaysfigure}
\restoregeometry

\newpage
\begin{table}[H]
\begin{threeparttable}
\centering
\renewcommand{\arraystretch}{1.3}
\footnotesize
$\begin{array}{c c c c l}
\toprule
\phantom{dd}\text{dim}(\mathcal{C})\phantom{dd} & \phantom{Fix}\text{Fixture}\phantom{Fix} & \phantom{CFT}\text{6d SCFT}\phantom{CFTT} & \phantom{nv}(n_v, n_h)\phantom{nv} & \text{Flavor Symmetry}\phantom{FlavorSym}
\\
\midrule
21 & [0, 0]_I & 1\ 2\overset{\mathfrak{su}_{2}}{2}\overset{\mathfrak{g}_{2}}{3}1\overset{\mathfrak{f}_{4}}{5}1\overset{\mathfrak{g}_{2}}{3}\overset{\mathfrak{su}_{2}}{2}2\ 1 & (831, 1080) & (\mathfrak{e}_8)_{60} \oplus (\mathfrak{e}_8)_{60}   
\\\midrule

20 & [0, A_{1}]_I & 1\ 2\overset{\mathfrak{su}_{2}}{2}\overset{\mathfrak{g}_{2}}{3}1\overset{\mathfrak{f}_{4}}{5}1\overset{\mathfrak{g}_{2}}{3}\overset{\mathfrak{su}_{2}}{2}1 & (772, 992) & (\mathfrak{e}_8)_{60} \oplus (\mathfrak{e}_7)_{48}   
\\\midrule

 \multirow{2}{*}{19} & [0, 2A_{1}]_I & 1\ 2\overset{\mathfrak{su}_{2}}{2}\overset{\mathfrak{g}_{2}}{3}1\overset{\mathfrak{f}_{4}}{5}1\overset{\mathfrak{g}_{2}}{3}\overset{\mathfrak{su}_{2}}{1} &  (725, 928) & (\mathfrak{e}_8)_{60} \oplus (\mathfrak{so}_{13})_{40}   
 \\\cline{2-5}
 & [A_{1}, A_{1}]_I & 1\overset{\mathfrak{su}_{2}}{2}\overset{\mathfrak{g}_{2}}{3}1\overset{\mathfrak{f}_{4}}{5}1\overset{\mathfrak{g}_{2}}{3}\overset{\mathfrak{su}_{2}}{2}1 &  (713, 904) & (\mathfrak{e}_7)_{48} \oplus (\mathfrak{e}_7)_{48}   
 \\\midrule

\multirow{3}{*}{18} & [0, 3A_{1}]_I & 1\ 2\overset{\mathfrak{su}_{2}}{2}\overset{\mathfrak{g}_{2}}{3}1\overset{\mathfrak{f}_{4}}{5}1\overset{\mathfrak{g}_{2}}{3}1 &  (686, 879) & (\mathfrak{e}_8)_{60} \oplus (\mathfrak{f}_4)_{36} \oplus (\mathfrak{su}_{2})_{31}   
\\\cline{2-5}
 & [0, A_{2}]_I & 1\ 2\overset{\mathfrak{su}_{2}}{2}\overset{\mathfrak{g}_{2}}{3}1\overset{\mathfrak{f}_{4}}{5}1\overset{\mathfrak{su}_{3}}{3}1 &  (656, 848) & (\mathfrak{e}_8)_{60} \oplus (\mathfrak{e}_6)_{36}   
 \\\cline{2-5}
 & [A_{1}, 2A_{1}]_I & 1\overset{\mathfrak{su}_{2}}{2}\overset{\mathfrak{g}_{2}}{3}1\overset{\mathfrak{f}_{4}}{5}1\overset{\mathfrak{g}_{2}}{3}\overset{\mathfrak{su}_{2}}{1} &  (666, 840) & (\mathfrak{e}_7)_{48} \oplus (\mathfrak{so}_{13})_{40}   
 \\\midrule

\multirow{6}{*}{17} & [0, 4A_{1}]_I & 1\overset{\phantom{\mathfrak{su}_{2}}}{2}\overset{\mathfrak{su}_{2}}{2}\overset{\mathfrak{g}_{2}}{3}1\overset{\mathfrak{f}_{4}}{5}1\overset{\mathfrak{g}_{2}}{2} &  (651, 836) & (\mathfrak{e}_8)_{60} \oplus (\mathfrak{usp}_{8})_{31}   
\\\cline{2-5}
 & [0, A_{2}+A_{1}]_I &  1\overset{\phantom{\mathfrak{su}_{2}}}{2}\overset{\mathfrak{su}_{2}}{2}\overset{\mathfrak{g}_{2}}{3}1\overset{\mathfrak{f}_{4}}{5}1\overset{\mathfrak{su}_{3}}{2} & (621, 802) & (\mathfrak{e}_8)_{60} \oplus (\mathfrak{su}_6)_{30}   
 \\\cline{2-5}
 & [A_{1}, 3A_{1}]_I & 1\overset{\mathfrak{su}_{2}}{2}\overset{\mathfrak{g}_{2}}{3}1\overset{\mathfrak{f}_{4}}{5}1\overset{\mathfrak{g}_{2}}{3}1 &  (627, 791) & (\mathfrak{e}_7)_{48} \oplus (\mathfrak{f}_4)_{36} \oplus (\mathfrak{su}_{2})_{31}   
 \\\cline{2-5}
 & [A_{1}, A_{2}]_I & 1\overset{\mathfrak{su}_{2}}{2}\overset{\mathfrak{g}_{2}}{3}1\overset{\mathfrak{f}_{4}}{5}1\overset{\mathfrak{su}_{3}}{3}1 &  (597, 760) & (\mathfrak{e}_7)_{48} \oplus  (\mathfrak{e}_6)_{36}   
 \\\cline{2-5}
 & [2A_{1}, 2A_{1}]_I & \overset{\mathfrak{su}_{2}}{1}\overset{\mathfrak{g}_{2}}{3}1\overset{\mathfrak{f}_{4}}{5}1\overset{\mathfrak{g}_{2}}{3}\overset{\mathfrak{su}_{2}}{1} &  (619, 776) & (\mathfrak{so}_{13})_{40} \oplus (\mathfrak{so}_{13})_{40}   
 \\\midrule

\multirow{6}{*}{16} & [0, A_{2}+2A_{1}]_I & 1\ 2\overset{\mathfrak{su}_{2}}{2}\overset{\mathfrak{g}_{2}}{3}1\overset{\mathfrak{f}_{4}}{5}1\overset{\mathfrak{su}_{2}}{2} &  (592, 768) & (\mathfrak{e}_8)_{60} \oplus (\mathfrak{so}_{7})_{28}\oplus (\mathfrak{su_2})_{144}   
\\\cline{2-5}
 & [A_{1}, 4A_{1}]_I & 1\overset{\mathfrak{su}_{2}}{2}\overset{\mathfrak{g}_{2}}{3}1\overset{\mathfrak{f}_{4}}{5}1\overset{\mathfrak{g}_{2}}{2} &  (592, 748) & (\mathfrak{e}_7)_{48} \oplus (\mathfrak{usp}_{8})_{31}   
 \\\cline{2-5}
 &[A_{1}, A_{2}+A_{1}]_I & 1\overset{\mathfrak{su}_{2}}{2}\overset{\mathfrak{g}_{2}}{3}1\overset{\mathfrak{f}_{4}}{5}1\overset{\mathfrak{su}_{3}}{2} &  (562, 714) & (\mathfrak{e}_7)_{48} \oplus (\mathfrak{su}_{6})_{30}   
 \\\cline{2-5}
 & [2A_{1}, 3A_{1}]_I &  \overset{\mathfrak{su}_{2}}{1}\overset{\mathfrak{g}_{2}}{3}1\overset{\mathfrak{f}_{4}}{5}1\overset{\mathfrak{g}_{2}}{3}1 & (580, 727) & (\mathfrak{f}_4)_{36} \oplus (\mathfrak{so}_{13})_{40} \oplus (\mathfrak{su}_{2})_{31}   
 \\\cline{2-5}
 & [2A_{1}, A_{2}]_I & \overset{\mathfrak{su}_{2}}{1}\overset{\mathfrak{g}_{2}}{3}1\overset{\mathfrak{f}_{4}}{5}1\overset{\mathfrak{su}_{3}}{3}1 &  (550, 696) & (\mathfrak{e}_6)_{36} \oplus (\mathfrak{so}_{13})_{40}   
 \\\midrule

\multirow{13}{*}{15} & [0, A_{2}+3A_{1}]_I & 1\ 2\overset{\mathfrak{su}_{2}}{2}\overset{\mathfrak{g}_{2}}{3}1\overset{\mathfrak{f}_{4}}{5}1\ 2 & (565, 737) & (\mathfrak{e}_8)_{60} \oplus (\mathfrak{g}_2)_{48} \oplus (\mathfrak{su}_{2})_{25}   
\\\cline{2-5}
 & [0, 2A_2]_I & \begin{array}{c}
1\overset{\phantom{\mathfrak{su}_{2}}}{2}\overset{\mathfrak{su}_{2}}{2}\overset{\mathfrak{g}_{2}}{3}1\overset{\mathfrak{f}_{4}}{5}1 \\[-9pt]
\phantom{1\overset{\phantom{\mathfrak{su}_{2}}}{2}\overset{\mathfrak{su}_{2}}{2}\overset{\mathfrak{g}_{2}}{3}1}\overset{\phantom{\mathfrak{f}_{4}}}{1}\phantom{1} \\
\end{array} &  (541,712) &  (\mathfrak{e}_{8})_{60} \oplus (\mathfrak{g}_{2})_{24} \oplus (\mathfrak{g}_{2})_{24}    
\\\cline{2-5}
 & [0, A_{3}]_I & 1\overset{\phantom{\mathfrak{su}_2}}{2}\overset{\mathfrak{su}_2}{2}\overset{\mathfrak{g}_{2}}{3}1\overset{\mathfrak{so}_{9}}{4}\overset{\mathfrak{su}_{2}}{1} & (505,680) &  (\mathfrak{e}_{8})_{60} \oplus(\mathfrak{so}_{11})_{28}   
 \\\cline{2-5}
 & [A_{1}, A_{2}+2A_{1}]_I & 1\overset{\mathfrak{su}_{2}}{2}\overset{\mathfrak{g}_{2}}{3}1\overset{\mathfrak{f}_{4}}{5}1\overset{\mathfrak{su}_{2}}{2} & (533, 680) & (\mathfrak{e}_7)_{48} \oplus (\mathfrak{so}_{7})_{28}\oplus (\mathfrak{su_2})_{144}   
 \\\cline{2-5}
 & [2A_{1}, 4A_{1}]_I & \overset{\mathfrak{su}_{2}}{1}\overset{\mathfrak{g}_{2}}{3}1\overset{\mathfrak{f}_{4}}{5}1\overset{\mathfrak{g}_{2}}{2} &  (545, 684) & (\mathfrak{so}_{13})_{40} \oplus(\mathfrak{usp}_{8})_{31}   
 \\\cline{2-5}
 & [2A_{1}, A_{2}+A_{1}]_I & \overset{\mathfrak{su}_{2}}{1}\overset{\mathfrak{g}_{2}}{3}1\overset{\mathfrak{f}_{4}}{5}1\overset{\mathfrak{su}_{3}}{2} & (515, 650) & (\mathfrak{so}_{13})_{40} \oplus(\mathfrak{su}_{6})_{30}   
 \\\cline{2-5}
 & [3A_{1}, 3A_{1}]_I & 1\overset{\mathfrak{g}_{2}}{3}1\overset{\mathfrak{f}_{4}}{5}1\overset{\mathfrak{g}_{2}}{3}1 &  (541, 678) & (\mathfrak{f}_{4})_{36}^{\oplus 2}  \oplus (\mathfrak{su}_{2})_{31}^{\oplus 2}    
 \\\cline{2-5}
 & [3A_{1}, A_{2}]_I & 1\overset{\mathfrak{g}_{2}}{3}1\overset{\mathfrak{f}_{4}}{5}1\overset{\mathfrak{su}_{3}}{3}1 &  (511, 647) & (\mathfrak{f}_{4})_{36} \oplus(\mathfrak{e}_{6})_{36} \oplus(\mathfrak{su}_{2})_{31}   
 \\\cline{2-5}
 &  [A_{2}, A_{2}]_I & 1\overset{\mathfrak{su}_{3}}{3}1\overset{\mathfrak{f}_{4}}{5}1\overset{\mathfrak{su}_{3}}{3}1 & (481, 616) & (\mathfrak{e}_{6})_{36} \oplus(\mathfrak{e}_{6})_{36}   
 \\\bottomrule
\end{array}$
\end{threeparttable}
\end{table}
\newpage
\begin{table}[H]
\begin{threeparttable}
\centering
\renewcommand{\arraystretch}{1.1}
\footnotesize
$\begin{array}{c c c c l}
\toprule
\phantom{dd}\text{dim}(\mathcal{C})\phantom{dd} & \phantom{F}\text{Fixture}\phantom{F} & \phantom{d}\text{6d SCFT}\phantom{d} & \phantom{nv}(n_v, n_h)\phantom{nv} & \text{Flavor Symmetry}
\\
\midrule
\multirow{14}{*}{14} & [0, 2A_{2}+A_{1}]_I & 1\ 2\overset{\mathfrak{su}_{2}}{2}\overset{\mathfrak{g}_{2}}{3}1\overset{\mathfrak{f}_{4}}{4}1 &  (518, 686) & (\mathfrak{e}_{8})_{60} \oplus(\mathfrak{g}_{2})_{24} \oplus(\mathfrak{su}_{2})_{62}  
\\\cline{2-5}
 & [0, A_{3}+A_{1}]_I & 1\ 2\overset{\mathfrak{su}_{2}}{2}\overset{\mathfrak{g}_{2}}{3}1\overset{\mathfrak{so}_{9}}{4}1 &  (478, 645) & (\mathfrak{e}_{8})_{60} \oplus(\mathfrak{so}_{7})_{24} \oplus(\mathfrak{su}_{2})_{21}  
 \\\cline{2-5}
 & [0, D_{4}(a_{1})]_I & 1\ 2\overset{\mathfrak{su}_{2}}{2}\overset{\mathfrak{g}_{2}}{3}1\overset{\mathfrak{so}_{8}}{4}1 &  (458, 624) & (\mathfrak{e}_{8})_{60} \oplus(\mathfrak{so}_{8})_{24}   
 \\\cline{2-5}
 & [A_{1}, A_{2}+3A_{1}]_I & 1\overset{\mathfrak{su}_{2}}{2}\overset{\mathfrak{g}_{2}}{3}1\overset{\mathfrak{f}_{4}}{5}12 &  (506, 649) & (\mathfrak{e}_{7})_{48} \oplus(\mathfrak{g}_{2})_{48} \oplus(\mathfrak{su}_{2})_{25}   
 \\\cline{2-5}
 & [A_{1}, 2A_{2}]_I & \begin{array}{c}
 1\overset{\mathfrak{su}_{2}}{2}\overset{\mathfrak{g}_{2}}{3}1\overset{\mathfrak{f}_4}{5}1 \\[-9pt]
 \phantom{1\overset{\mathfrak{su}_{2}}{2}\overset{\mathfrak{g}_{2}}{3}1}\overset{\phantom{\mathfrak{f}_4}}{1}\phantom{1} \\
 \end{array}  &  (482, 624) &  (\mathfrak{e}_{7})_{48} \oplus (\mathfrak{g}_{2})_{24} \oplus (\mathfrak{g}_{2})_{24}
 \\\cline{2-5}
 & [3A_{1}, 4A_{1}]_I & 1\overset{\mathfrak{g}_{2}}{3}1\overset{\mathfrak{f}_{4}}{5}1\overset{\mathfrak{g}_{2}}{2} & (506, 635) & (\mathfrak{f}_{4})_{36} \oplus(\mathfrak{su}_{2})_{31} \oplus(\mathfrak{usp}_{8})_{31}   
 \\\cline{2-5}
 & [A_{2}, 4A_{1}]_I & 1\overset{\mathfrak{su}_{3}}{3}1\overset{\mathfrak{f}_{4}}{5}1\overset{\mathfrak{g}_{2}}{2} &  (476, 604) & (\mathfrak{e}_{6})_{36} \oplus(\mathfrak{usp}_{8})_{31}    
 \\\cline{2-5}
 & [2A_{1}, A_{2}+2A_{1}]_I & \overset{\mathfrak{su}_{2}}{1}\overset{\mathfrak{g}_{2}}{3}1\overset{\mathfrak{f}_{4}}{5}1\overset{\mathfrak{su}_{2}}{2} & (486, 616) & (\mathfrak{su}_{2})_{144} \oplus(\mathfrak{so}_{13})_{40} \oplus(\mathfrak{so}_{7})_{28}   
 \\\cline{2-5}
 & [3A_{1}, A_{2}+A_{1}]_I & 1\overset{\mathfrak{g}_{2}}{3}1\overset{\mathfrak{f}_{4}}{5}1\overset{\mathfrak{su}_{3}}{2} &  (476, 601) & (\mathfrak{f}_{4})_{36} \oplus(\mathfrak{su}_{2})_{31} \oplus(\mathfrak{su}_{6})_{30}    
 \\\cline{2-5}
 & [A_{2}, A_{2}+A_{1}]_I & 1\overset{\mathfrak{su}_{3}}{3}1\overset{\mathfrak{f}_{4}}{5}1\overset{\mathfrak{su}_{3}}{2} &  (446, 570) & (\mathfrak{e}_{6})_{36} \oplus(\mathfrak{su}_{6})_{30}   
 \\\cline{2-5}
 & [A_{1}, A_{3}]_I & 1\overset{\mathfrak{su}_{2}}{2}\overset{\mathfrak{g}_{2}}{3}1\overset{\mathfrak{so}_{9}}{4}\overset{\mathfrak{su}_2}{1} & (446, 592) & (\mathfrak{e}_{7})_{48} \oplus(\mathfrak{so}_{11})_{28}   
 \\\midrule

\multirow{20}{*}{13} & [0, 2A_{2}+2A_{1}]_I & 1\ 2\overset{\mathfrak{su}_{2}}{2}\overset{\mathfrak{g}_{2}}{3}1\overset{\mathfrak{f}_{4}}{3} &  (495, 660) & (\mathfrak{e}_{8})_{60} \oplus(\mathfrak{usp}_{4})_{62}   
\\\cline{2-5}
& [0, A_{3}+2A_{1}]_I & 1\ 2\overset{\mathfrak{su}_{2}}{2}\overset{\mathfrak{g}_{2}}{3}1\overset{\mathfrak{so}_{9}}{3} &  (455, 618) & (\mathfrak{e}_{8})_{60} \oplus(\mathfrak{usp}_{4})_{21} \oplus(\mathfrak{su}_{2})_{40}   
\\\cline{2-5}
& [0, D_{4}(a_{1})+A_{1}]_I & 1\ 2\overset{\mathfrak{su}_{2}}{2}\overset{\mathfrak{g}_{2}}{3}1\overset{\mathfrak{so}_{8}}{3} &  (435, 596) & (\mathfrak{e}_{8})_{60} \oplus (\mathfrak{su}_{2})_{20}^{\oplus 3}    
\\\cline{2-5}
& [2A_{1}, A_{2}+3A_{1}]_I &  \overset{\mathfrak{su}_{2}}{1}\overset{\mathfrak{g}_{2}}{3}1\overset{\mathfrak{f}_{4}}{5}1\overset{\phantom{\mathfrak{su}_{2}}}{2} & (459, 585) & (\mathfrak{g}_{2})_{48} \oplus(\mathfrak{su}_{2})_{25} \oplus(\mathfrak{so}_{13})_{40}   
\\\cline{2-5}
& [3A_{1}, A_{2}+2A_{1}]_I & 1\overset{\mathfrak{g}_{2}}{3}1\overset{\mathfrak{f}_{4}}{5}1\overset{\mathfrak{su}_{2}}{2} & (447, 567) & (\mathfrak{f}_{4})_{36} \oplus(\mathfrak{su}_{2})_{144} \oplus(\mathfrak{su}_{2})_{31} \oplus(\mathfrak{so}_{7})_{28}   
\\\cline{2-5}
& [A_{2}, A_{2}+2A_{1}]_I & 1\overset{\mathfrak{su}_{3}}{3}1\overset{\mathfrak{f}_{4}}{5}1\overset{\mathfrak{su}_{2}}{2} &  (417, 536) & (\mathfrak{e}_{6})_{36} \oplus(\mathfrak{su}_{2})_{144} \oplus(\mathfrak{so}_{7})_{28}   
\\\cline{2-5}
& [2A_{1}, 2A_{2}]_I & \begin{array}{c}
\overset{\mathfrak{su}_{2}}{1}\overset{\mathfrak{g}_{2}}{3}1\overset{\mathfrak{f}_4}{5}1 \\[-9pt]
\phantom{\overset{\mathfrak{su}_{2}}{1}\overset{\mathfrak{g}_{2}}{3}1}\overset{\phantom{\mathfrak{f}_4}}{1}\phantom{1}
\end{array} &  (435, 560) &  (\mathfrak{g}_{2})_{24} \oplus (\mathfrak{g}_{2})_{24} \oplus(\mathfrak{so}_{13})_{40}   
\\\cline{2-5}
& [4A_{1}, 4A_{1}]_I & \overset{\mathfrak{g}_{2}}{2}1\overset{\mathfrak{f}_{4}}{5}1\overset{\mathfrak{g}_{2}}{2} & (471, 592) & (\mathfrak{usp}_{8})_{31} \oplus (\mathfrak{usp}_{8})_{31}   
\\\cline{2-5}
& [4A_{1}, A_{2}+A_{1}]_I & \overset{\mathfrak{g}_{2}}{2}1\overset{\mathfrak{f}_{4}}{5}1\overset{\mathfrak{su}_{3}}{2} &  (441, 558) & (\mathfrak{usp}_{8})_{31} \oplus(\mathfrak{su}_{6})_{30}   
\\\cline{2-5}
& [A_{2}+A_{1}, A_{2}+A_{1}]_I & \overset{\mathfrak{su}_{3}}{2}1\overset{\mathfrak{f}_{4}}{5}1\overset{\mathfrak{su}_{3}}{2} &  (411, 524) & (\mathfrak{su}_{6})_{30} \oplus (\mathfrak{su}_{6})_{30}   
\\\cline{2-5}
& [A_{1}, 2A_{2}+A_{1}]_I & 1\overset{\mathfrak{su}_{2}}{2}\overset{\mathfrak{g}_{2}}{3}1\overset{\mathfrak{f}_{4}}{4}1 &  (459, 598) & (\mathfrak{e}_{7})_{48} \oplus(\mathfrak{g}_{2})_{24} \oplus(\mathfrak{su}_{2})_{62}   
\\\cline{2-5}
& [A_{1}, A_{3}+A_{1}]_I  & 1\overset{\mathfrak{su}_{2}}{2}\overset{\mathfrak{g}_{2}}{3}1\overset{\mathfrak{so}_{9}}{4}1 & (419, 557) & (\mathfrak{e}_{7})_{48} \oplus(\mathfrak{so}_{7})_{24} \oplus(\mathfrak{su}_{2})_{21}   
\\\cline{2-5}
& [A_{1}, D_{4}(a_{1})]_I & 1\overset{\mathfrak{su}_{2}}{2}\overset{\mathfrak{g}_{2}}{3}1\overset{\mathfrak{so}_{8}}{4}1 & (399, 536) & (\mathfrak{e}_{7})_{48} \oplus(\mathfrak{so}_{8})_{24}   
\\\cline{2-5}
& [2A_{1}, A_{3}]_I & \overset{\mathfrak{su}_{2}}{1}\overset{\mathfrak{g}_{2}}{3}1\overset{\mathfrak{so}_9}{4}\overset{\mathfrak{su}_2}{1} &  (399, 528) &  (\mathfrak{so}_{13})_{40} \oplus(\mathfrak{so}_{11})_{28}   
\\\midrule

\multirow{6}{*}{12} & [0, A_{3}+A_{2}]_I & 1\ 2\overset{\mathfrak{su}_{2}}{2}\overset{\mathfrak{g}_{2}}{3}1\overset{\mathfrak{so}_{7}}{3} &  (416, 576) & (\mathfrak{e}_{8})_{60} \oplus(\mathfrak{usp}_{4})_{20}   
\\\cline{2-5}
 & [A_{1}, 2A_{2}+2A_{1}]_I & 1\overset{\mathfrak{su}_{2}}{2}\overset{\mathfrak{g}_{2}}{3}1\overset{\mathfrak{f}_{4}}{3} &  (436, 572) & (\mathfrak{e}_{7})_{48} \oplus(\mathfrak{usp}_{4})_{62}   
 \\\cline{2-5}
 & [A_{1}, A_{3}+2A_{1}]_I & 1\overset{\mathfrak{su}_{2}}{2}\overset{\mathfrak{g}_{2}}{3}1\overset{\mathfrak{so}_{9}}{3} &  (396, 530) & (\mathfrak{e}_{7})_{48} \oplus(\mathfrak{usp}_{4})_{21} \oplus(\mathfrak{su}_{2})_{40}   
 \\\cline{2-5}
 & [A_{1}, D_{4}(a_{1})+A_{1}]_I & 1\overset{\mathfrak{su}_{2}}{2}\overset{\mathfrak{g}_{2}}{3}1\overset{\mathfrak{so}_{8}}{3} &  (376, 508) & (\mathfrak{e}_{7})_{48} \oplus (\mathfrak{su}_{2})_{20}^{\oplus 3}   
 \\\cline{2-5}
 & [3A_{1}, A_{2}+3A_{1}]_I & 1\overset{\mathfrak{g}_{2}}{3}1\overset{\mathfrak{f}_{4}}{5}12 &  (420, 536) & (\mathfrak{f}_{4})_{36} \oplus(\mathfrak{g}_{2})_{48} \oplus(\mathfrak{su}_{2})_{25} \oplus(\mathfrak{su}_{2})_{31}   
 \\\bottomrule
\end{array}$
\end{threeparttable}
\end{table}
\newpage
\begin{table}[H]
\begin{threeparttable}
\centering
\renewcommand{\arraystretch}{1.1}
\footnotesize
$\begin{array}{c c c c l}
\toprule
\phantom{d}\text{dim}(\mathcal{C})\phantom{d} & \phantom{F}\text{Fixture}\phantom{F} & \phantom{d}\text{6d SCFT}\phantom{d} & \phantom{nv}(n_v, n_h)\phantom{nv} & \text{Flavor Symmetry}
\\
\midrule

\multirow{13}{*}{12} & [A_{2}, A_{2}+3A_{1}]_I & 1\overset{\mathfrak{su}_{3}}{3}1\overset{\mathfrak{f}_{4}}{5}1\ 2 &  (390, 505) & (\mathfrak{e}_{6})_{36} \oplus(\mathfrak{g}_{2})_{48} \oplus(\mathfrak{su}_{2})_{25}   
\\\cline{2-5}
 & [A_{2}+A_{1}, A_{2}+2A_{1}]_I & \overset{\mathfrak{su}_{3}}{2}1\overset{\mathfrak{f}_{4}}{5}1\overset{\mathfrak{su}_{2}}{2} &  (382, 490) & (\mathfrak{su}_{2})_{144} \oplus(\mathfrak{su}_{6})_{30} \oplus(\mathfrak{so}_{7})_{28}   
 \\\cline{2-5}
 & [4A_{1}, A_{2}+2A_{1}]_I & \overset{\mathfrak{g}_{2}}{2}1\overset{\mathfrak{f}_{4}}{5}1\overset{\mathfrak{su}_{2}}{2} &  (412, 524) & (\mathfrak{su}_{2})_{144} \oplus(\mathfrak{usp}_{8})_{31} \oplus(\mathfrak{so}_{7})_{28}   
 \\\cline{2-5}
 & [3A_{1}, 2A_{2}]_I & \begin{array}{c}
 1\overset{\mathfrak{g}_{2}}{3}1\overset{\mathfrak{f}_4}{5}1 \\[-9pt]
 \phantom{1\overset{\mathfrak{g}_{2}}{3}1}\overset{\phantom{\mathfrak{f}_4}}{1}\phantom{1} \\
 \end{array} &  (396, 511) & (\mathfrak{f}_{4})_{36} \oplus (\mathfrak{g}_{2})_{24}^{\oplus 2} \oplus(\mathfrak{su}_{2})_{31}   
 \\\cline{2-5}
 & [A_2, 2A_{2}]_I & \begin{array}{c}
 1\overset{\mathfrak{su}_{3}}{3}1\overset{\mathfrak{f}_4}{5}1 \\[-9pt]
 \phantom{1\overset{\mathfrak{su}_{3}}{3}1}\overset{\phantom{\mathfrak{f}_4}}{1}\phantom{1} \\
 \end{array} &  (366, 480) &  (\mathfrak{e}_{6})_{36} \oplus (\mathfrak{g}_{2})_{24}  \oplus  (\mathfrak{g}_{2})_{24} 
 \\\cline{2-5}
 & [2A_{1}, 2A_{2}+A_{1}]_I & \overset{\mathfrak{su}_{2}}{1}\overset{\mathfrak{g}_{2}}{3}1\overset{\mathfrak{f}_{4}}{4}1 &  (412, 534) & (\mathfrak{g}_{2})_{24} \oplus(\mathfrak{so}_{13})_{40} \oplus(\mathfrak{su}_{2})_{62}   
 \\\cline{2-5}
 & [2A_{1}, A_{3}+A_{1}]_I & \overset{\mathfrak{su}_{2}}{1}\overset{\mathfrak{g}_{2}}{3}1\overset{\mathfrak{so}_{9}}{4}1 & (372, 493) & (\mathfrak{so}_{7})_{24} \oplus(\mathfrak{so}_{13})_{40} \oplus(\mathfrak{su}_{2})_{21}   
 \\\cline{2-5}
 & [2A_{1}, D_{4}(a_{1})]_I & \overset{\mathfrak{su}_{2}}{1}\overset{\mathfrak{g}_{2}}{3}1\overset{\mathfrak{so}_{8}}{4}1 & (352, 472) & (\mathfrak{so}_{8})_{24} \oplus(\mathfrak{so}_{13})_{40}    
 \\\cline{2-5}
 & [3A_{1}, A_{3}]_I & 1\overset{\mathfrak{g}_{2}}{3}1\overset{\mathfrak{so}_9}{4}\overset{\mathfrak{su}_2}{1} &  (360, 479) & (\mathfrak{f}_{4})_{36} \oplus(\mathfrak{su}_{2})_{31} \oplus(\mathfrak{so}_{11})_{28}   
 \\\cline{2-5}
 & [A_2, A_{3}]_I & 1\overset{\mathfrak{su}_{3}}{3}1\overset{\mathfrak{so}_9}{4}\overset{\mathfrak{su}_2}{1} &  (330, 448) & (\mathfrak{e}_{6})_{36} \oplus(\mathfrak{so}_{11})_{28}   
 \\\midrule

\multirow{28}{*}{11} & [0, A_{3}+A_{2}+A_{1}]_I & 1\overset{\phantom{\mathfrak{su}_{2}}}{2}\overset{\mathfrak{su}_{2}}{2}\overset{\mathfrak{g}_{2}}{3}1\overset{\mathfrak{g}_{2}}{3} &  (397, 555) & (\mathfrak{e}_{8})_{60} \oplus(\mathfrak{su}_{2})_{384} \oplus(\mathfrak{su}_{2})_{19}   
\\\cline{2-5}
& [0, D_{4}(a_{1})+A_{2}]_I &  1\overset{\phantom{\mathfrak{su}_{2}}}{2}\overset{\mathfrak{su}_{2}}{2}\overset{\mathfrak{g}_{2}}{3}1\overset{\mathfrak{su}_{3}}{3} &  (379, 536) & (\mathfrak{e}_{8})_{60} \oplus(\mathfrak{su}_{3})_{96}   
\\\cline{2-5}
& [A_{1}, A_{3}+A_{2}]_I & 1\overset{\mathfrak{su}_{2}}{2}\overset{\mathfrak{g}_{2}}{3}1\overset{\mathfrak{so}_{7}}{3} &  (357, 488) & (\mathfrak{e}_{7})_{48} \oplus(\mathfrak{usp}_{4})_{20}   
\\\cline{2-5}
& [2A_{1}, 2A_{2}+2A_{1}]_I & \overset{\mathfrak{su}_{2}}{1}\overset{\mathfrak{g}_{2}}{3}1\overset{\mathfrak{f}_{4}}{3} &  (389, 508) & (\mathfrak{so}_{13})_{40} \oplus(\mathfrak{usp}_{4})_{62}   
\\\cline{2-5}
& [2A_{1}, A_{3}+2A_{1}]_I & \overset{\mathfrak{su}_{2}}{1}\overset{\mathfrak{g}_{2}}{3}1\overset{\mathfrak{so}_{9}}{3} &  (349, 466) & (\mathfrak{so}_{13})_{40} \oplus(\mathfrak{usp}_{4})_{21} \oplus(\mathfrak{su}_{2})_{40}   
\\\cline{2-5}
& [2A_{1}, D_{4}(a_{1})+A_{1}]_I & \overset{\mathfrak{su}_{2}}{1}\overset{\mathfrak{g}_{2}}{3}1\overset{\mathfrak{so}_{8}}{3} &  (329, 444) & (\mathfrak{so}_{13})_{40} \oplus (\mathfrak{su}_{2})_{20}^{\oplus 3}   
\\\cline{2-5}
& [0, A_4]_I &  1\overset{\phantom{\mathfrak{su}_{2}}}{2}\overset{\mathfrak{su}_{2}}{2}\overset{\mathfrak{su}_{3}}{2}\overset{\mathfrak{su}_{4}}{2} & (321, 480) & (\mathfrak{e}_{8})_{60} \oplus(\mathfrak{su}_{5})_{20}   
\\\cline{2-5}
& [4A_{1}, A_{2}+3A_{1}]_I & \overset{\mathfrak{g}_{2}}{2}1\overset{\mathfrak{f}_{4}}{5}1\overset{\phantom{\mathfrak{su}_{2}}}{2} &  (385, 493) & (\mathfrak{g}_{2})_{48} \oplus(\mathfrak{su}_{2})_{25} \oplus(\mathfrak{usp}_{8})_{31}   
\\\cline{2-5}
&  [A_{2}+A_{1}, A_{2}+3A_{1}]_I & \overset{\mathfrak{su}_{3}}{2}1\overset{\mathfrak{f}_{4}}{5}1\overset{\phantom{\mathfrak{su}_{2}}}{2} & (355, 459) & (\mathfrak{g}_{2})_{48} \oplus(\mathfrak{su}_{2})_{25} \oplus(\mathfrak{su}_{6})_{30}   
\\\cline{2-5}
& [A_{2}+2A_{1}, A_{2}+2A_{1}]_I & \overset{\mathfrak{su}_{2}}{2}1\overset{\mathfrak{f}_{4}}{5}1\overset{\mathfrak{su}_{2}}{2} &  (353, 456) & (\mathfrak{su}_{2})_{144}^{\oplus 2} \oplus (\mathfrak{so}_{7})_{28}^{\oplus 2}   
\\\cline{2-5}
& [4A_1, 2A_2]_I & \begin{array}{c}
 \overset{\mathfrak{g}_2}{2}1\overset{\mathfrak{f}_4}{5}1 \\[-9pt]
 \phantom{\overset{\mathfrak{g}_2}{2}1}\overset{\phantom{\mathfrak{f}_4}}{1}\phantom{1} \\
 \end{array} &  (361, 468) &  (\mathfrak{g}_{2})_{24} \oplus (\mathfrak{g}_{2})_{24} \oplus(\mathfrak{usp}_{8})_{31}   
 \\\cline{2-5}   
& [A_2 +A_1, 2A_2]_I & \begin{array}{c}
 \overset{\mathfrak{su}_3}{2}1\overset{\mathfrak{f}_4}{5}1 \\[-9pt]
 \phantom{\overset{\mathfrak{g}_2}{2}1}\overset{\phantom{\mathfrak{f}_4}}{1}\phantom{1} \\
 \end{array} &  (331, 434) &  (\mathfrak{g}_{2})_{24} \oplus (\mathfrak{g}_{2})_{24} \oplus(\mathfrak{su}_{6})_{30}   
 \\\cline{2-5}
& [4A_1, A_3]_I & \overset{\mathfrak{g}_2}{2}1\overset{\mathfrak{so}_9}{4}\overset{\mathfrak{su}_2}{1} &  (325, 436) &  (\mathfrak{usp}_{8})_{31} \oplus(\mathfrak{so}_{11})_{28}   
\\\cline{2-5}
& [A_2 + A_1, A_3]_I & \overset{\mathfrak{su}_3}{2}1\overset{\mathfrak{so}_9}{4}\overset{\mathfrak{su}_2}{1} &  (295, 402) &  (\mathfrak{su}_{6})_{30} \oplus(\mathfrak{so}_{11})_{28}   
\\\cline{2-5}
& [3A_{1}, 2A_{2}+A_{1}]_I & 1\overset{\mathfrak{g}_{2}}{3}1\overset{\mathfrak{f}_{4}}{4}1 &  (373, 485) & (\mathfrak{f}_{4})_{36} \oplus(\mathfrak{g}_{2})_{24} \oplus(\mathfrak{su}_{2})_{31} \oplus(\mathfrak{su}_{2})_{62}
\\\cline{2-5}
& [3A_{1}, A_{3}+A_{1}]_I &   1\overset{\mathfrak{g}_{2}}{3}1\overset{\mathfrak{so}_{9}}{4}1 & (333, 444) & (\mathfrak{f}_{4})_{36} \oplus(\mathfrak{so}_{7})_{24} \oplus(\mathfrak{su}_{2})_{31} \oplus(\mathfrak{su}_{2})_{21}   
\\\cline{2-5}
& [3A_{1}, D_{4}(a_{1})]_I &  1\overset{\mathfrak{g}_{2}}{3}1\overset{\mathfrak{so}_{8}}{4}1 & (313, 423) &  (\mathfrak{f}_{4})_{36} \oplus(\mathfrak{so}_{8})_{24} \oplus(\mathfrak{su}_{2})_{31}   
\\\cline{2-5}
& [A_{2}, 2A_{2}+A_{1}]_I & 1\overset{\mathfrak{su}_{3}}{3}1\overset{\mathfrak{f}_{4}}{4}1 &  (343, 454) & (\mathfrak{e}_{6})_{36} \oplus(\mathfrak{g}_{2})_{24} \oplus(\mathfrak{su}_{2})_{62}   
\\\cline{2-5}
& [A_{2}, A_{3}+A_{1}]_I & 1\overset{\mathfrak{su}_{3}}{3}1\overset{\mathfrak{so}_{9}}{4}1 &  (303, 413) & (\mathfrak{e}_{6})_{36} \oplus(\mathfrak{so}_{7})_{24} \oplus(\mathfrak{su}_{2})_{21}   
\\\cline{2-5}
& [A_{2}, D_{4}(a_{1})]_I & 1\overset{\mathfrak{su}_{3}}{3}1\overset{\mathfrak{so}_{8}}{4}1 &  (283, 392) & (\mathfrak{e}_{6})_{36} \oplus(\mathfrak{so}_{8})_{24}   
\\\bottomrule
\end{array}$
\end{threeparttable}
\end{table}
\newpage
\begin{table}[H]
\begin{threeparttable}
\centering
\renewcommand{\arraystretch}{1.3}
\footnotesize
$\begin{array}{c c c c l}
\toprule
\phantom{d}\text{dim}(\mathcal{C})\phantom{d} & \phantom{i}\text{Fixture}\phantom{i} & \text{6d SCFT} & \phantom{n}(n_v, n_h)\phantom{n} & \text{Flavor Symmetry}
\\
\midrule
\multirow{25}{*}{10} & [A_{1}, A_{3}+A_{2}+A_{1}]_I & 1\overset{\mathfrak{su}_{2}}{2}\overset{\mathfrak{g}_{2}}{3}1\overset{\mathfrak{g}_{2}}{3} &  (338, 467) & (\mathfrak{e}_{7})_{48} \oplus(\mathfrak{su}_{2})_{384} \oplus(\mathfrak{su}_{2})_{19}   
\\\cline{2-5}
 & [A_{1}, D_{4}(a_{1})+A_{2}]_I & 1\overset{\mathfrak{su}_{2}}{2}\overset{\mathfrak{g}_{2}}{3}1\overset{\mathfrak{su}_{3}}{3} &  (320, 448) & (\mathfrak{e}_{7})_{48} \oplus(\mathfrak{su}_{3})_{96}   
 \\\cline{2-5}
 & [3A_{1}, 2A_{2}+2A_{1}]_I & 1\overset{\mathfrak{g}_{2}}{3}1\overset{\mathfrak{f}_{4}}{3} &  (350, 459) & (\mathfrak{f}_{4})_{36} \oplus(\mathfrak{su}_{2})_{31} \oplus(\mathfrak{usp}_{4})_{62}   
 \\\cline{2-5}
 & [3A_{1}, A_{3}+2A_{1}]_I & 1\overset{\mathfrak{g}_{2}}{3}1\overset{\mathfrak{so}_{9}}{3} &  (310, 417) & (\mathfrak{f}_{4})_{36} \oplus(\mathfrak{su}_{2})_{31} \oplus(\mathfrak{usp}_{4})_{21} \oplus(\mathfrak{su}_{2})_{40}   
 \\\cline{2-5}
 & [3A_{1}, D_{4}(a_{1})+A_{1}]_I & 1\overset{\mathfrak{g}_{2}}{3}1\overset{\mathfrak{so}_{8}}{3} &  (290, 395) & (\mathfrak{f}_{4})_{36} \oplus(\mathfrak{su}_{2})_{31} \oplus (\mathfrak{su}_{2})_{20}^{\oplus 3}   
 \\\cline{2-5}
 & [A_{2}, 2A_{2}+2A_{1}]_I & 1\overset{\mathfrak{su}_{3}}{3}1\overset{\mathfrak{f}_{4}}{3} &  (320, 428) & (\mathfrak{e}_{6})_{36} \oplus(\mathfrak{usp}_{4})_{62}   
 \\\cline{2-5}
 & [A_{2}, A_{3}+2A_{1}]_I & 1\overset{\mathfrak{su}_{3}}{3}1\overset{\mathfrak{so}_{9}}{3} &  (280, 386) & (\mathfrak{e}_{6})_{36} \oplus(\mathfrak{usp}_{4})_{21} \oplus(\mathfrak{su}_{2})_{40}  
 \\\cline{2-5}
 & [A_{2}, D_{4}(a_{1})+A_{1}]_I & 1\overset{\mathfrak{su}_{3}}{3}1\overset{\mathfrak{so}_{8}}{3} &  (260, 364) & (\mathfrak{e}_{6})_{36} \oplus (\mathfrak{su}_{2})_{20}^{\oplus 3}   
 \\\cline{2-5}
 & [2A_{1}, A_{3}+A_{2}]_I & \overset{\mathfrak{su}_{2}}{1}\overset{\mathfrak{g}_{2}}{3}1\overset{\mathfrak{so}_{7}}{3} &  (310, 424) & (\mathfrak{so}_{13})_{40} \oplus(\mathfrak{usp}_{4})_{20}   
 \\\cline{2-5}
 & [0, A_4 + A_1]_I & 1\overset{\phantom{\mathfrak{su}_{2}}}{2}\overset{\mathfrak{su}_{2}}{2}\overset{\mathfrak{su}_{3}}{2}\overset{\mathfrak{su}_{3}}{2} &  (302, 457) & (\mathfrak{e}_{8})_{60} \oplus(\mathfrak{su}_{3})_{18} 
 \\\cline{2-5}
 & [A_1, A_4]_I & 1\overset{\mathfrak{su}_{2}}{2}\overset{\mathfrak{su}_{3}}{2}\overset{\mathfrak{su}_{4}}{2} &  (262, 392) & (\mathfrak{e}_{7})_{48} \oplus(\mathfrak{so}_{2})_{40} \oplus(\mathfrak{su}_{5})_{20}   
 \\\cline{2-5}
 & [A_{2}+2A_{1}, A_{2}+3A_{1}]_I & \overset{\mathfrak{su}_{2}}{2}1\overset{\mathfrak{f}_{4}}{5}1\overset{\phantom{\mathfrak{su}_{2}}}{2} &  (326, 425) & (\mathfrak{su}_{2})_{144} \oplus(\mathfrak{g}_{2})_{48} \oplus(\mathfrak{su}_{2})_{25} \oplus(\mathfrak{so}_{7})_{28}   
 \\\cline{2-5} 
 & [A_2 + 2A_1, 2A_2]_I & \begin{array}{c}
 \overset{\mathfrak{su}_2}{2}1\overset{\mathfrak{f}_4}{5}1 \\[-9pt]
 \phantom{\overset{\mathfrak{su}_2}{2}1}\overset{\phantom{\mathfrak{f}_4}}{1}\phantom{1} \\
 \end{array} &  (302, 400) & (\mathfrak{su}_{2})_{144} \oplus(\mathfrak{g}_{2})_{24} \oplus(\mathfrak{g}_{2})_{24} \oplus(\mathfrak{so}_{7})_{28}   
 \\\cline{2-5}
 & [4A_{1}, 2A_{2}+A_{1}]_I & \overset{\mathfrak{g}_{2}}{2}1\overset{\mathfrak{f}_{4}}{4}1 &   (338, 442) & (\mathfrak{g}_{2})_{24} \oplus(\mathfrak{usp}_{8})_{31} \oplus(\mathfrak{su}_{2})_{62}   
 \\\cline{2-5}
 & [4A_{1}, A_{3}+A_{1}]_I & \overset{\mathfrak{g}_{2}}{2}1\overset{\mathfrak{so}_{9}}{4}1 &  (298, 401) & (\mathfrak{so}_{7})_{24} \oplus(\mathfrak{usp}_{8})_{31} \oplus(\mathfrak{su}_{2})_{21}   
 \\\cline{2-5}
 & [4A_{1}, D_{4}(a_{1})]_I & \overset{\mathfrak{g}_{2}}{2}1\overset{\mathfrak{so}_{8}}{4}1 &  (278, 380) & (\mathfrak{so}_{8})_{24} \oplus(\mathfrak{usp}_{8})_{31}   
 \\\cline{2-5}
 & [A_{2} +2A_1, A_3]_I & \overset{\mathfrak{su}_2}{2}1\overset{\mathfrak{so}_9}{4}\overset{\mathfrak{su}_2}{1} &  (266, 368) & (\mathfrak{su}_{2})_{48} \oplus(\mathfrak{su}_{2})_{96} \oplus(\mathfrak{so}_{7})_{28} \oplus(\mathfrak{so}_{11})_{28} 
 \\\cline{2-5}
 & [A_{2}+A_{1}, 2A_{2}+A_{1}]_I & \overset{\mathfrak{su}_{3}}{2}1\overset{\mathfrak{f}_{4}}{4}1 &  (308, 408) &  (\mathfrak{g}_{2})_{24} \oplus(\mathfrak{su}_{6})_{30} \oplus(\mathfrak{su}_{2})_{62}   
 \\\cline{2-5}
 & [A_{2}+A_{1}, A_{3}+A_{1}]_I &  \overset{\mathfrak{su}_{3}}{2}1\overset{\mathfrak{so}_{9}}{4}1 & (268, 367) & (\mathfrak{so}_{7})_{24} \oplus(\mathfrak{su}_{6})_{30} \oplus(\mathfrak{su}_{2})_{21}   
 \\\cline{2-5}
 & [A_{2}+A_{1}, D_{4}(a_{1})]_I & \overset{\mathfrak{su}_{3}}{2}1\overset{\mathfrak{so}_{8}}{4}1 &  (248, 346) & (\mathfrak{so}_{8})_{24} \oplus(\mathfrak{su}_{6})_{30}   
 \\\midrule

\multirow{11}{*}{9} & [A_{2}+3A_{1}, A_{2}+3A_{1}]_I & \overset{\phantom{\mathfrak{su}_{2}}}{2}1\overset{\mathfrak{f}_{4}}{5}1\overset{\phantom{\mathfrak{su}_{2}}}{2} &  (299, 394) & (\mathfrak{g}_{2})_{48}^{\oplus 2} \oplus (\mathfrak{su}_{2})_{25}^{\oplus 2}    
\\\cline{2-5}
 & [0, D_4]_I & 1\overset{\phantom{\mathfrak{su}_{2}}}{2}\overset{\mathfrak{su}_{2}}{2}\overset{\mathfrak{g}_{2}}{3}1 \sqcup 1 &  (227, 392) &  (\mathfrak{e}_{8})_{12} \oplus  (\mathfrak{e}_{8})_{48} \oplus  (\mathfrak{f}_{4})_{24} 
 \\\cline{2-5}
 & [0, 2A_3]_I & 1\overset{\phantom{\mathfrak{su}_{2}}}{2}\overset{\mathfrak{su}_{2}}{2}\overset{\mathfrak{g}_{2}}{2}\overset{\mathfrak{su}_{2}}{2} &  (315, 470) & (\mathfrak{e}_{8})_{60} \oplus(\mathfrak{usp}_{4})_{31}   
 \\\cline{2-5}
 & [0, A_4 + 2A_1]_I & 1\overset{\phantom{\mathfrak{su}_{2}}}{2}\overset{\mathfrak{su}_{2}}{2}\overset{\mathfrak{su}_{3}}{2}\overset{\mathfrak{su}_{2}}{2} &  (285, 438) & (\mathfrak{e}_{8})_{60} \oplus(\mathfrak{su}_{2})_{30} 
 \\\cline{2-5}
 & [A_1, A_4 + A_1]_I & 1\overset{\mathfrak{su}_{2}}{2}\overset{\mathfrak{su}_{3}}{2}\overset{\mathfrak{su}_{3}}{2} &  (243, 369) & (\mathfrak{e}_{7})_{48} \oplus(\mathfrak{su}_{3})_{18} 
 \\\cline{2-5}
 & [2A_{1}, A_{3}+A_{2}+A_{1}]_I & \overset{\mathfrak{su}_{2}}{1}\overset{\mathfrak{g}_{2}}{3}1\overset{\mathfrak{g}_{2}}{3} &  (291, 403) & (\mathfrak{su}_{2})_{384} \oplus(\mathfrak{so}_{13})_{40} \oplus(\mathfrak{su}_{2})_{19}    
 \\\cline{2-5}
 & [2A_{1}, D_{4}(a_{1})+A_{2}]_I & \overset{\mathfrak{su}_{2}}{1}\overset{\mathfrak{g}_{2}}{3}1\overset{\mathfrak{su}_{3}}{3} &  (273, 384) & (\mathfrak{su}_{3})_{96} \oplus(\mathfrak{so}_{13})_{40}   
 \\\cline{2-5}
 & [3A_{1}, A_{3}+A_{2}]_I & 1\overset{\mathfrak{g}_{2}}{3}1\overset{\mathfrak{so}_{7}}{3} &  (271, 375) & (\mathfrak{f}_{4})_{36} \oplus(\mathfrak{su}_{2})_{31} \oplus(\mathfrak{usp}_{4})_{20}   
 \\\cline{2-5}
 & [A_{2}, A_{3}+A_{2}]_I & 1\overset{\mathfrak{su}_{3}}{3}1\overset{\mathfrak{so}_{7}}{3} &  (241, 344) & (\mathfrak{e}_{6})_{36} \oplus(\mathfrak{su}_{2})_{96} \oplus(\mathfrak{usp}_{4})_{20}
 \\\bottomrule
\end{array}$
\end{threeparttable}
\end{table}
\newpage
\begin{table}[H]
\begin{threeparttable}
\centering
\renewcommand{\arraystretch}{1.3}
\footnotesize
$\begin{array}{c c c c l}
\toprule
\phantom{d}\text{dim}(\mathcal{C})\phantom{d} & \text{Fixture} & \text{6d SCFT} & (n_v, n_h) & \text{Flavor Symmetry} 
\\
\midrule
\multirow{20}{*}{9} & [A_2 + 3A_1, 2A_2]_I & \begin{array}{c}
 2\overset{\phantom{\mathfrak{su}_2}}{1}\overset{\mathfrak{f}_4}{5}1 \\[-9pt]
 \phantom{2\overset{\phantom{\mathfrak{su}_2}}{1}}\overset{\phantom{\mathfrak{f}_4}}{1}\phantom{1} \\
 \end{array} &  (275, 369) & (\mathfrak{g}_{2})_{48} \oplus(\mathfrak{su}_{2})_{25} \oplus(\mathfrak{g}_{2})_{24} \oplus(\mathfrak{g}_{2})_{24}   
 \\\cline{2-5}
 & [A_{2}+2A_{1}, 2A_{2}+A_{1}]_I & \overset{\mathfrak{su}_{2}}{2}1\overset{\mathfrak{f}_{4}}{4}1 &  (279, 374) & (\mathfrak{su}_{2})_{144} \oplus(\mathfrak{g}_{2})_{24} \oplus(\mathfrak{so}_{7})_{28} \oplus(\mathfrak{su}_{2})_{62}   
 \\\cline{2-5}
 & [A_{2}+2A_{1}, A_{3}+A_{1}]_I & \overset{\mathfrak{su}_{2}}{2}1\overset{\mathfrak{so}_{9}}{4}1 &  (239, 333) & \begin{aligned}(\mathfrak{su}_{2})_{48} &\oplus(\mathfrak{su}_{2})_{96} \oplus(\mathfrak{so}_{7})_{24} \cr &\oplus(\mathfrak{so}_{7})_{28} \oplus(\mathfrak{su}_{2})_{21}\end{aligned}   
 \\\cline{2-5}
 & [A_{2}+2A_{1}, D_{4}(a_{1})]_I & \overset{\mathfrak{su}_{2}}{2}1\overset{\mathfrak{so}_{8}}{4}1 &  (219, 312) & (\mathfrak{su}_{2})_{48}^{\oplus 3}  \oplus(\mathfrak{so}_{8})_{24} \oplus(\mathfrak{so}_{7})_{28}   
 \\\cline{2-5}
 & [A_{2} + 3A_1, A_3]_I & 2\overset{\phantom{\mathfrak{su}_{2}}}{1}\overset{\mathfrak{so}_9}{4}\overset{\mathfrak{su}_2}{1} &  (239, 337) & (\mathfrak{so}_{7})_{48} \oplus(\mathfrak{su}_{2})_{25} \oplus(\mathfrak{so}_{11})_{28}   
 \\\cline{2-5}
 & [4A_{1}, 2A_{2}+2A_{1}]_I & \overset{\mathfrak{g}_{2}}{2}1\overset{\mathfrak{f}_{4}}{3} &  (315, 416) & (\mathfrak{usp}_{8})_{31} \oplus(\mathfrak{usp}_{4})_{62}   
 \\\cline{2-5} 
 & [4A_{1}, A_{3}+2A_{1}]_I & \overset{\mathfrak{g}_{2}}{2}1\overset{\mathfrak{so}_{9}}{3} &  (275, 374) & (\mathfrak{usp}_{8})_{31} \oplus(\mathfrak{usp}_{4})_{21} \oplus(\mathfrak{su}_{2})_{40}    
 \\\cline{2-5}
 & [4A_{1}, D_{4}(a_{1})+A_{1}]_I & \overset{\mathfrak{g}_{2}}{2}1\overset{\mathfrak{so}_{8}}{3} &  (255, 352) & (\mathfrak{usp}_{8})_{31} \oplus(\mathfrak{su}_{2})_{20}^{\oplus 3}   
 \\\cline{2-5}
 & [A_{2}+A_{1}, 2A_{2}+2A_{1}]_I & \overset{\mathfrak{su}_{3}}{2}1\overset{\mathfrak{f}_{4}}{3} &  (285, 382) & (\mathfrak{su}_{6})_{30} \oplus(\mathfrak{usp}_{4})_{62}   
 \\\cline{2-5}
 & [A_{2}+A_{1}, A_{3}+2A_{1}]_I & \overset{\mathfrak{su}_{3}}{2}1\overset{\mathfrak{so}_{9}}{3} &  (245, 340) & (\mathfrak{su}_{6})_{30} \oplus(\mathfrak{usp}_{4})_{21} \oplus(\mathfrak{su}_{2})_{40} 
 \\\cline{2-5}
 & [A_{2}+A_{1}, D_{4}(a_{1})+A_{1}]_I & \overset{\mathfrak{su}_{3}}{2}1\overset{\mathfrak{so}_{8}}{3} &  (225, 318) & (\mathfrak{su}_{6})_{30} \oplus(\mathfrak{su}_{2})_{20}^{\oplus 3}
 \\\cline{2-5}
 & [2A_1, A_4]_I & \overset{\mathfrak{su}_{2}}{1}\overset{\mathfrak{su}_{3}}{2}\overset{\mathfrak{su}_{4}}{2} &  (215, 328) &(\mathfrak{so}_{14})_{40} \oplus(\mathfrak{su}_{5})_{20} 
 \\\cline{2-5}
 & [2A_{2}, 2A_2]_I & \begin{array}{c}
 \overset{\phantom{\mathfrak{so}}}{1}\overset{\phantom{\mathfrak{so}}}{1}\\[-9pt]
 \overset{\,\mathfrak{f}_4}{5} \\[-9pt]
 \overset{\phantom{\mathfrak{so}}}{1}\overset{\phantom{\mathfrak{so}}}{1}\\[4pt]
 \end{array} &  (251, 344) & (\mathfrak{g}_{2})_{24}^{\oplus 4}
 \\\cline{2-5}
 & [2A_{2}, A_3]_I & \begin{array}{c}
 1\overset{\mathfrak{so}_9}{4}\overset{\mathfrak{su}_2}{1} \\[-9pt]
 \phantom{1}\overset{\phantom{\mathfrak{so}_9}}{1}\phantom{\overset{\mathfrak{su}_2}{1}} \\
 \end{array} &  (215, 312) & (\mathfrak{so}_{7})_{24} \oplus (\mathfrak{so}_{7})_{24} \oplus(\mathfrak{so}_{11})_{28}   
 \\\midrule

\multirow{18}{*}{8} & [0, A_4 + A_2]_I & 1\overset{\phantom{\mathfrak{su}_{2}}}{2}\overset{\mathfrak{su}_{2}}{2}\overset{\mathfrak{su}_{2}}{2}\overset{\mathfrak{su}_{2}}{2} &  (256, 408) & (\mathfrak{e}_{8})_{60} \oplus (\mathfrak{su}_{2})_{16} \oplus (\mathfrak{su}_{2})_{200}  
\\\cline{2-5}
 & [0, D_4+A_1]_I & 1\overset{\phantom{\mathfrak{su}_{2}}}{2}\overset{\mathfrak{su}_{2}}{2}\overset{\mathfrak{g}_{2}}{2} \sqcup 1 &  (204, 361) & (\mathfrak{e}_{8})_{12} \oplus (\mathfrak{e}_{8})_{48} \oplus (\mathfrak{usp}_{6})_{19} \\\cline{2-5}
 & [0, D_5(a_1)]_I & 1\overset{\phantom{\mathfrak{su}_{2}}}{2}\overset{\mathfrak{su}_{2}}{2}\overset{\mathfrak{su}_{3}}{2} \sqcup 1 &  (186, 340) & (\mathfrak{e}_{8})_{12} \oplus (\mathfrak{e}_{8})_{48} \oplus (\mathfrak{su}_{4})_{18} \\\cline{2-5}
 & [A_1, 2A_3]_I & 1\overset{\mathfrak{su}_{2}}{2}\overset{\mathfrak{g}_{2}}{2}\overset{\mathfrak{su}_{2}}{2} &  (256, 382) & (\mathfrak{e}_{7})_{48} \oplus(\mathfrak{usp}_{4})_{31}   
 \\\cline{2-5}
 & [A_1, A_4 + 2A_1]_I & 1\overset{\mathfrak{su}_{2}}{2}\overset{\mathfrak{su}_{3}}{2}\overset{\mathfrak{su}_{2}}{2} &  (226, 350) & (\mathfrak{e}_{7})_{48} \oplus(\mathfrak{su}_{2})_{30} 
 \\\cline{2-5}
 & [3A_{1}, A_{3}+A_{2}+A_{1}]_I & 1\overset{\mathfrak{g}_{2}}{3}1\overset{\mathfrak{g}_{2}}{3} & (252, 354) & (\mathfrak{f}_{4})_{36} \oplus(\mathfrak{su}_{2})_{384} \oplus(\mathfrak{su}_{2})_{31} \oplus(\mathfrak{su}_{2})_{19}   
 \\\cline{2-5}
 & [3A_{1}, D_{4}(a_{1})+A_{2}]_I & 1\overset{\mathfrak{g}_{2}}{3}1\overset{\mathfrak{su}_{3}}{3} &  (234, 335) & (\mathfrak{f}_{4})_{36} \oplus(\mathfrak{su}_{3})_{96} \oplus(\mathfrak{su}_{2})_{31}   
 \\\cline{2-5}
 & [A_{2}, A_{3}+A_{2}+A_{1}]_I & 1\overset{\mathfrak{su}_{3}}{3}1\overset{\mathfrak{g}_{2}}{3} &  (222, 323) & (\mathfrak{e}_{6})_{36} \oplus(\mathfrak{su}_{3})_{96} \oplus(\mathfrak{su}_{2})_{19}    
 \\\cline{2-5}
 & [A_{2}, D_{4}(a_{1})+A_{2}]_I & 1\overset{\mathfrak{su}_{3}}{3}1\overset{\mathfrak{su}_{3}}{3} &  (204, 304) & (\mathfrak{e}_{6})_{36} \oplus(\mathfrak{su}_{3})_{48} \oplus(\mathfrak{su}_{3})_{48} 
 \\\cline{2-5}
 & [A_{2}+3A_{1}, 2A_{2}+A_{1}]_I & \overset{\phantom{\mathfrak{su}_{2}}}{2}1\overset{\mathfrak{f}_{4}}{4}1 &  (252, 343) &  (\mathfrak{g}_{2})_{48} \oplus(\mathfrak{su}_{2})_{25} \oplus(\mathfrak{g}_{2})_{24} \oplus(\mathfrak{su}_{2})_{62}   
 \\\cline{2-5}
 & [A_{2}+3A_{1}, A_{3}+A_{1}]_I & \overset{\phantom{\mathfrak{su}_{2}}}{2}1\overset{\mathfrak{so}_{9}}{4}1 &  (212, 302) & (\mathfrak{so}_{7})_{48} \oplus(\mathfrak{su}_{2})_{25} \oplus(\mathfrak{so}_{7})_{24} \oplus(\mathfrak{su}_{2})_{21}    
 \\\cline{2-5}
 & [A_{2}+3A_{1}, D_{4}(a_{1})]_I & \overset{\phantom{\mathfrak{su}_{2}}}{2}1\overset{\mathfrak{so}_{8}}{4}1 &  (192, 281) & (\mathfrak{so}_{8})_{48} \oplus(\mathfrak{su}_{2})_{25} \oplus(\mathfrak{so}_{8})_{24}   
 \\\cline{2-5}
 & [A_{2}+2A_{1}, 2A_{2}+2A_{1}]_I & \overset{\mathfrak{su}_{2}}{2}1\overset{\mathfrak{f}_{4}}{3} &  (256, 348) & (\mathfrak{su}_{2})_{144} \oplus(\mathfrak{so}_{7})_{28} \oplus(\mathfrak{usp}_{4})_{62}   
 \\\cline{2-5}
 & [A_{2}+2A_{1}, A_{3}+2A_{1}]_I & \overset{\mathfrak{su}_{2}}{2}1\overset{\mathfrak{so}_{9}}{3} &  (216, 306) & \begin{aligned}(\mathfrak{su}_{2})_{48} &\oplus(\mathfrak{su}_{2})_{96} \oplus(\mathfrak{so}_{7})_{28} \cr &\oplus(\mathfrak{usp}_{4})_{21} \oplus(\mathfrak{su}_{2})_{40} \end{aligned}    
 \\\bottomrule
\end{array}$
\end{threeparttable}
\end{table}
\newpage
\begin{table}[H]
\begin{threeparttable}
\centering
\renewcommand{\arraystretch}{1.3}
\footnotesize
$\begin{array}{c c c c l}
\toprule
\text{dim}(\mathcal{C}) & \text{Fixture} & \text{6d SCFT} & \phantom{n}(n_v, n_h)\phantom{n} & \text{Flavor Symmetry}
\\
\midrule
\multirow{12}{*}{8}  & [A_{2}+2A_{1}, D_{4}(a_{1})+A_{1}]_I & \overset{\mathfrak{su}_{2}}{2}1\overset{\mathfrak{so}_{8}}{3} &  (196, 284) & (\mathfrak{su}_{2})_{48}^{\oplus 3} \oplus(\mathfrak{so}_{7})_{28} \oplus (\mathfrak{su}_{2})_{20}^{\oplus 3}    
\\\cline{2-5}
 & [4A_{1}, A_{3}+A_{2}]_I & \overset{\mathfrak{g}_{2}}{2}1\overset{\mathfrak{so}_{7}}{3} &  (236, 332) & (\mathfrak{usp}_{8})_{31} \oplus(\mathfrak{usp}_{4})_{20}   
 \\\cline{2-5}
 & [A_{2}+A_{1}, A_{3}+A_{2}]_I & \overset{\mathfrak{su}_{3}}{2}1\overset{\mathfrak{so}_{7}}{3} &  (206, 298) & (\mathfrak{su}_{2})_{96} \oplus(\mathfrak{su}_{6})_{30} \oplus(\mathfrak{usp}_{4})_{20} 
 \\\cline{2-5}
 & [2A_1, A_4 + A_1]_I & \overset{\mathfrak{su}_{2}}{1}\overset{\mathfrak{su}_{3}}{2}\overset{\mathfrak{su}_{3}}{2} &  (196, 305) & (\mathfrak{so}_{14})_{40} \oplus(\mathfrak{su}_{3})_{18} 
 \\\cline{2-5}
 & [3A_1, A_4]_I & 1\overset{\mathfrak{su}_{3}}{2}\overset{\mathfrak{su}_{4}}{2} &  (176, 278) & (\mathfrak{e}_{6})_{36} \oplus(\mathfrak{su}_{2})_{30} \oplus(\mathfrak{su}_{5})_{20} 
 \\\cline{2-5}
 & [A_3, 2A_2+A_1]_M & \overset{\mathfrak{su}_{2}}{1}\overset{\mathfrak{so}_{9}}{3}1 &  (192, 285) & (\mathfrak{so}_{7})_{24} \oplus(\mathfrak{so}_{11})_{28} \oplus(\mathfrak{su}_{2})_{21} \oplus(\mathfrak{su}_{2})_{40} 
 \\\cline{2-5}
 & [2A_{2}, 2A_{2} + A_1]_I & \begin{array}{c}
 1\overset{\mathfrak{f}_4}{4}1 \\[-9pt]
 \phantom{1}\overset{\phantom{\mathfrak{f}_4}}{1}\phantom{1} \\
 \end{array} &  (228, 318) & (\mathfrak{g}_{2})_{24}^{\oplus 3} \oplus(\mathfrak{su}_{2})_{62}   
 \\\cline{2-5}
 & [2A_{2}, A_3 + A_1]_I & \begin{array}{c}
 1\overset{\mathfrak{so}_9}{4}1 \\[-9pt]
 \phantom{1}\overset{\phantom{\mathfrak{so}_9}}{1}\phantom{1} \\
 \end{array} &  (188, 277) & (\mathfrak{so}_{7})_{24}^{\oplus 3} \oplus(\mathfrak{su}_{2})_{21}   
 \\\cline{2-5}
 & [2A_{2}, D_4(a_1)]_I & \begin{array}{c}
 1\overset{\mathfrak{so}_8}{4}1 \\[-9pt]
 \phantom{1}\overset{\phantom{\mathfrak{so}_8}}{1}\phantom{1} \\
 \end{array} &  (168, 256) & (\mathfrak{so}_{8})_{24}^{\oplus 3}   
 \\\midrule
\multirow{28}{*}{7} & [0, A_4 + A_2 + A_1]_I & 1\overset{\phantom{\mathfrak{su}_{2}}}{2}\overset{\mathfrak{su}_{2}}{2}\overset{\mathfrak{su}_{2}}{2}\overset{\phantom{\mathfrak{su}_{2}}}{2} &  (241, 392) &  (\mathfrak{e}_{8})_{60} \oplus (\mathfrak{su}_{2})_{200}  
\\\cline{2-5}
 & [A_1, A_4 + A_2]_I & 1\overset{\mathfrak{su}_{2}}{2}\overset{\mathfrak{su}_{2}}{2}\overset{\mathfrak{su}_{2}}{2} &  (197, 320) & (\mathfrak{e}_{7})_{48} \oplus (\mathfrak{su}_{2})_{16} \oplus (\mathfrak{su}_{2})_{160} \oplus (\mathfrak{su}_{2})_{40} 
 \\\cline{2-5}
 & [0, D_5(a_1) + A_1]_I & 1\overset{\phantom{\mathfrak{su}_{2}}}{2}\overset{\mathfrak{su}_{2}}{2}\overset{\mathfrak{su}_{2}}{2} \sqcup 1 &  (169, 320) &  (\mathfrak{e}_{8})_{12} \oplus (\mathfrak{e}_{8})_{48} \oplus (\mathfrak{su}_{2})_{16} \oplus (\mathfrak{su}_{2})_{112}
 \\\cline{2-5}
 & [2A_1, 2A_3]_I & \overset{\mathfrak{su}_{2}}{1}\overset{\mathfrak{g}_{2}}{2}\overset{\mathfrak{su}_{2}}{2} &  (209, 318) & (\mathfrak{so}_{13})_{40} \oplus(\mathfrak{usp}_{4})_{31}   
 \\\cline{2-5}
 & [2A_1, A_4 + 2A_1]_I & \overset{\mathfrak{su}_{2}}{1}\overset{\mathfrak{su}_{3}}{2}\overset{\mathfrak{su}_{2}}{2} &  (179, 286) & (\mathfrak{so}_{14})_{40} \oplus(\mathfrak{su}_{2})_{30}  
 \\\cline{2-5}

 & [3A_1, A_4+A_1]_M & 1\overset{\mathfrak{su}_{3}}{2}\overset{\mathfrak{su}_{3}}{2} &  (157, 255) & (\mathfrak{e}_{6})_{36} \oplus(\mathfrak{su}_{3})_{30} \oplus(\mathfrak{su}_{3})_{18} 
 \\\cline{2-5}
 & [A_{2}+3A_{1}, 2A_{2}+2A_{1}]_I & \overset{\phantom{\mathfrak{su}_{2}}}{2}1\overset{\mathfrak{f}_{4}}{3} &  (229, 317) &  (\mathfrak{g}_{2})_{48} \oplus(\mathfrak{su}_{2})_{25} \oplus(\mathfrak{usp}_{4})_{62}   
 \\\cline{2-5}
 & [A_{2}+3A_{1}, A_{3}+2A_{1}]_I & \overset{\phantom{\mathfrak{su}_{2}}}{2}1\overset{\mathfrak{so}_{9}}{3} & (189, 275) &  (\mathfrak{so}_{7})_{48} \oplus(\mathfrak{su}_{2})_{25} \oplus(\mathfrak{usp}_{4})_{21} \oplus(\mathfrak{su}_{2})_{40}   
 \\\cline{2-5}
 & [A_{2}+3A_{1}, D_{4}(a_{1})+A_{1}]_I & \overset{\phantom{\mathfrak{su}_{2}}}{2}1\overset{\mathfrak{so}_{8}}{3} &  (169, 253) &  (\mathfrak{so}_{8})_{48} \oplus(\mathfrak{su}_{2})_{25} \oplus (\mathfrak{su}_{2})_{20}^{\oplus 3}   
 \\\cline{2-5}
 & [4A_{1}, A_{3}+A_{2}+A_{1}]_I & \overset{\mathfrak{g}_{2}}{2}1\overset{\mathfrak{g}_{2}}{3} &  (217, 311) & (\mathfrak{su}_{2})_{384} \oplus(\mathfrak{usp}_{8})_{31} \oplus(\mathfrak{su}_{2})_{19}   
 \\\cline{2-5}
 & [4A_{1}, D_{4}(a_{1})+A_{2}]_I & \overset{\mathfrak{g}_{2}}{2}1\overset{\mathfrak{su}_{3}}{3} &  (199, 292) & (\mathfrak{su}_{3})_{96} \oplus(\mathfrak{usp}_{8})_{31}   
 \\\cline{2-5}
 & [A_{2}+A_{1}, A_{3}+A_{2}+A_{1}]_I & \overset{\mathfrak{su}_{3}}{2}1\overset{\mathfrak{g}_{2}}{3} &  (187, 277) & (\mathfrak{su}_{3})_{96} \oplus(\mathfrak{su}_{6})_{30} \oplus(\mathfrak{su}_{2})_{19}   
 \\\cline{2-5}
 & [A_{2}+A_{1}, D_{4}(a_{1})+A_{2}]_I & \overset{\mathfrak{su}_{3}}{2}1\overset{\mathfrak{su}_{3}}{3} &  (169, 258) & (\mathfrak{su}_{3})_{48} \oplus(\mathfrak{su}_{3})_{48} \oplus(\mathfrak{su}_{6})_{30}  
 \\\cline{2-5}
 & [A_{2}+2A_{1}, A_{3}+A_{2}]_I & \overset{\mathfrak{su}_{2}}{2}1\overset{\mathfrak{so}_{7}}{3} &  (177, 264) & (\mathfrak{su}_{2})_{48} \oplus(\mathfrak{usp}_{4})_{48} \oplus(\mathfrak{so}_{7})_{28} \oplus(\mathfrak{usp}_{4})_{20}   
 \\\cline{2-5}
 & \begin{gathered}[][2A_{2}, 2A_{2}+2A_{1}]_I \\[-3pt] [2A_{2}+A_{1}, 2A_{2}+A_{1}]_I \end{gathered} & 1\overset{\mathfrak{f}_{4}}{3}1 &  (205, 292) & (\mathfrak{g}_{2})_{24} \oplus(\mathfrak{g}_{2})_{24} \oplus(\mathfrak{usp}_{4})_{62}   
 \\\cline{2-5}
 & \begin{gathered}[][2A_{2}, A_{3}+2A_{1}]_I \\[-3pt] [2A_2+A_1, A_3+A_1]_M \end{gathered} & 1\overset{\mathfrak{so}_{9}}{3}1 &  (165, 250) & (\mathfrak{so}_{7})_{24}^{\oplus 2} \oplus(\mathfrak{usp}_{4})_{21} \oplus(\mathfrak{su}_{2})_{40}   
 \\\cline{2-5}
 & \begin{gathered}[] [2A_2, D_4(a_1) + A_1]_I \\[-3pt] [2A_2+ A_1, D_4(a_1)]_M \end{gathered} & 1\overset{\mathfrak{so}_{8}}{3}1 &  (145, 228) & (\mathfrak{so}_{8})_{24}^{\oplus 2} \oplus (\mathfrak{su}_{2})_{20}^{\oplus 3}   
 \\\cline{2-5}
 & [A_3, 2A_2+2A_1]_M & \overset{\mathfrak{su}_{2}}{1}\overset{\mathfrak{so}_{9}}{2} &  (169, 258) & (\mathfrak{so}_{11})_{28} \oplus(\mathfrak{usp}_{4})_{21} \oplus(\mathfrak{usp}_{4})_{40} 
 \\\cline{2-5}
 & [4A_1, A_4]_M & \overset{\mathfrak{su}_{3}}{1}\overset{\mathfrak{su}_{4}}{2} &  (141, 232) & (\mathfrak{su}_{8})_{30} \oplus(\mathfrak{su}_{5})_{20}  
 \\\bottomrule
\end{array}$
\end{threeparttable}
\end{table}
\newpage
\begin{table}[H]
\begin{threeparttable}
\centering
\renewcommand{\arraystretch}{1.3}
\footnotesize
$\begin{array}{c c c c l}
\toprule
\text{dim}(\mathcal{C}) & \phantom{i}\text{Fixture}\phantom{i} & \text{6d SCFT} & (n_v, n_h) & \text{Flavor Symmetry}
\\
\midrule
\multirow{20}{*}{6} & [A_1, A_4 + A_2 + A_1]_I & 1\overset{\mathfrak{su}_{2}}{2}\overset{\mathfrak{su}_{2}}{2}2 & (182, 304) & (\mathfrak{e}_{7})_{48} \oplus(\mathfrak{su}_{2})_{40} \oplus(\mathfrak{su}_{2})_{160}  
\\\cline{2-5}
 & [3A_1, 2A_3]_M & 1\overset{\mathfrak{g}_{2}}{2}\overset{\mathfrak{su}_{2}}{2} & (170, 269) & (\mathfrak{f}_{4})_{36} \oplus(\mathfrak{usp}_{6})_{31}    
 \\\cline{2-5}
 & \begin{gathered}[] [3A_1, A_4+2A_1]_M \\[-3pt] [A_2, 2A_3]_M \end{gathered} & 1\overset{\mathfrak{su}_{3}}{2}\overset{\mathfrak{su}_{2}}{2} & (140, 236) & (\mathfrak{e}_{6})_{36} \oplus(\mathfrak{su}_{4})_{30} 
 \\\cline{2-5}
 & [2A_1, A_4 + A_2]_I & \overset{\mathfrak{su}_{2}}{1}\overset{\mathfrak{su}_{2}}{2}\overset{\mathfrak{su}_{2}}{2} & (150, 256) & (\mathfrak{so}_{16})_{40} \oplus(\mathfrak{su}_{2})_{16} \oplus(\mathfrak{su}_{2})_{120} 
 \\\cline{2-5}
 & [A_{2}+2A_{1}, D_{4}(a_{1})+A_{2}]_I & \overset{\mathfrak{su}_{2}}{2}1\overset{\mathfrak{su}_{3}}{3} & (140, 224) & (\mathfrak{su}_{6})_{48} \oplus(\mathfrak{so}_{7})_{28}   
 \\\cline{2-5}
 & [A_{2}+3A_{1}, A_{3}+A_{2}]_I & \overset{\phantom{\mathfrak{su}_{2}}}{2}1\overset{\mathfrak{so}_{7}}{3} & (150, 233) & (\mathfrak{so}_{9})_{48} \oplus(\mathfrak{su}_{2})_{25} \oplus(\mathfrak{usp}_{4})_{20}   
 \\\cline{2-5}
 & [A_{2}+2A_{1}, A_{3}+A_{2}+A_{1}]_I & \overset{\mathfrak{su}_{2}}{2}1\overset{\mathfrak{g}_{2}}{3} & (158, 243) & (\mathfrak{usp}_{6})_{48} \oplus(\mathfrak{so}_{7})_{28} \oplus(\mathfrak{su}_{2})_{19}  
 \\\cline{2-5}
 & [2A_{2}, A_{3}+A_{2}]_I & 1\overset{\mathfrak{so}_{7}}{3}1 & (126, 208) & (\mathfrak{so}_{9})_{24} \oplus(\mathfrak{so}_{9})_{24} \oplus(\mathfrak{usp}_{4})_{20} 
 \\\cline{2-5}
 & [2A_{2}+A_{1}, 2A_{2}+2A_{1}]_I & 1\overset{\mathfrak{f}_{4}}{2} & (182, 266) & (\mathfrak{g}_{2})_{24} \oplus(\mathfrak{usp}_{6})_{62}  
 \\\cline{2-5}
 & \begin{gathered}[][A_3+A_1, 2A_2+2A_1]_M \\ [2A_2+A_1, A_3+2A_1]_M \end{gathered} & 1\overset{\mathfrak{so}_{9}}{2} & (142, 223) & (\mathfrak{so}_{7})_{24} \oplus(\mathfrak{usp}_{6})_{21} \oplus(\mathfrak{usp}_{4})_{40} 
 \\\cline{2-5}
 & \begin{gathered}[][D_4(a_1), 2A_2+2A_1]_M \\ [2A_2+A_1, D_4(a_1)+A_1]_M \end{gathered} & 1\overset{\mathfrak{so}_{8}}{2} & (122, 200) & (\mathfrak{so}_{8})_{24} \oplus (\mathfrak{usp}_{4})_{20}^{\oplus 3} 
 \\\cline{2-5}
 & [4A_1, A_4+A_1]_M & \overset{\mathfrak{su}_{3}}{1}\overset{\mathfrak{su}_{3}}{2} & (122, 209) & (\mathfrak{su}_{9})_{30} \oplus(\mathfrak{su}_{3})_{18} 
 \\\cline{2-5}
 & [0, D_4 + A_2]_I & 1\overset{\phantom{\mathfrak{su}_{2}}}{2}\overset{\mathfrak{su}_{2}}{2}\overset{\phantom{\mathfrak{su}_{2}}}{2} \sqcup 1 & (154, 304) & (\mathfrak{e}_8)_{12} \oplus (\mathfrak{e}_8)_{48} \oplus (\mathfrak{su}_3)_{28} 
 \\\cline{2-5}
 &  [0, A_5]_I & 1\overset{\phantom{\mathfrak{su}_{2}}}{2}\overset{\mathfrak{su}_{2}}{2} \sqcup 1\overset{\phantom{\mathfrak{su}_{2}}}{2} & (118, 269) &  (\mathfrak{e}_8)_{24} \oplus (\mathfrak{su}_2)_{13} \oplus (\mathfrak{e}_8)_{36} \oplus  (\mathfrak{g}_2)_{16}   
 \\\cline{2-5}
 & [0, E_6(a_3)]_I & 1\overset{\phantom{\mathfrak{su}_{2}}}{2}\overset{\mathfrak{su}_{2}}{2} \sqcup 1 \sqcup 1 & 
(106, 256) & (\mathfrak{e}_8)_{12} \oplus (\mathfrak{e}_8)_{12} \oplus (\mathfrak{e}_8)_{36} \oplus  (\mathfrak{g}_2)_{16}  
\\\midrule

\multirow{17}{*}{5} & [0, A_4 + A_3]_I & 1\overset{\phantom{\mathfrak{su}_{2}}}{2}\overset{\phantom{\mathfrak{su}_{2}}}{2}\overset{\phantom{\mathfrak{su}_{2}}}{2}\overset{\phantom{\mathfrak{su}_{2}}}{2} & (175, 324) & (\mathfrak{e}_{8})_{60} \oplus(\mathfrak{su}_{2})_{124} 
\\\cline{2-5}
 & [0, D_5(a_1) + A_2]_I & 1\overset{\phantom{\mathfrak{su}_{2}}}{2}\overset{\phantom{\mathfrak{su}_{2}}}{2}\overset{\phantom{\mathfrak{su}_{2}}}{2} \sqcup \overset{\phantom{\mathfrak{su}_{2}}}{1} & (127, 275) & (\mathfrak{e}_{8})_{12} \oplus (\mathfrak{e}_{8})_{48} \oplus(\mathfrak{su}_{2})_{75}   
 \\\cline{2-5}
 & [2A_1, A_4 + A_2 + A_1]_I & \overset{\mathfrak{su}_{2}}{1}\overset{\mathfrak{su}_{2}}{2}2 & (135, 240) & (\mathfrak{so}_{16})_{40} \oplus(\mathfrak{su}_{2})_{120} 
 \\\cline{2-5}
 & [3A_1, A_4+A_2]_M & 1\overset{\mathfrak{su}_{2}}{2}\overset{\mathfrak{su}_{2}}{2} & (111, 204) & (\mathfrak{e}_{7})_{36} \oplus(\mathfrak{su}_{2})_{28} \oplus(\mathfrak{su}_{2})_{16} \oplus(\mathfrak{su}_{2})_{84}  
 \\\cline{2-5}
 & [A_{2}+3A_{1}, A_{3}+A_{2}+A_{1}]_I & \overset{\phantom{\mathfrak{su}_{2}}}{2}1\overset{\mathfrak{g}_{2}}{3} & (131, 212) & (\mathfrak{f}_{4})_{48} \oplus(\mathfrak{su}_{2})_{25} \oplus(\mathfrak{su}_{2})_{19}   
 \\\cline{2-5}
 & [A_{2}+3A_{1}, D_{4}(a_{1})+A_{2}]_I & \overset{\phantom{\mathfrak{su}_{2}}}{2}1\overset{\mathfrak{su}_{3}}{3} & (113, 193) & (\mathfrak{e}_{6})_{48} \oplus(\mathfrak{su}_{2})_{25}    
 \\\cline{2-5}
 & [4A_1, 2A_3]_M & \overset{\mathfrak{g}_{2}}{1}\overset{\mathfrak{su}_{2}}{2} & (135, 226) & (\mathfrak{usp}_{12})_{31}   
 \\\cline{2-5}
 & \begin{gathered}[] [4A_1, A_4+2A_1]_M \\[-3pt] [A_2 + A_1, 2A_3]_I \end{gathered} & \overset{\mathfrak{su}_{3}}{1}\overset{\mathfrak{su}_{2}}{2} & (105, 190) & (\mathfrak{su}_{10})_{30} 
 \\\cline{2-5}
 & [2A_{2}, A_{3}+A_{2}+A_{1}]_I & 1\overset{\mathfrak{g}_{2}}{3}1 & (107, 187) & (\mathfrak{f}_{4})_{24} \oplus(\mathfrak{f}_{4})_{24} \oplus(\mathfrak{su}_{2})_{19}   
 \\\cline{2-5}
 & [2A_{2}, D_{4}(a_{1})+A_{2}]_I & 1\overset{\mathfrak{su}_{3}}{3}1 & (89, 168) & (\mathfrak{e}_{6})_{24} \oplus (\mathfrak{e}_{6})_{24}   
 \\\cline{2-5}
 & [2A_2+A_1, A_3+A_2]_M & 1\overset{\mathfrak{so}_{7}}{2} & (103, 179) & (\mathfrak{so}_{9})_{24} \oplus(\mathfrak{su}_{2})_{19} \oplus(\mathfrak{usp}_{8})_{20}   
 \\\cline{2-5}
 & [2A_{2}+2A_{1}, 2A_{2}+2A_{1}]_I & \overset{\mathfrak{f}_{4}}{1} & (159, 240) & (\mathfrak{usp}_{8})_{62}   
 \\\cline{2-5}
 & [2A_2+2A_1, A_3+2A_1]_M & \overset{\mathfrak{so}_{9}}{1} & (119, 196) & (\mathfrak{usp}_{8})_{21} \oplus(\mathfrak{usp}_{6})_{40}   
 \\\cline{2-5}
 & [2A_2+2A_1, D_4(a_1)+A_1]_M & \overset{\mathfrak{so}_{8}}{1} & (99, 172) & (\mathfrak{usp}_{6})_{20}^{\oplus 3}   
 \\\bottomrule
\end{array}$
\end{threeparttable}
\end{table}
\newpage
\begin{table}[H]
\begin{threeparttable}
\centering
\renewcommand{\arraystretch}{1.3}
\footnotesize
$\begin{array}{c c c c l}
\toprule
\text{dim}(\mathcal{C}) & \text{Fixture} & \text{6d SCFT} & (n_v, n_h) & \text{Flavor Symmetry}
\\
\midrule
\multirow{6}{*}{5}  & [0, A_5 + A_1]_I & 1\overset{\phantom{\mathfrak{su}_{2}}}{2}\overset{\phantom{\mathfrak{su}_{2}}}{2} \sqcup \overset{\phantom{\mathfrak{su}_{2}}}{1}\overset{\phantom{\mathfrak{su}_{2}}}{2} & (103, 251) &  \begin{aligned}(\mathfrak{e}_{8})_{36} &\oplus(\mathfrak{su}_{2})_{38} \cr  &\oplus (\mathfrak{e}_{8})_{24} \oplus(\mathfrak{su}_{2})_{13}\end{aligned}   
\\\cline{2-5}
 & [0, E_6(a_3) + A_1]_I & 1\overset{\phantom{\mathfrak{su}_{2}}}{2}\overset{\phantom{\mathfrak{su}_{2}}}{2} \sqcup \overset{\phantom{\mathfrak{su}_{2}}}{1} \sqcup \overset{\phantom{\mathfrak{su}_{2}}}{1} & (91, 238) &  (\mathfrak{e}_{8})_{36} \oplus(\mathfrak{su}_{2})_{38} \oplus (\mathfrak{e}_{8})_{12}^{\oplus 2}   
 \\\cline{2-5}
 & [0, D_6(a_2)]_I & 1\overset{\phantom{\mathfrak{su}_{2}}}{2} \sqcup \overset{\phantom{\mathfrak{su}_{2}}}{1}\overset{\phantom{\mathfrak{su}_{2}}}{2} \sqcup \overset{\phantom{\mathfrak{su}_{2}}}{1}& (79, 226) &  (\mathfrak{e}_{8})_{12} \oplus (\mathfrak{e}_{8})_{24}^{\oplus 2} \oplus (\mathfrak{su}_{2})_{13}^{\oplus 2}
 \\\cline{2-5}
 & [0, E_7(a_5)]_I & 1\overset{\phantom{\mathfrak{su}_{2}}}{2} \sqcup \overset{\phantom{\mathfrak{su}_{2}}}{1} \sqcup \overset{\phantom{\mathfrak{su}_{2}}}{1} \sqcup \overset{\phantom{\mathfrak{su}_{2}}}{1} & (67, 213) &  (\mathfrak{e}_{8})_{12}^{\oplus 3} \oplus (\mathfrak{e}_{8})_{24} \oplus(\mathfrak{su}_{2})_{13}   
 \\\cline{2-5}
 & [0, E_8(a_7)]_I & \overset{\phantom{\mathfrak{su}_{2}}}{1} \sqcup \overset{\phantom{\mathfrak{su}_{2}}}{1} \sqcup \overset{\phantom{\mathfrak{su}_{2}}}{1} \sqcup \overset{\phantom{\mathfrak{su}_{2}}}{1} \sqcup \overset{\phantom{\mathfrak{su}_{2}}}{1} & (55, 200) &  (\mathfrak{e}_{8})_{12}^{\oplus 5}   
 \\\midrule

\multirow{7}{*}{4} & [A_1, A_4 + A_3]_I & 1\overset{\phantom{\mathfrak{su}_{2}}}{2}2\overset{\phantom{\mathfrak{su}_{2}}}{2} & (116, 235) & (\mathfrak{e}_{8})_{48} \oplus(\mathfrak{su}_{2})_{75}   
\\\cline{2-5}
 & [3A_1, A_4+A_2+A_1]_M & 1\overset{\mathfrak{su}_{2}}{2}2 & (96, 188) & (\mathfrak{e}_{7})_{36} \oplus(\mathfrak{g}_{2})_{28}   
 \\\cline{2-5}
 & \begin{gathered}[] [4A_1, A_4+A_2]_M \\ [A_2+2A_1, 2A_3]_M \end{gathered} & \overset{\mathfrak{su}_{2}}{1}\overset{\mathfrak{su}_{2}}{2} & (76, 152) & (\mathfrak{so}_{16})_{28} \oplus (\mathfrak{su}_{2})_{16} \oplus(\mathfrak{su}_{2})_{56}  
 \\\cline{2-5}
 & [2A_2+A_1, A_3+A_2+A_1]_M & 1\overset{\mathfrak{g}_{2}}{2} & (84, 156) & (\mathfrak{f}_{4})_{24} \oplus(\mathfrak{usp}_{8})_{19}   
 \\\cline{2-5}
 & [2A_2+A_1, D_4(a_1)+A_2]_M & 1\overset{\mathfrak{su}_{3}}{2} & (66, 134) & (\mathfrak{e}_{6})_{24} \oplus(\mathfrak{su}_{6})_{18}   
 \\\cline{2-5} 
 & [2A_2+2A_1, A_3+A_2]_M & \overset{\mathfrak{so}_{7}}{1} & (80, 150) & (\mathfrak{usp}_{4})_{19} \oplus(\mathfrak{usp}_{12})_{20}   
 \\\midrule
 
\multirow{6}{*}{3} & [A_1, A_4 + A_3]_I & 1\overset{\phantom{\mathfrak{su}_{2}}}{2}2 & (69, 158) &  (\mathfrak{e}_{8})_{36} \oplus(\mathfrak{su}_{2})_{38}   
\\\cline{2-5}
 & [4A_1, A_4+A_2+A_1]_M & \overset{\mathfrak{su}_{2}}{1}2 & (61, 136) & (\mathfrak{so}_{19})_{28}   
 \\\cline{2-5}
 & \begin{gathered}[] [A_2+3A_1, 2A_3]_M \\[-3pt] [D_5, (A_1,E_7)]_I \end{gathered} & 1\overset{\mathfrak{su}_{2}}{2} & (49, 112) & (\mathfrak{e}_{7})_{24} \oplus(\mathfrak{so}_{7})_{16}   
 \\\cline{2-5}
 & [2A_2+2A_1, A_3+A_2+A_1]_M & \overset{\mathfrak{g}_{2}}{1} & (61, 125) & (\mathfrak{usp}_{14})_{19}   
 \\\cline{2-5}
 & [2A_2+2A_1, D_4(a_1)+A_2]_M & \overset{\mathfrak{su}_{3}}{1} & (43, 100) & (\mathfrak{su}_{12})_{18}   
 \\\midrule

\multirow{2}{*}{2} & [D_5+A_1, (A_1,E_7)]_I & 1\overset{\phantom{\mathfrak{su}_2}}{2} & (34, 93) & (\mathfrak{e}_{8})_{24} \oplus(\mathfrak{su}_{2})_{13}   
\\\cline{2-5}
 & [D_6(a_1), (A_1,E_7), E_8(a_1)]_I & \overset{\phantom{\mathfrak{su}_{2}}}{1} \sqcup \overset{\phantom{\mathfrak{su}_{2}}}{1} & (22, 80) & (\mathfrak{e}_{8})_{12} \oplus (\mathfrak{e}_{8})_{12}   
 \\\midrule

 \multirow{1}{*}{1} & \begin{gathered}[] [E_6, (D_4,F_4)]_I \\[-3pt] [E_6(a_1), (A_2,E_6)]_I \\[-3pt] [A_6, (0,E_7), ]_I\\[-3pt] [E_7(a_4), (A_1,E_7)]_M \end{gathered} & \overset{\phantom{\mathfrak{su}_2}}{1} & (11, 40) & (\mathfrak{e}_{8})_{12}   
 \\\bottomrule
\end{array}$
\end{threeparttable}
\vspace{2mm}
\caption{In this table, we show the relationship between the 4d $\mathcal{N}=2$ SCFTs obtained as $T^2$ compactifications of nilpotent Higgsings of minimal $(\mathfrak{e}_8, \mathfrak{e}_8)$ conformal matter, $\mathcal{T}_{\mathfrak{e}_8}\{Y_1, Y_2\}\langle T^2\rangle$, and the class $\mathcal{S}$ theories $\mathcal{S}_{\mathfrak{e}_8}\langle C_{0,3}\rangle\{Y_1, Y_2, Y_\text{simple}\}$. The information contained in each column is as described in the caption of Table \ref{tbl:E6E6}. From the class $\mathcal{S}$ perspective the determination of the flavor symmetry requires knowledge of the Hall--Littlewood index, which was determined for $\mathfrak{e}_8$ fixtures in \cite{Chcaltana:2018zag}.}
\label{tbl:E8E8}
\end{table}

\section{Discussion}\label{sec:discussion}

We have now established the correspondence 
\begin{equation}
    \mathcal{S}_\mathfrak{g}\langle C_{0,3} \rangle \{Y_1, Y_2, Y_\text{simple}\} = \mathcal{T}_\mathfrak{g}\{Y_1, Y_2\}\langle T^2 \rangle
\end{equation}
between two different 6d constructions for 4d $\mathcal{N}=2$ SCFTs. The left hand side is the class $\mathcal{S}$ construction that describes 4d $\mathcal{N}=2$ SCFTs obtained via 6d $(2,0)$ SCFTs compactified on a punctured Riemann surface.
The right hand side is the construction of 4d $\mathcal{N}=2$ SCFTs from a 6d $(1,0)$ SCFT compactified on a $T^2$. What we have shown is that these two constructions from two different 6d origins give rise to identical 4d theories.

Based on this result, we demonstrate how such a plurality of origins has immediate practical applications, such as determining whether the 4d SCFT has a product structure or a flavor symmetry enhancement. Such properties can be hidden in the 6d $(2,0)$ construction but are manifest from the 6d $(1,0)$ point of view. Furthermore, knowing that two different 6d SCFTs can yield the same 4d theory after compactification leads to an interesting web of AGT-esque correspondences which we highlight here.

We want to make a remark that our approach in this paper is based on field-theoretic principles, and we expect that this correspondence uplifts to a relationship between compactification spaces in string theory. We discuss how the correspondence is related to mirror symmetry between Calabi--Yau threefolds. In each case, the correspondence sheds some light on diverse aspects of 4d $\mathcal{N}=2$ SCFTs and therefore opens interesting directions for future work.

\subsection{Enhanced flavor symmetry}\label{sec:fenc}

From the perspective of the 6d $(1,0)$ SCFT origin, the non-Abelian part of the superconformal flavor symmetry is manifest: it is the flavor symmetry of the 6d SCFT. On the other hand, from the class $\mathcal{S}$ point-of-view one observes only the manifest flavor group as in equation \eqref{eqn:manifest-flavor}, and the full superconformal flavor symmetry is obtained by calculating the first terms of the Hall--Littlewood index. We emphasize that this often subtle computation of the Hall--Littlewood index is rendered unnecessary by the existence of the 6d $(1,0)$ origin.

From the 6d $(2,0)$ perspective, when the flavor symmetry is enhanced, the total flavor central charge is determined in terms of the central charges of the manifest flavor symmetries and the index of the embedding \cite{Argyres:2007cn}. There are cases where computing the flavor central charge requires additional care. For example, if the manifest flavor symmetry is $(\mathfrak{u}_1)_{k^\prime}$ and enhances into an $(\mathfrak{su}_2)_k$, then the lack of knowledge about $k^\prime$ prevents us from determining $k$. This can sometimes be circumvented by utilizing S-duality, which was done in \cite{Chacaltana:2014jba,Chacaltana:2017boe,Chcaltana:2018zag} for the $\mathfrak{e}_6$, $\mathfrak{e}_7$, and $\mathfrak{e}_8$ fixtures; however, such a technique does not work in all cases. 

In contrast, using the relationship to $T^2$ compactifications of 6d $(1,0)$ SCFTs explored in this paper, where the flavor central charge is manifest, we were able to determine the previously unknown flavor central charges for various fixtures. These flavor charges are gathered in Table \ref{tbl:enclevels}.

\begin{table}[H]
\centering
\begin{threeparttable}
\renewcommand{\arraystretch}{1.3}
\begin{tabular}{c|c}
    \toprule
        Fixture & $k$\\
    \midrule 
        $[A_2 + 2A_1, A_3, E_8(a_1)]$ & $ 48$\\
        $[A_2, A_3 + A_2, E_8(a_1)]$ & $ 96$\\
        $[A_{2}, D_{4}(a_{1})+A_{2}, E_8(a_1)]$ & $ 48$\\
        $[0, D_{4}+A_{1}, E_8(a_1)]$ & $ 48$\\
        $[A_{2}+A_{1}, A_{3}+A_{2}, E_8(a_1)]$ & $ 96$\\
        $[A_3, 2A_2+A_1, E_8(a_1)]$ & $ 21$\\
        $[A_{2}+A_{1}, D_{4}(a_{1})+A_{2}, E_8(a_1)]$ & $48$\\
        $[A_3, 2A_2+2A_1, E_8(a_1)]$ & $ 21$\\
        $[A_1, A_4 + A_2, E_8(a_1)]$ & $ 40$\\
        $[0, D_{5}(a_1)+A_{1}, E_8(a_1)]$ & $ 48$\\
        $[A_1, A_4 + A_2 + A_1,E_8(a_1)]$ & $ 40$\\
        $[D_4(a_1), 2A_2+2A_1, E_8(a_1)]$ & $k_1 = k_2 = 20$\\
        $[0, D_4 + A_2, E_8(a_1)]$ & $ 12$\\
    \bottomrule
    \end{tabular}
\end{threeparttable}
\caption{Predicted value of the enhanced flavor central charge for a set of class $\mathcal{S}$ theories of type $\mathfrak{e}_8$, which were previously determined only up to a free parameter in \cite{Chcaltana:2018zag}. The 6d $(1,0)$ perspective makes the flavor central charge $k$ manifest in terms of an 't Hooft anomaly coefficient associated to the 6d flavor symmetry.}
\label{tbl:enclevels}
\end{table}

\subsection{Product theories in class \texorpdfstring{$\mathcal{S}$}{S}}\label{sec:product}

When considering a fixture of class $\mathcal{S}$, 
\begin{equation}
        \mathcal{S}_\mathfrak{g}\langle C_{0,3} \rangle \{Y_1, Y_2, Y_3\} \,,
\end{equation}
it is an ongoing problem to determine whether the interacting part of the 4d $\mathcal{N}=2$ SCFT is an irreducible SCFT, or if it is a product of several SCFTs. A careful manipulation of the superconformal indices may be used in some cases to extract the number of stress-tensor multiplets in the fixture, which determines whether the theory is a product or not \cite{Distler:2017xba}. An alternative and computationally tractable criterion for the identification of product SCFTs is presented based on unitarity bounds and applied to the fixtures of the $\mathfrak{e}_7$ theory in \cite{Distler:2018gbc}. 

Once again, the correspondence we established in this paper reveals a particularly illuminating perspective: by utilizing its 6d $(1,0)$ origin, it offers a straightforward principle to determine whether the class $\mathcal{S}$ theory is a product or not. If the parent 6d $(1,0)$ theory is a product of multiple disconnected SCFTs, then the 4d $\mathcal{N}=2$ SCFT obtained after $T^2$ compactification is also a product SCFT. We verify that each product theory appearing in \cite{Distler:2017xba} and \cite{Distler:2018gbc} for the $\mathfrak{e}_6$ and $\mathfrak{e}_7$ fixtures arises from a $T^2$ compactification of a 6d $(1,0)$ SCFT that is a product. For $\mathfrak{e}_8$ fixtures involving a simple puncture, we find fourteen product SCFTs; it would be interesting to check the product structure directly from the class $\mathcal{S}$ perspective. 

Intriguingly, product 6d $(1,0)$ SCFTs always arise through E-string nucleation, as described in Section \ref{sec:def}. Due to this feature, when a 6d $(1,0)$ SCFT compactified on a $T^2$ gives rise to a product 4d $\mathcal{N}=2$ SCFTs, then all but one of the factors in the product are copies of Minahan--Nemeschansky theories.

\subsection{Connections to the AGT correspondence}\label{sec:agt}

Theories of class $\mathcal{S}$ are central to the AGT correspondence \cite{Alday:2009aq}, which describes a deep relationship between 4d $\mathcal{N}=2$ theories and 2d topological theories.\footnote{For a recent review of the AGT correspondence, see \cite{LeFloch:2020uop}.} The AGT correspondence arises from constructing both the 4d $\mathcal{N}=2$ theories and 2d topological theories via compactifying 6d $(2,0)$ SCFTs. To be specific, let us consider the partition function of the 6d $(2,0)$ SCFT of type $\mathfrak{g}$ on a product manifold
\begin{equation}
    Z_\mathfrak{g}^{(2,0)}(M_4 \times C_{g,n}) \,,
\end{equation}
where $M_4$ is an arbitrary four-manifold and $C_{g,n}$ is a punctured Riemann surface. With appropriate topological twists, the partition function does not depend on the volume of $M_4$ or $C_{g,n}$, and we can thus consider evaluating $Z_\mathfrak{g}^{(2,0)}$ in two different perspectives. 

We can first consider compactifying the 6d $(2,0)$ theory on $C_{g,n}$ to obtain a 4d $\mathcal{N}=2$ SCFT, and then find the partition function of the associated 4d theory on $M_4$. This is a theory of class $\mathcal{S}$ with the partition function:
\begin{equation}\label{eqn:20Z}
    Z_\mathfrak{g}^{(2,0)}\langle C_{g,n} \rangle (M_4) \,.
\end{equation}

Conversely, we can instead consider compactifying the 6d $(2,0)$ theory on a four-manifold $M_4$ to obtain an effective 2d theory and then evaluate its partition function on $C_{g,n}$:
\begin{equation}
    Z_\mathfrak{g}^{(2,0)}\langle M_4 \rangle (C_{g,n}) \,.
\end{equation}
An example of such a 2d theory is the Toda CFT associated to the algebra $\mathfrak{g}$ when $M_4 = S^4$. Since the partition function does not depend on the volumes of $M_4$ and $C_{g,n}$, we are led to the equality
\begin{equation}
    Z_\mathfrak{g}^{(2,0)}\langle C_{g,n} \rangle (M_4) = Z_\mathfrak{g}^{(2,0)}\langle M_4 \rangle (C_{g,n}) \,.
\end{equation}

In an analogous vein, we can generalize the above argument and consider the partition functions of the 6d $(1,0)$ theories $\mathcal{T}_\mathfrak{g}\{Y_1, Y_2\}$ on a product manifold
\begin{equation}\label{eqn:10ZUV}
    Z_{\mathcal{T}_\mathfrak{g}\{Y_1, Y_2\}}^{(1,0)} (M_4 \times T^2) \,,
\end{equation}
suitably twisted to be independent of the volumes of the torus and the four-manifold $M_4$. Following the argument above, we can again construct two theories from compactifications of the 6d $(1,0)$ theory on either manifold: a 4d $\mathcal{N}=2$ SCFT or a 2d theory. At the level of the partition functions, we have
\begin{equation}\label{eqn:10Z}
    Z_{\mathcal{T}_\mathfrak{g}\{Y_1, Y_2\}}^{(1,0)}\langle T^2 \rangle (M_4) = 
    Z_{\mathcal{T}_\mathfrak{g}\{Y_1, Y_2\}}^{(1,0)}\langle M_4 \rangle (T^2)\,.
\end{equation}

In this paper, we showed a set of 4d $\mathcal{N}=2$ SCFTs that can be obtained from both $(1,0)$ and $(2,0)$ origins in 6d; thus, the partition functions can be computed from both perspectives as in equations \eqref{eqn:20Z} and \eqref{eqn:10Z}. Putting these together, we obtain equalities between the four quantities:
\begin{equation}\label{eqn:Zeq}
    Z_\mathfrak{g}^{(2,0)}\langle M_4 \rangle (C_{g,n}) = Z_\mathfrak{g}^{(2,0)}\langle C_{g,n} \rangle (M_4) = 
    Z_{\mathcal{T}_\mathfrak{g}\{Y_1, Y_2\}}^{(1,0)}\langle T^2 \rangle (M_4) =
    Z_{\mathcal{T}_\mathfrak{g}\{Y_1, Y_2\}}^{(1,0)}\langle M_4 \rangle (T^2) \,.
\end{equation}
The rightmost partition function is that of the particular 2d theory, obtained from $\mathcal{T}_\mathfrak{g}\{Y_1, Y_2\}$ compactified on $M_4$, evaluated on $T^2$. These 2d theories are not known in general, but there are some known examples of 6d $(1,0)$ theories compactified on four-manifolds \cite{Apruzzi:2016nfr}. It would be valuable to understand how the equation \eqref{eqn:Zeq} can be utilized to learn more about various theories appearing therein and beyond.

\subsection{Mirror symmetry and a geometric dictionary}\label{sec:mirror}

From the 6d $(1,0)$ perspective, each 6d SCFT is engineered from a non-compact elliptically fibered Calabi--Yau manifold $Y_3$. When the resulting SCFT is further compactified on a $T^2$, one can utilize the duality between F-theory and type IIA, via M-theory, to obtain type IIA on $Y_3$. Using mirror symmetry, we can translate this into type IIB on $\widehat{Y}_3$, the mirror manifold to $Y_3$. Starting from the singular geometry associated to minimal $(\mathfrak{g}, \mathfrak{g})$ conformal matter, the Riemann surface and the puncture data appearing in the 6d $(2,0)$ description can be read off from the mirror manifold \cite{DelZotto:2015rca}. In this way, a geometric dictionary can be constructed where the features coming from the 6d $(1,0)$ origin are encoded in $Y_3$ and those associated to the 6d $(2,0)$ origin are obtained from the mirror $\widehat{Y}_3$.

To illustrate how some of the class $\mathcal{S}$ features are contained in the mirror of the 6d $(1,0)$ geometry, we consider the 4d $\mathcal{N} = 2$ theory obtained as $\mathcal{T}_{\mathfrak{e}_6}\{Y_\text{full}, Y_\text{full}\}\langle T^2 \rangle$ and equivalently as $\mathcal{S}_{\mathfrak{e}_6}\langle C_{0,3} \rangle \{Y_\text{full}, Y_\text{full}, Y_\text{simple}\}$. The tensor branch geometry of minimal $(\mathfrak{e}_6, \mathfrak{e}_6)$ conformal matter is captured by the configuration
\begin{equation}
    1 \overset{\mathfrak{su}_3}{3} 1.
\end{equation}
The geometry associated to the SCFT at the origin of the tensor branch is obtained by shrinking the curves to zero volume. In this case, this singular geometry can be described as an orbifold
\begin{equation}
    (T^2 \times \mathbb{C}^2)/\Gamma,
\end{equation}
where the quotients $\Gamma \subset U(1) \times SU(2) \subset SU(3)$ are Abelian and their elements can be expressed as eigenvalues of a $3 \times 3$ matrix. For the singular geometry associated to minimal $(\mathfrak{e}_6, \mathfrak{e}_6)$ conformal matter, $\Gamma$ is generated by the two elements $(a; a^{-1}, 1)$ and $(1; b, b^{-1})$ where $a$ and $b$ are primitive roots of unity satisfying $a^3 = b^3 = 1$.

The mirror manifold was determined directly from the orbifold action in \cite{DelZotto:2015rca}, and can be expressed as a Landau--Ginzburg model where the superpotential (after some tuning) can be written as
\begin{equation}
    W = x_1^3 + x_2^2(x_3 + y_1 + y_2) + (x_3 + y_1 + y_2)^2 x_3 \,.
\end{equation}
By going into the appropriate patch, $x_3 = 1$, one obtains a local threefold described by
\begin{equation}\label{eqn:blah}
    f = x_1^3 + x_2^2\rho + \rho^2 = 0 \,,
\end{equation}
where
\begin{equation}
    \rho = 1 + y_1 + y_2 \,.
\end{equation}
Recalling that the $y_i$ are coordinates on $\mathbb{P}^1$, the hypersurface $\rho = 0$ describes a sphere with three punctures, at $y_1 = 0$, $y_1 = -1$, and $y_1 = \infty$. One can explore the geometry around the puncture $y_1 = 0$ by deforming and making appropriate coordinate changes to rewrite the equation \eqref{eqn:blah} as
\begin{equation}
  \begin{aligned}
    f &= \widetilde{w}^2 + \widetilde{x}_1^3 + \widetilde{x}_2^2\rho + \rho^2 + \left( \frac{m_1}{y_1^2} + \frac{m_1^\prime}{y_1}\right) \widetilde{x}_1 \widetilde{x}_2^2  + \left( \frac{m_2}{y_1^5} + \frac{m_2^\prime}{y_1^4}\right) \widetilde{x}_1 \widetilde{x}_2 \cr 
    &\quad + \left( \frac{m_3}{y_1^6} + \frac{u_1}{y_1^5} +  \frac{m_3^\prime}{y_1^4}\right) \widetilde{x}_2^2
    + \left( \frac{m_4}{y_1^8} + \frac{u_2}{y_1^7} +  \frac{m_4^\prime}{y_1^4}\right) \widetilde{x}_1 \cr
    &\quad + \left( \frac{m_5}{y_1^9} + \frac{u_3}{y_1^8} +  \frac{m_5^\prime}{y_1^7}\right) \widetilde{x}_2 + \left( \frac{m_6}{y_1^{12}} + \frac{u_4}{y_1^{11}} +  \frac{m_6^\prime}{y_1^9}\right) = 0 \,.
  \end{aligned}
\end{equation}
The subleading pole in each term captures the contribution to the Coulomb branch from the puncture at $y_1 = 0$, and we can directly observe that this pole structure is $(1,4,5,7,8,11)$. This is identical to the pole structure associated to a full $\mathfrak{e}_6$ puncture in \cite{Chacaltana:2014jba}. Since the puncture at $y_1 = -1$ is equivalent to the puncture at $y_2 = 0$, the pole structure at $y_1=0$ and $y_2=0$ are identical by symmetry. A similar analysis around $y_1 = \infty$ reveals a pole structure $(1,1,2,2,2,3)$, which is identical to the pole structure of the simple puncture. Then, we see directly from the mirror geometry a sphere with two full punctures and a simple puncture. We are thus observing the data of the 6d $(2,0)$ origin, the punctured Riemann surface, directly from the mirror geometry of the elliptically fibered Calabi--Yau manifold associated to the 6d $(1,0)$ origin.

It would be interesting to extend this approach to construct the mirror manifolds to the geometries describing the 6d $(1,0)$ theories $\mathcal{T}_\mathfrak{g}\{Y_1, Y_2\}$ that were explored in this paper and to observe the puncture data for the 6d $(2,0)$ description explicitly. This would provide the string-theoretic underpinning for the plurality of 6d origins (as in equation \eqref{eqn:big}) established in this paper.

\subsection*{Acknowledgements}

We thank Philip Argyres, Jacques Distler, Jonathan Heckman, and Mario Martone for helpful discussions. F.B.~thanks the Instituto de F\'isica Te\'orica de Madrid for hospitality. C.L.~thanks the Center for Theoretical Physics of the Universe at the Institute for Basic Science for hospitality during the final stages of this work.
F.B.~is supported by the Swiss National Science Foundation (SNSF) grant number P400P2\_194341.
M.J.K.~is supported by a Sherman Fairchild Postdoctoral Fellowship and the National Research Foundation of Korea (NRF) grants NRF-2020R1C1C1007591 and NRF-2020R1A4A3079707. This material is based upon work supported by the U.S. Department of Energy, Office of Science, Office of High Energy Physics, under Award Number DE-SC0011632.
The work of C.L.~was supported by a University Research Foundation grant at the University of Pennsylvania and DOE (HEP) Award DE-SC0021484.

\bibliography{references}{}

\providecommand{\href}[2]{#2}\begingroup\raggedright\begin{thebibliography}{10}

\bibitem{Alday:2009aq}
L.~F. Alday, D.~Gaiotto, and Y.~Tachikawa, ``{Liouville Correlation Functions
  from Four-dimensional Gauge Theories},''
  \href{http://dx.doi.org/10.1007/s11005-010-0369-5}{{\em Lett. Math. Phys.}
  {\bfseries 91} (2010) 167--197},
  \href{http://arxiv.org/abs/0906.3219}{{\ttfamily arXiv:0906.3219 [hep-th]}}.

\bibitem{AlvarezGaume:1983ig}
L.~Alvarez-Gaume and E.~Witten, ``{Gravitational Anomalies},''
  \href{http://dx.doi.org/10.1016/0550-3213(84)90066-X}{{\em Nucl. Phys. B}
  {\bfseries 234} (1984) 269}.

\bibitem{Apruzzi:2020eqi}
F.~Apruzzi, M.~Fazzi, J.~J. Heckman, T.~Rudelius, and H.~Y. Zhang, ``{General
  prescription for global $U(1)$s in 6D SCFTs},''
  \href{http://dx.doi.org/10.1103/PhysRevD.101.086023}{{\em Phys. Rev. D}
  {\bfseries 101} no.~8, (2020) 086023},
  \href{http://arxiv.org/abs/2001.10549}{{\ttfamily arXiv:2001.10549
  [hep-th]}}.

\bibitem{Apruzzi:2016nfr}
F.~Apruzzi, F.~Hassler, J.~J. Heckman, and I.~V. Melnikov, ``{From 6D SCFTs to
  Dynamic GLSMs},'' \href{http://dx.doi.org/10.1103/PhysRevD.96.066015}{{\em
  Phys. Rev. D} {\bfseries 96} no.~6, (2017) 066015},
  \href{http://arxiv.org/abs/1610.00718}{{\ttfamily arXiv:1610.00718
  [hep-th]}}.

\bibitem{Apruzzi:2018oge}
F.~Apruzzi, J.~J. Heckman, D.~R. Morrison, and L.~Tizzano, ``{4D Gauge Theories
  with Conformal Matter},''
  \href{http://dx.doi.org/10.1007/JHEP09(2018)088}{{\em JHEP} {\bfseries 09}
  (2018) 088}, \href{http://arxiv.org/abs/1803.00582}{{\ttfamily
  arXiv:1803.00582 [hep-th]}}.

\bibitem{Argyres:2007cn}
P.~C. Argyres and N.~Seiberg, ``{S-duality in N=2 supersymmetric gauge
  theories},'' \href{http://dx.doi.org/10.1088/1126-6708/2007/12/088}{{\em
  JHEP} {\bfseries 12} (2007) 088},
  \href{http://arxiv.org/abs/0711.0054}{{\ttfamily arXiv:0711.0054 [hep-th]}}.

\bibitem{Arras:2016evy}
P.~Arras, A.~Grassi, and T.~Weigand, ``{Terminal Singularities, Milnor Numbers,
  and Matter in F-theory},''
  \href{http://dx.doi.org/10.1016/j.geomphys.2017.09.001}{{\em J. Geom. Phys.}
  {\bfseries 123} (2018) 71--97},
  \href{http://arxiv.org/abs/1612.05646}{{\ttfamily arXiv:1612.05646
  [hep-th]}}.

\bibitem{MR417306}
P.~Bala and R.~W. Carter, ``Classes of unipotent elements in simple algebraic
  groups. {I},'' \href{http://dx.doi.org/10.1017/S0305004100052403}{{\em Math.
  Proc. Cambridge Philos. Soc.} {\bfseries 79} no.~3, (1976) 401--425}.
  \url{https://doi.org/10.1017/S0305004100052403}.

\bibitem{MR417307}
P.~Bala and R.~W. Carter, ``Classes of unipotent elements in simple algebraic
  groups. {II},'' \href{http://dx.doi.org/10.1017/S0305004100052610}{{\em Math.
  Proc. Cambridge Philos. Soc.} {\bfseries 80} no.~1, (1976) 1--17}.
  \url{https://doi.org/10.1017/S0305004100052610}.

\bibitem{Baume:2020ure}
F.~Baume, J.~J. Heckman, and C.~Lawrie, ``{6D SCFTs, 4D SCFTs, Conformal
  Matter, and Spin Chains},''
  \href{http://dx.doi.org/10.1016/j.nuclphysb.2021.115401}{{\em Nucl. Phys. B}
  {\bfseries 967} (2021) 115401},
  \href{http://arxiv.org/abs/2007.07262}{{\ttfamily arXiv:2007.07262
  [hep-th]}}.

\bibitem{Beccaria:2015ypa}
M.~Beccaria and A.~A. Tseytlin, ``{Conformal anomaly c-coefficients of
  superconformal 6d theories},''
  \href{http://dx.doi.org/10.1007/JHEP01(2016)001}{{\em JHEP} {\bfseries 01}
  (2016) 001}, \href{http://arxiv.org/abs/1510.02685}{{\ttfamily
  arXiv:1510.02685 [hep-th]}}.

\bibitem{Beem:2013sza}
C.~Beem, M.~Lemos, P.~Liendo, W.~Peelaers, L.~Rastelli, and B.~C. van Rees,
  ``{Infinite Chiral Symmetry in Four Dimensions},''
  \href{http://dx.doi.org/10.1007/s00220-014-2272-x}{{\em Commun. Math. Phys.}
  {\bfseries 336} no.~3, (2015) 1359--1433},
  \href{http://arxiv.org/abs/1312.5344}{{\ttfamily arXiv:1312.5344 [hep-th]}}.

\bibitem{Beem:2014zpa}
C.~Beem, M.~Lemos, P.~Liendo, L.~Rastelli, and B.~C. van Rees, ``{The $
  \mathcal{N}=2 $ superconformal bootstrap},''
  \href{http://dx.doi.org/10.1007/JHEP03(2016)183}{{\em JHEP} {\bfseries 03}
  (2016) 183}, \href{http://arxiv.org/abs/1412.7541}{{\ttfamily arXiv:1412.7541
  [hep-th]}}.

\bibitem{Bershadsky:1997sb}
M.~Bershadsky and C.~Vafa, ``{Global anomalies and geometric engineering of
  critical theories in six-dimensions},''
  \href{http://arxiv.org/abs/hep-th/9703167}{{\ttfamily arXiv:hep-th/9703167}}.

\bibitem{Bertolini:2015bwa}
M.~Bertolini, P.~R. Merkx, and D.~R. Morrison, ``{On the global symmetries of
  6D superconformal field theories},''
  \href{http://dx.doi.org/10.1007/JHEP07(2016)005}{{\em JHEP} {\bfseries 07}
  (2016) 005}, \href{http://arxiv.org/abs/1510.08056}{{\ttfamily
  arXiv:1510.08056 [hep-th]}}.

\bibitem{Bobev:2017uzs}
N.~Bobev and P.~M. Crichigno, ``{Universal RG Flows Across Dimensions and
  Holography},'' \href{http://dx.doi.org/10.1007/JHEP12(2017)065}{{\em JHEP}
  {\bfseries 12} (2017) 065}, \href{http://arxiv.org/abs/1708.05052}{{\ttfamily
  arXiv:1708.05052 [hep-th]}}.

\bibitem{Bobev:2016phc}
N.~Bobev, G.~Dibitetto, F.~F. Gautason, and B.~Truijen, ``{Holography, Brane
  Intersections and Six-dimensional SCFTs},''
  \href{http://dx.doi.org/10.1007/JHEP02(2017)116}{{\em JHEP} {\bfseries 02}
  (2017) 116}, \href{http://arxiv.org/abs/1612.06324}{{\ttfamily
  arXiv:1612.06324 [hep-th]}}.

\bibitem{Chacaltana:2010ks}
O.~Chacaltana and J.~Distler, ``{Tinkertoys for Gaiotto Duality},''
  \href{http://dx.doi.org/10.1007/JHEP11(2010)099}{{\em JHEP} {\bfseries 11}
  (2010) 099}, \href{http://arxiv.org/abs/1008.5203}{{\ttfamily arXiv:1008.5203
  [hep-th]}}.

\bibitem{Chacaltana:2012zy}
O.~Chacaltana, J.~Distler, and Y.~Tachikawa, ``{Nilpotent orbits and
  codimension-two defects of 6d N=(2,0) theories},''
  \href{http://dx.doi.org/10.1142/S0217751X1340006X}{{\em Int. J. Mod. Phys. A}
  {\bfseries 28} (2013) 1340006},
  \href{http://arxiv.org/abs/1203.2930}{{\ttfamily arXiv:1203.2930 [hep-th]}}.

\bibitem{Chacaltana:2014jba}
O.~Chacaltana, J.~Distler, and A.~Trimm, ``{Tinkertoys for the E$_{6}$
  theory},'' \href{http://dx.doi.org/10.1007/JHEP09(2015)007}{{\em JHEP}
  {\bfseries 09} (2015) 007}, \href{http://arxiv.org/abs/1403.4604}{{\ttfamily
  arXiv:1403.4604 [hep-th]}}.

\bibitem{Chacaltana:2017boe}
O.~Chacaltana, J.~Distler, A.~Trimm, and Y.~Zhu, ``{Tinkertoys for the E$_{7}$
  theory},'' \href{http://dx.doi.org/10.1007/JHEP05(2018)031}{{\em JHEP}
  {\bfseries 05} (2018) 031}, \href{http://arxiv.org/abs/1704.07890}{{\ttfamily
  arXiv:1704.07890 [hep-th]}}.

\bibitem{Chcaltana:2018zag}
O.~Chacaltana, J.~Distler, A.~Trimm, and Y.~Zhu, ``{Tinkertoys for the $E_8$
  Theory},'' \href{http://arxiv.org/abs/1802.09626}{{\ttfamily arXiv:1802.09626
  [hep-th]}}.

\bibitem{Cheung:1997id}
Y.-K.~E. Cheung, O.~J. Ganor, and M.~Krogh, ``{Correlators of the global
  symmetry currents of 4-D and 6-D superconformal theories},''
  \href{http://dx.doi.org/10.1016/S0550-3213(98)00139-4}{{\em Nucl. Phys. B}
  {\bfseries 523} (1998) 171--192},
  \href{http://arxiv.org/abs/hep-th/9710053}{{\ttfamily arXiv:hep-th/9710053}}.

\bibitem{MR1251060}
D.~H. Collingwood and W.~M. McGovern, {\em Nilpotent orbits in semisimple {L}ie
  algebras}.
\newblock Van Nostrand Reinhold Mathematics Series. Van Nostrand Reinhold Co.,
  New York, 1993.

\bibitem{Cordova:2015fha}
C.~Cordova, T.~T. Dumitrescu, and K.~Intriligator, ``{Anomalies,
  renormalization group flows, and the a-theorem in six-dimensional (1, 0)
  theories},'' \href{http://dx.doi.org/10.1007/JHEP10(2016)080}{{\em JHEP}
  {\bfseries 10} (2016) 080}, \href{http://arxiv.org/abs/1506.03807}{{\ttfamily
  arXiv:1506.03807 [hep-th]}}.

\bibitem{Cordova:2020tij}
C.~Cordova, T.~T. Dumitrescu, and K.~Intriligator, ``{2-Group Global Symmetries
  and Anomalies in Six-Dimensional Quantum Field Theories},''
  \href{http://arxiv.org/abs/2009.00138}{{\ttfamily arXiv:2009.00138
  [hep-th]}}.

\bibitem{Cremonesi:2015bld}
S.~Cremonesi and A.~Tomasiello, ``{6d holographic anomaly match as a continuum
  limit},'' \href{http://dx.doi.org/10.1007/JHEP05(2016)031}{{\em JHEP}
  {\bfseries 05} (2016) 031}, \href{http://arxiv.org/abs/1512.02225}{{\ttfamily
  arXiv:1512.02225 [hep-th]}}.

\bibitem{Cvetic:2012xn}
M.~Cvetic, T.~W. Grimm, and D.~Klevers, ``{Anomaly Cancellation And Abelian
  Gauge Symmetries In F-theory},''
  \href{http://dx.doi.org/10.1007/JHEP02(2013)101}{{\em JHEP} {\bfseries 02}
  (2013) 101}, \href{http://arxiv.org/abs/1210.6034}{{\ttfamily arXiv:1210.6034
  [hep-th]}}.

\bibitem{DelZotto:2014fia}
M.~Del~Zotto, J.~J. Heckman, D.~R. Morrison, and D.~S. Park, ``{6D SCFTs and
  Gravity},'' \href{http://dx.doi.org/10.1007/JHEP06(2015)158}{{\em JHEP}
  {\bfseries 06} (2015) 158}, \href{http://arxiv.org/abs/1412.6526}{{\ttfamily
  arXiv:1412.6526 [hep-th]}}.

\bibitem{DelZotto:2014hpa}
M.~Del~Zotto, J.~J. Heckman, A.~Tomasiello, and C.~Vafa, ``{6d Conformal
  Matter},'' \href{http://dx.doi.org/10.1007/JHEP02(2015)054}{{\em JHEP}
  {\bfseries 02} (2015) 054}, \href{http://arxiv.org/abs/1407.6359}{{\ttfamily
  arXiv:1407.6359 [hep-th]}}.

\bibitem{DelZotto:2015rca}
M.~Del~Zotto, C.~Vafa, and D.~Xie, ``{Geometric engineering, mirror symmetry
  and $ 6{\mathrm{d}}_{\left(1,0\right)}\to
  4{\mathrm{d}}_{\left(\mathcal{N}=2\right)} $},''
  \href{http://dx.doi.org/10.1007/JHEP11(2015)123}{{\em JHEP} {\bfseries 11}
  (2015) 123}, \href{http://arxiv.org/abs/1504.08348}{{\ttfamily
  arXiv:1504.08348 [hep-th]}}.

\bibitem{Distler:2018gbc}
J.~Distler and B.~Ergun, ``{Product SCFTs for the $E_7$ Theory},''
  \href{http://arxiv.org/abs/1803.02425}{{\ttfamily arXiv:1803.02425
  [hep-th]}}.

\bibitem{Distler:2020tub}
J.~Distler, B.~Ergun, and A.~Shehper, ``{Distinguishing $d = 4 N$ = 2 SCFTs},''
  \href{http://arxiv.org/abs/2012.15249}{{\ttfamily arXiv:2012.15249
  [hep-th]}}.

\bibitem{Distler:2017xba}
J.~Distler, B.~Ergun, and F.~Yan, ``{Product SCFTs in Class-S},''
  \href{http://arxiv.org/abs/1711.04727}{{\ttfamily arXiv:1711.04727
  [hep-th]}}.

\bibitem{Esole:2017qeh}
M.~Esole, R.~Jagadeesan, and M.~J. Kang, ``{The Geometry of G$_2$, Spin(7), and
  Spin(8)-models},'' \href{http://arxiv.org/abs/1709.04913}{{\ttfamily
  arXiv:1709.04913 [hep-th]}}.

\bibitem{Esole:2019asj}
M.~Esole, R.~Jagadeesan, and M.~J. Kang, ``{48 Crepant Paths to
  $\text{SU}(2)\!\times\!\text{SU}(3)$},''
  \href{http://arxiv.org/abs/1905.05174}{{\ttfamily arXiv:1905.05174
  [hep-th]}}.

\bibitem{Esole:2017rgz}
M.~Esole, P.~Jefferson, and M.~J. Kang, ``{The Geometry of F$_4$-Models},''
  \href{http://arxiv.org/abs/1704.08251}{{\ttfamily arXiv:1704.08251
  [hep-th]}}.

\bibitem{Esole:2018csl}
M.~Esole and M.~J. Kang, ``{Flopping and slicing: SO(4) and Spin(4)-models},''
  \href{http://dx.doi.org/10.4310/ATMP.2019.v23.n4.a2}{{\em Adv. Theor. Math.
  Phys.} {\bfseries 23} no.~4, (2019) 1003--1066},
  \href{http://arxiv.org/abs/1802.04802}{{\ttfamily arXiv:1802.04802
  [hep-th]}}.

\bibitem{Esole:2018mqb}
M.~Esole and M.~J. Kang, ``{The Geometry of the SU(2)$\times$ G$_2$-model},''
  \href{http://dx.doi.org/10.1007/JHEP02(2019)091}{{\em JHEP} {\bfseries 02}
  (2019) 091}, \href{http://arxiv.org/abs/1805.03214}{{\ttfamily
  arXiv:1805.03214 [hep-th]}}.

\bibitem{Esole:2020tby}
M.~Esole and M.~J. Kang, ``{Matter representations from geometry: under the
  spell of Dynkin},'' \href{http://arxiv.org/abs/2012.13401}{{\ttfamily
  arXiv:2012.13401 [hep-th]}}.

\bibitem{Esole:2017hlw}
M.~Esole, M.~J. Kang, and S.-T. Yau, ``{Mordell--Weil Torsion, Anomalies, and
  Phase Transitions},'' \href{http://arxiv.org/abs/1712.02337}{{\ttfamily
  arXiv:1712.02337 [hep-th]}}.

\bibitem{Esole:2015xfa}
M.~Esole and S.-H. Shao, ``{M-theory on Elliptic Calabi--Yau Threefolds and 6d
  Anomalies},'' \href{http://arxiv.org/abs/1504.01387}{{\ttfamily
  arXiv:1504.01387 [hep-th]}}.

\bibitem{Gaiotto:2009we}
D.~Gaiotto, ``{N=2 dualities},''
  \href{http://dx.doi.org/10.1007/JHEP08(2012)034}{{\em JHEP} {\bfseries 08}
  (2012) 034}, \href{http://arxiv.org/abs/0904.2715}{{\ttfamily arXiv:0904.2715
  [hep-th]}}.

\bibitem{Gaiotto:2009hg}
D.~Gaiotto, G.~W. Moore, and A.~Neitzke, ``{Wall-crossing, Hitchin Systems, and
  the WKB Approximation},'' \href{http://arxiv.org/abs/0907.3987}{{\ttfamily
  arXiv:0907.3987 [hep-th]}}.

\bibitem{Gaiotto:2012uq}
D.~Gaiotto and S.~S. Razamat, ``{Exceptional Indices},''
  \href{http://dx.doi.org/10.1007/JHEP05(2012)145}{{\em JHEP} {\bfseries 05}
  (2012) 145}, \href{http://arxiv.org/abs/1203.5517}{{\ttfamily arXiv:1203.5517
  [hep-th]}}.

\bibitem{Gaiotto:2014lca}
D.~Gaiotto and A.~Tomasiello, ``{Holography for (1,0) theories in six
  dimensions},'' \href{http://dx.doi.org/10.1007/JHEP12(2014)003}{{\em JHEP}
  {\bfseries 12} (2014) 003}, \href{http://arxiv.org/abs/1404.0711}{{\ttfamily
  arXiv:1404.0711 [hep-th]}}.

\bibitem{Ganor:1996mu}
O.~J. Ganor and A.~Hanany, ``{Small E(8) instantons and tensionless noncritical
  strings},'' \href{http://dx.doi.org/10.1016/0550-3213(96)00243-X}{{\em Nucl.
  Phys. B} {\bfseries 474} (1996) 122--140},
  \href{http://arxiv.org/abs/hep-th/9602120}{{\ttfamily arXiv:hep-th/9602120}}.

\bibitem{Giacomelli:2020jel}
S.~Giacomelli, C.~Meneghelli, and W.~Peelaers, ``{New N=2 superconformal field
  theories from S-folds},'' \href{http://arxiv.org/abs/2007.00647}{{\ttfamily
  arXiv:2007.00647 [hep-th]}}.

\bibitem{Grassi:2011hq}
A.~Grassi and D.~R. Morrison, ``{Anomalies and the Euler characteristic of
  elliptic Calabi--Yau threefolds},''
  \href{http://dx.doi.org/10.4310/CNTP.2012.v6.n1.a2}{{\em Commun. Num. Theor.
  Phys.} {\bfseries 6} (2012) 51--127},
  \href{http://arxiv.org/abs/1109.0042}{{\ttfamily arXiv:1109.0042 [hep-th]}}.

\bibitem{Green:1984sg}
M.~B. Green and J.~H. Schwarz, ``{Anomaly Cancellation in Supersymmetric D=10
  Gauge Theory and Superstring Theory},''
  \href{http://dx.doi.org/10.1016/0370-2693(84)91565-X}{{\em Phys. Lett. B}
  {\bfseries 149} (1984) 117--122}.

\bibitem{Green:1984bx}
M.~B. Green, J.~H. Schwarz, and P.~C. West, ``{Anomaly Free Chiral Theories in
  Six-Dimensions},'' \href{http://dx.doi.org/10.1016/0550-3213(85)90222-6}{{\em
  Nucl. Phys. B} {\bfseries 254} (1985) 327--348}.

\bibitem{Grimm:2012yq}
T.~W. Grimm and W.~Taylor, ``{Structure in 6D and 4D N=1 supergravity theories
  from F-theory},'' \href{http://dx.doi.org/10.1007/JHEP10(2012)105}{{\em JHEP}
  {\bfseries 10} (2012) 105}, \href{http://arxiv.org/abs/1204.3092}{{\ttfamily
  arXiv:1204.3092 [hep-th]}}.

\bibitem{Hassler:2019eso}
F.~Hassler, J.~J. Heckman, T.~B. Rochais, T.~Rudelius, and H.~Y. Zhang,
  ``{T-Branes, String Junctions, and 6D SCFTs},''
  \href{http://dx.doi.org/10.1103/PhysRevD.101.086018}{{\em Phys. Rev. D}
  {\bfseries 101} no.~8, (2020) 086018},
  \href{http://arxiv.org/abs/1907.11230}{{\ttfamily arXiv:1907.11230
  [hep-th]}}.

\bibitem{Heckman:2020otd}
J.~J. Heckman, ``{Qubit Construction in 6D SCFTs},''
  \href{http://dx.doi.org/10.1016/j.physletb.2020.135891}{{\em Phys. Lett. B}
  {\bfseries 811} (2020) 135891},
  \href{http://arxiv.org/abs/2007.08545}{{\ttfamily arXiv:2007.08545
  [hep-th]}}.

\bibitem{Heckman:2015bfa}
J.~J. Heckman, D.~R. Morrison, T.~Rudelius, and C.~Vafa, ``{Atomic
  Classification of 6D SCFTs},''
  \href{http://dx.doi.org/10.1002/prop.201500024}{{\em Fortsch. Phys.}
  {\bfseries 63} (2015) 468--530},
  \href{http://arxiv.org/abs/1502.05405}{{\ttfamily arXiv:1502.05405
  [hep-th]}}.

\bibitem{Heckman:2013pva}
J.~J. Heckman, D.~R. Morrison, and C.~Vafa, ``{On the Classification of 6D
  SCFTs and Generalized ADE Orbifolds},''
  \href{http://dx.doi.org/10.1007/JHEP05(2014)028}{{\em JHEP} {\bfseries 05}
  (2014) 028}, \href{http://arxiv.org/abs/1312.5746}{{\ttfamily arXiv:1312.5746
  [hep-th]}}. [Erratum: JHEP 06, 017 (2015)].

\bibitem{Heckman:2018jxk}
J.~J. Heckman and T.~Rudelius, ``{Top Down Approach to 6D SCFTs},''
  \href{http://dx.doi.org/10.1088/1751-8121/aafc81}{{\em J. Phys. A} {\bfseries
  52} no.~9, (2019) 093001}, \href{http://arxiv.org/abs/1805.06467}{{\ttfamily
  arXiv:1805.06467 [hep-th]}}.

\bibitem{Heckman:2016ssk}
J.~J. Heckman, T.~Rudelius, and A.~Tomasiello, ``{6D RG Flows and Nilpotent
  Hierarchies},'' \href{http://dx.doi.org/10.1007/JHEP07(2016)082}{{\em JHEP}
  {\bfseries 07} (2016) 082}, \href{http://arxiv.org/abs/1601.04078}{{\ttfamily
  arXiv:1601.04078 [hep-th]}}.

\bibitem{Heckman:2018pqx}
J.~J. Heckman, T.~Rudelius, and A.~Tomasiello, ``{Fission, Fusion, and 6D RG
  Flows},'' \href{http://dx.doi.org/10.1007/JHEP02(2019)167}{{\em JHEP}
  {\bfseries 02} (2019) 167}, \href{http://arxiv.org/abs/1807.10274}{{\ttfamily
  arXiv:1807.10274 [hep-th]}}.

\bibitem{Intriligator:2014eaa}
K.~Intriligator, ``{6d, $ \mathcal{N}=\left(1,\;0\right) $ Coulomb branch
  anomaly matching},'' \href{http://dx.doi.org/10.1007/JHEP10(2014)162}{{\em
  JHEP} {\bfseries 10} (2014) 162},
  \href{http://arxiv.org/abs/1408.6745}{{\ttfamily arXiv:1408.6745 [hep-th]}}.

\bibitem{Katz:1996xe}
S.~H. Katz and C.~Vafa, ``{Matter from geometry},''
  \href{http://dx.doi.org/10.1016/S0550-3213(97)00280-0}{{\em Nucl. Phys. B}
  {\bfseries 497} (1997) 146--154},
  \href{http://arxiv.org/abs/hep-th/9606086}{{\ttfamily arXiv:hep-th/9606086}}.

\bibitem{Kim:2018lfo}
H.-C. Kim, S.~S. Razamat, C.~Vafa, and G.~Zafrir, ``{Compactifications of ADE
  conformal matter on a torus},''
  \href{http://dx.doi.org/10.1007/JHEP09(2018)110}{{\em JHEP} {\bfseries 09}
  (2018) 110}, \href{http://arxiv.org/abs/1806.07620}{{\ttfamily
  arXiv:1806.07620 [hep-th]}}.

\bibitem{Kim:2018bpg}
H.-C. Kim, S.~S. Razamat, C.~Vafa, and G.~Zafrir, ``{D-type Conformal Matter
  and SU/USp Quivers},'' \href{http://dx.doi.org/10.1007/JHEP06(2018)058}{{\em
  JHEP} {\bfseries 06} (2018) 058},
  \href{http://arxiv.org/abs/1802.00620}{{\ttfamily arXiv:1802.00620
  [hep-th]}}.

\bibitem{LeFloch:2020uop}
B.~Le~Floch, ``{A slow review of the AGT correspondence},''
  \href{http://arxiv.org/abs/2006.14025}{{\ttfamily arXiv:2006.14025
  [hep-th]}}.

\bibitem{Lee:2018ihr}
S.-J. Lee, D.~Regalado, and T.~Weigand, ``{6d SCFTs and U(1) Flavour
  Symmetries},'' \href{http://dx.doi.org/10.1007/JHEP11(2018)147}{{\em JHEP}
  {\bfseries 11} (2018) 147}, \href{http://arxiv.org/abs/1803.07998}{{\ttfamily
  arXiv:1803.07998 [hep-th]}}.

\bibitem{Mekareeya:2017sqh}
N.~Mekareeya, K.~Ohmori, H.~Shimizu, and A.~Tomasiello, ``{Small instanton
  transitions for M5 fractions},''
  \href{http://dx.doi.org/10.1007/JHEP10(2017)055}{{\em JHEP} {\bfseries 10}
  (2017) 055}, \href{http://arxiv.org/abs/1707.05785}{{\ttfamily
  arXiv:1707.05785 [hep-th]}}.

\bibitem{Mekareeya:2017jgc}
N.~Mekareeya, K.~Ohmori, Y.~Tachikawa, and G.~Zafrir, ``{E$_{8}$ instantons on
  type-A ALE spaces and supersymmetric field theories},''
  \href{http://dx.doi.org/10.1007/JHEP09(2017)144}{{\em JHEP} {\bfseries 09}
  (2017) 144}, \href{http://arxiv.org/abs/1707.04370}{{\ttfamily
  arXiv:1707.04370 [hep-th]}}.

\bibitem{Mekareeya:2016yal}
N.~Mekareeya, T.~Rudelius, and A.~Tomasiello, ``{T-branes, Anomalies and Moduli
  Spaces in 6D SCFTs},'' \href{http://dx.doi.org/10.1007/JHEP10(2017)158}{{\em
  JHEP} {\bfseries 10} (2017) 158},
  \href{http://arxiv.org/abs/1612.06399}{{\ttfamily arXiv:1612.06399
  [hep-th]}}.

\bibitem{Merkx:2017jey}
P.~R. Merkx, ``{Classifying Global Symmetries of 6D SCFTs},''
  \href{http://dx.doi.org/10.1007/JHEP03(2018)163}{{\em JHEP} {\bfseries 03}
  (2018) 163}, \href{http://arxiv.org/abs/1711.05155}{{\ttfamily
  arXiv:1711.05155 [hep-th]}}.

\bibitem{Monnier:2018nfs}
S.~Monnier and G.~W. Moore, ``{Remarks on the Green\textendash{}Schwarz Terms
  of Six-Dimensional Supergravity Theories},''
  \href{http://dx.doi.org/10.1007/s00220-019-03341-7}{{\em Commun. Math. Phys.}
  {\bfseries 372} no.~3, (2019) 963--1025},
  \href{http://arxiv.org/abs/1808.01334}{{\ttfamily arXiv:1808.01334
  [hep-th]}}.

\bibitem{Monnier:2017oqd}
S.~Monnier, G.~W. Moore, and D.~S. Park, ``{Quantization of anomaly
  coefficients in 6D $\mathcal{N}=(1,0)$ supergravity},''
  \href{http://dx.doi.org/10.1007/JHEP02(2018)020}{{\em JHEP} {\bfseries 02}
  (2018) 020}, \href{http://arxiv.org/abs/1711.04777}{{\ttfamily
  arXiv:1711.04777 [hep-th]}}.

\bibitem{Morrison:2016djb}
D.~R. Morrison and T.~Rudelius, ``{F-theory and Unpaired Tensors in 6D SCFTs
  and LSTs},'' \href{http://dx.doi.org/10.1002/prop.201600069}{{\em Fortsch.
  Phys.} {\bfseries 64} (2016) 645--656},
  \href{http://arxiv.org/abs/1605.08045}{{\ttfamily arXiv:1605.08045
  [hep-th]}}.

\bibitem{Morrison:2012np}
D.~R. Morrison and W.~Taylor, ``{Classifying bases for 6D F-theory models},''
  \href{http://dx.doi.org/10.2478/s11534-012-0065-4}{{\em Central Eur. J.
  Phys.} {\bfseries 10} (2012) 1072--1088},
  \href{http://arxiv.org/abs/1201.1943}{{\ttfamily arXiv:1201.1943 [hep-th]}}.

\bibitem{Morrison:1996na}
D.~R. Morrison and C.~Vafa, ``{Compactifications of F-theory on Calabi--Yau
  threefolds. 1},'' \href{http://dx.doi.org/10.1016/0550-3213(96)00242-8}{{\em
  Nucl. Phys. B} {\bfseries 473} (1996) 74--92},
  \href{http://arxiv.org/abs/hep-th/9602114}{{\ttfamily arXiv:hep-th/9602114}}.

\bibitem{Morrison:1996pp}
D.~R. Morrison and C.~Vafa, ``{Compactifications of F-theory on Calabi--Yau
  threefolds. 2.},'' \href{http://dx.doi.org/10.1016/0550-3213(96)00369-0}{{\em
  Nucl. Phys. B} {\bfseries 476} (1996) 437--469},
  \href{http://arxiv.org/abs/hep-th/9603161}{{\ttfamily arXiv:hep-th/9603161}}.

\bibitem{Ohmori:2015tka}
K.~Ohmori and H.~Shimizu, ``{$S^1/T^2$ compactifications of 6d $
  \mathcal{N}=\left(1,\;0\right) $ theories and brane webs},''
  \href{http://dx.doi.org/10.1007/JHEP03(2016)024}{{\em JHEP} {\bfseries 03}
  (2016) 024}, \href{http://arxiv.org/abs/1509.03195}{{\ttfamily
  arXiv:1509.03195 [hep-th]}}.

\bibitem{Ohmori:2014pca}
K.~Ohmori, H.~Shimizu, and Y.~Tachikawa, ``{Anomaly polynomial of E-string
  theories},'' \href{http://dx.doi.org/10.1007/JHEP08(2014)002}{{\em JHEP}
  {\bfseries 08} (2014) 002}, \href{http://arxiv.org/abs/1404.3887}{{\ttfamily
  arXiv:1404.3887 [hep-th]}}.

\bibitem{Ohmori:2014kda}
K.~Ohmori, H.~Shimizu, Y.~Tachikawa, and K.~Yonekura, ``{Anomaly polynomial of
  general 6d SCFTs},'' \href{http://dx.doi.org/10.1093/ptep/ptu140}{{\em PTEP}
  {\bfseries 2014} no.~10, (2014) 103B07},
  \href{http://arxiv.org/abs/1408.5572}{{\ttfamily arXiv:1408.5572 [hep-th]}}.

\bibitem{Ohmori:2015pia}
K.~Ohmori, H.~Shimizu, Y.~Tachikawa, and K.~Yonekura, ``{6d
  $\mathcal{N}=\left(1,\;0\right) $ theories on S$^{1}$ /T$^{2}$ and class S
  theories: part II},'' \href{http://dx.doi.org/10.1007/JHEP12(2015)131}{{\em
  JHEP} {\bfseries 12} (2015) 131},
  \href{http://arxiv.org/abs/1508.00915}{{\ttfamily arXiv:1508.00915
  [hep-th]}}.

\bibitem{Ohmori:2015pua}
K.~Ohmori, H.~Shimizu, Y.~Tachikawa, and K.~Yonekura\sout{}, ``{6d
  $\mathcal{N}=(1,0)$ theories on $T^2$ and class S theories: Part I},''
  \href{http://dx.doi.org/10.1007/JHEP07(2015)014}{{\em JHEP} {\bfseries 07}
  (2015) 014}, \href{http://arxiv.org/abs/1503.06217}{{\ttfamily
  arXiv:1503.06217 [hep-th]}}.

\bibitem{Ohmori:2018ona}
K.~Ohmori, Y.~Tachikawa, and G.~Zafrir, ``{Compactifications of 6d $N = (1, 0)$
  SCFTs with non-trivial Stiefel-Whitney classes},''
  \href{http://dx.doi.org/10.1007/JHEP04(2019)006}{{\em JHEP} {\bfseries 04}
  (2019) 006}, \href{http://arxiv.org/abs/1812.04637}{{\ttfamily
  arXiv:1812.04637 [hep-th]}}.

\bibitem{Osborn:1993cr}
H.~Osborn and A.~Petkou, ``{Implications of conformal invariance in field
  theories for general dimensions},''
  \href{http://dx.doi.org/10.1006/aphy.1994.1045}{{\em Annals Phys.} {\bfseries
  231} (1994) 311--362}, \href{http://arxiv.org/abs/hep-th/9307010}{{\ttfamily
  arXiv:hep-th/9307010}}.

\bibitem{Park:2011ji}
D.~S. Park, ``{Anomaly Equations and Intersection Theory},''
  \href{http://dx.doi.org/10.1007/JHEP01(2012)093}{{\em JHEP} {\bfseries 01}
  (2012) 093}, \href{http://arxiv.org/abs/1111.2351}{{\ttfamily arXiv:1111.2351
  [hep-th]}}.

\bibitem{Park:2011wv}
D.~S. Park and W.~Taylor, ``{Constraints on 6D Supergravity Theories with
  Abelian Gauge Symmetry},''
  \href{http://dx.doi.org/10.1007/JHEP01(2012)141}{{\em JHEP} {\bfseries 01}
  (2012) 141}, \href{http://arxiv.org/abs/1110.5916}{{\ttfamily arXiv:1110.5916
  [hep-th]}}.

\bibitem{Pasquetti:2019hxf}
S.~Pasquetti, S.~S. Razamat, M.~Sacchi, and G.~Zafrir, ``{Rank $Q$ E-string on
  a torus with flux},''
  \href{http://dx.doi.org/10.21468/SciPostPhys.8.1.014}{{\em SciPost Phys.}
  {\bfseries 8} no.~1, (2020) 014},
  \href{http://arxiv.org/abs/1908.03278}{{\ttfamily arXiv:1908.03278
  [hep-th]}}.

\bibitem{Sadov:1996zm}
V.~Sadov, ``{Generalized Green-Schwarz mechanism in F-theory},''
  \href{http://dx.doi.org/10.1016/0370-2693(96)01134-3}{{\em Phys. Lett. B}
  {\bfseries 388} (1996) 45--50},
  \href{http://arxiv.org/abs/hep-th/9606008}{{\ttfamily arXiv:hep-th/9606008}}.

\bibitem{Sagnotti:1992qw}
A.~Sagnotti, ``{A Note on the Green--Schwarz mechanism in open string
  theories},'' \href{http://dx.doi.org/10.1016/0370-2693(92)90682-T}{{\em Phys.
  Lett. B} {\bfseries 294} (1992) 196--203},
  \href{http://arxiv.org/abs/hep-th/9210127}{{\ttfamily arXiv:hep-th/9210127}}.

\bibitem{Seiberg:1996vs}
N.~Seiberg and E.~Witten, ``{Comments on string dynamics in six-dimensions},''
  \href{http://dx.doi.org/10.1016/0550-3213(96)00189-7}{{\em Nucl. Phys. B}
  {\bfseries 471} (1996) 121--134},
  \href{http://arxiv.org/abs/hep-th/9603003}{{\ttfamily arXiv:hep-th/9603003}}.

\bibitem{Shapere:2008zf}
A.~D. Shapere and Y.~Tachikawa, ``{Central charges of N=2 superconformal field
  theories in four dimensions},''
  \href{http://dx.doi.org/10.1088/1126-6708/2008/09/109}{{\em JHEP} {\bfseries
  09} (2008) 109}, \href{http://arxiv.org/abs/0804.1957}{{\ttfamily
  arXiv:0804.1957 [hep-th]}}.

\bibitem{Shimizu:2017kzs}
H.~Shimizu, Y.~Tachikawa, and G.~Zafrir, ``{Anomaly matching on the Higgs
  branch},'' \href{http://dx.doi.org/10.1007/JHEP12(2017)127}{{\em JHEP}
  {\bfseries 12} (2017) 127}, \href{http://arxiv.org/abs/1703.01013}{{\ttfamily
  arXiv:1703.01013 [hep-th]}}.

\bibitem{Tachikawa:2015bga}
Y.~Tachikawa, ``{A review of the $T_N$ theory and its cousins},''
  \href{http://dx.doi.org/10.1093/ptep/ptv098}{{\em PTEP} {\bfseries 2015}
  no.~11, (2015) 11B102}, \href{http://arxiv.org/abs/1504.01481}{{\ttfamily
  arXiv:1504.01481 [hep-th]}}.

\bibitem{Vafa:1996xn}
C.~Vafa, ``{Evidence for F-theory},''
  \href{http://dx.doi.org/10.1016/0550-3213(96)00172-1}{{\em Nucl. Phys. B}
  {\bfseries 469} (1996) 403--418},
  \href{http://arxiv.org/abs/hep-th/9602022}{{\ttfamily arXiv:hep-th/9602022}}.

\bibitem{Witten:1995zh}
E.~Witten, ``{Some comments on string dynamics},'' in {\em {STRINGS 95: Future
  Perspectives in String Theory}}, pp.~501--523.
\newblock 7, 1995.
\newblock \href{http://arxiv.org/abs/hep-th/9507121}{{\ttfamily
  arXiv:hep-th/9507121}}.

\bibitem{Witten:1996qb}
E.~Witten, ``{Phase transitions in M theory and F theory},''
  \href{http://dx.doi.org/10.1016/0550-3213(96)00212-X}{{\em Nucl. Phys. B}
  {\bfseries 471} (1996) 195--216},
  \href{http://arxiv.org/abs/hep-th/9603150}{{\ttfamily arXiv:hep-th/9603150}}.

\bibitem{Witten:1996qz}
E.~Witten, ``{Physical interpretation of certain strong coupling
  singularities},'' \href{http://dx.doi.org/10.1142/S0217732396002642}{{\em
  Mod. Phys. Lett. A} {\bfseries 11} (1996) 2649--2654},
  \href{http://arxiv.org/abs/hep-th/9609159}{{\ttfamily arXiv:hep-th/9609159}}.

\bibitem{Yankielowicz:2017xkf}
S.~Yankielowicz and Y.~Zhou, ``{Supersymmetric R\'enyi entropy and Anomalies in
  6d (1,0) SCFTs},'' \href{http://dx.doi.org/10.1007/JHEP04(2017)128}{{\em
  JHEP} {\bfseries 04} (2017) 128},
  \href{http://arxiv.org/abs/1702.03518}{{\ttfamily arXiv:1702.03518
  [hep-th]}}.

\bibitem{Zafrir:2018hkr}
G.~Zafrir, ``{On the torus compactifications of Z$_{2}$ orbifolds of E-string
  theories},'' \href{http://dx.doi.org/10.1007/JHEP10(2019)040}{{\em JHEP}
  {\bfseries 10} (2019) 040}, \href{http://arxiv.org/abs/1809.04260}{{\ttfamily
  arXiv:1809.04260 [hep-th]}}.

\end{thebibliography}\endgroup
\bibliographystyle{sortedbutpretty} 

\end{document}